\documentclass[aps,prb,twocolumn]{revtex4}
\usepackage{amsmath,amssymb,mathrsfs}
\usepackage{psfig}

\begin{document}

\title{Current-driven homogenization and effective medium parameters
  for finite samples}

\author{Vadim A. Markel} 
\affiliation{Departments of Radiology and
  Bioengineering and the Graduate Group in Applied Mathematics and
  Computational Science, University of Pennsylvania Philadelphia, PA
  19104}
\email{vmarkel@mail.med.upenn.edu}

\author{Igor Tsukerman} 
\affiliation{Department of Electrical and Computer Engineering,
  The University of Akron, OH 44325-3904}
\email{igor@uakron.edu}

\date{\today}

\begin{abstract}
  Reflection and refraction of electromagnetic waves by artificial
  periodic composites (metamaterials) can be accurately modeled by an
  effective medium theory only if the boundary of the medium is
  explicitly taken into account and the two effective parameters of
  the medium -- the index of refraction and the impedance -- are
  correctly determined. Theories that consider infinite periodic
  composites do not satisfy the above condition. As a result, they
  cannot model reflection and transmission by finite samples with the
  desired accuracy and are not useful for design of metamaterial-based
  devices. As an instructive case in point, we consider the
  ``current-driven'' homogenization theory, which has recently gained
  popularity. We apply this theory to the case of one-dimensional
  periodic medium wherein both exact and homogenization results can be
  obtained analytically in closed form.  We show that, beyond the
  well-understood zero-cell limit, the current-driven homogenization
  result is inconsistent with the exact reflection and transmission
  characteristics of the slab.
\end{abstract}

\maketitle

\section{Introduction}
\label{sec:intro}

In the past decade, interest in electromagnetic homogenization
theories has experienced a remarkable revival, especially when applied
to artificial periodic composites
(metamaterials)~\cite{simovski_09_1,simovski_11_1}. The ultimate goal
of any homogenization or effective medium theory (EMT) is to describe
reflection and refraction of waves by finite samples.  In the case of
homogeneous natural materials, an accurate description of this kind is
possible only if both the index of refraction and the impedance of the
material are known with sufficient precision. Correspondingly, the
majority of EMTs attempt to replace a periodic composite sample with a
sample of the same overall shape but spatially-uniform effective
refractive index and impedance and sharp boundaries, although in some
cases Drude transition layers are introduced or
considered~\cite{simovski_09_1}. However, when the EMTs are tested or
evaluated, the attention is frequently paid only to the physical
quantities that depend on the index of refraction alone but not on the
impedance. In particular, this is the case for all EMTs that consider
infinite composites and do not account for the boundary of the medium.
Still, these theories always predict {\em some} impedance, and the
question remains whether this prediction is applicable to finite
samples.

The analysis is relatively simple in the classical homogenization
limit $h\rightarrow 0$, where $h$ is the heterogeneity scale such as
the lattice period of a composite. Note that here we assume that all
physical characteristics of the constituents of the composite are
independent of $h$. We will refer to this kind of EMT as ``standard''.
Note that an alternative approach has been
proposed~\cite{felbacq_05_1,felbacq_05_2} in which the limit
$h\rightarrow 0$ is also taken but the permittivity of one of the
composite constituents is assumed to depend on $h$. This theory is of
a more general or, as we shall say, of the ``extended'' type.  The
fundamental differences between standard and extended theories have
been discussed by Bohren~\cite{bohren_86_1,bohren_09_1}. What is
important here is that standard EMTs do not mix the electric and
magnetic properties of the composite
constituents~\cite{wellander_03_1}. This means, in particular, that
the effective permeability obtained in a standard EMT is identically
equal to unity if the constituents of the composite are intrinsically
nonmagnetic. A closely related point is that, in standard theories,
the impedance of the medium can be inferred from the bulk behavior of
waves as long as we accept that the effective permeability is trivial.
It can be proved independently that, in the $h\rightarrow 0$ limit,
this choice of impedance is consistent with the exact Fresnel
reflection and refraction coefficients at a planar
boundary~\cite{markel_12_1}.  Thus, in a standard theory, both the
impedance and the refractive index are consistent with reflection and
refraction properties of a finite sample.

However, standard EMTs are typically viewed as inadequate in the
modern research of electromagnetic metamaterials because these
theories do not predict or describe the phenomenon of ``artificial
magnetism'', which has a number of potentially groundbreaking
applications~\cite{pendry_01_1}. This difficulty is not characteristic
of the extended theories. An extended EMT either does not employ the
limit $h\rightarrow 0$ or, otherwise, assumes mathematical dependence
between $h$ and other physical parameters of the composite. The main
question we consider in this paper is whether an extended EMT can
predict the refractive index and impedance simultaneously and in a
reasonable way.  Of course, a refractive index {\em per se}
(generally, tensorial and dependent on the direction of the Bloch wave
vector) can always be formally introduced for a Bloch wave. This can
be done even in the case when the composite is obviously not
electromagnetically homogeneous.  But all extended EMTs yield both a
refractive index and an impedance, and in the case of infinite
unbounded media there is no way to tell whether this homogenization
result is reasonable. In this paper, we present a case study by
comparing the so-called current-driven homogenization theory (which is
of extended type and is formulated for an infinite medium) to exact
results in a layered finite slab. Note that, although we analyze a
particular EMT, the central theme of this paper is related to the
fundamental difference between standard and extended EMTs.

There are, of course, many extended EMTs currently in circulation.
Theories of this kind have been first proposed by
Lewin~\cite{lewin_47_1} and
Khizhnyak~\cite{khizhnyak_57_1,khizhnyak_57_2,khizhnyak_59_1} but they
came to the fore more recently in the work of Niklasson {\em et
  al.}~\cite{niklasson_81_1}, Doyle~\cite{doyle_89_1}, and Waterman
and Pedersen~\cite{waterman_86_1}, who have generalized the classical
Maxwell-Garnett approximation to account for the magnetic dipole
moments of spherical particles (e.g., computed using Mie theory).
Although the extended Maxwell-Garnett approximation of
Refs.~\onlinecite{niklasson_81_1,doyle_89_1,waterman_86_1} applies
only to the dilute case, it has served as an important precursor of
several more generally applicable extended EMTs. Among these we can
mention the modified multiscale
approach~\cite{felbacq_05_1,felbacq_05_2}, Bloch analysis of
electromagnetic
lattices~\cite{cherednichenko_07_1,guenneau_07_1,guenneau_11_1,craster_11_1},
coarse-graining (averaging) of the electromagnetic fields using
curl-conforming and div-conforming
interpolants~\cite{tsukerman_11_1,pors_11_1,tsukerman_11_2}, and the
current-driven homogenization theory~\cite{silveirinha_07_1,alu_11_2}.
The latter approach has gained considerable traction
lately~\cite{silveirinha_09_1,costa_09_1,fietz_10_1,fietz_10_2,fietz_10_3,costa_11_1,silveirinha_11_1,alu_11_3,fietz_12_1,chebykin_11_1,chebykin_12_1}.
In this paper, we analyze this theory as an instructive case in point.

One of the co-authors has already published~\cite{markel_10_2} a
theoretical analysis of the current-driven excitation model (not
related to the theory of homogenization). However, since multiple
claims have been made that the current-driven homogenization approach
is rigorous, completely general and derived from first
principles~\cite{silveirinha_07_1,silveirinha_09_1,alu_11_2}, it
deserves additional scrutiny. Also, our previous analysis was mainly
theoretical and no numerical examples were given. But the best test of
any EMT is the test of its predictive power. It appears, therefore,
useful to investigate the predictions of current-driven homogenization
by using a simple exactly-solvable case of one-dimensional periodic
medium.

In fact, current-driven homogenization has been already applied to
such media~\cite{chebykin_11_1,chebykin_12_1}. However, the
transmission and reflection coefficients $T$ and $R$ of a layered slab
have not been studied in these references. Instead, the nonlocal
permittivity tensor $\Sigma(\omega,{\bf k})$ (defined below) was
computed numerically. Current-driven homogenization of
Refs.~\onlinecite{silveirinha_07_1,alu_11_2} entails an additional
step in which $\Sigma(\omega,{\bf k})$ is used to compute purely local
effective tensors $\epsilon$ and $\mu$ (in non-centrocymmetric media,
magneto-electric coupling parameters must also be introduced) and then
$T$ and $R$ according to the standard formulas [e.g., see equation
(\ref{T_R_theta_Zet}) below]. The nonlocal tensor $\Sigma(\omega,{\bf
  k})$ can be used for this purpose only when complemented with
additional boundary conditions (ABCs), and this computation has not
been done. In addition, Refs.~\onlinecite{chebykin_11_1,chebykin_12_1}
do not provide a closed-form expression for $\Sigma(\omega,{\bf k})$.

In what follows, we derive a closed-form expression for
$\Sigma(\omega,{\bf k})$ in the case of s-polarization. Consideration
of p-polarization is not mathematically difficult but is not needed
for our purposes. We follow the current-driven homogenization
methodology to derive closed-form expressions for the local tensors
$\epsilon$ and $\mu$. Then we use this result to compute $T$ and $R$
of layered slabs. In Sec.~\ref{sec:cd}, we summarize and discuss the
prescription of current-driven homogenization of
Refs.~\onlinecite{silveirinha_07_1,alu_11_2}. In Sec.~\ref{sec:exact}
we use this prescription to obtain closed-form expressions for the
case of a one-dimensional layered medium. In Sec.~\ref{sec:TR} we list
for reference the relevant formulas for the transmission and
reflection coefficients of layered and homogeneous slabs. Numerical
examples are given in Sec.~\ref{sec:num}. Here we compute local
effective medium parameters obtained by current-driven homogenization,
by the S-parameter retrieval method and by the classical (standard)
homogenization approach. We then use these results to compute $T$ and
$R$ and to compare the latter to the exact values for finite layered
slabs. In Sec.~\ref{sec:Bloch}, we present a Bloch-wave analysis of
current-driven homogenization.  Secs.~\ref{sec:disc} and \ref{sec:sum}
contain a discussion and a summary of the obtained results. Some
technical details of the derivations and method used in this paper are
given in the appendices.

\section{Current-driven homogenization}
\label{sec:cd}

The current-driven homogenization theory is formulated for an infinite
periodic medium and consists, essentially, of two steps.

In the first step, one derives or computes numerically the nonlocal
permittivity tensor $\Sigma(\omega,{\bf k})$, which is defined as a
coefficient between the appropriately averaged fields ${\bf D}({\bf
  r})$ and ${\bf E}({\bf r})$. The exact prescription for this
computation is given below. One could, potentially, stop at this point
and attempt to use $\Sigma(\omega,{\bf k})$ directly to compute the
physical quantities of interest. However, this computation is
difficult to perform due to the explicit dependence of $\Sigma$ on
${\bf k}$. At the very least, it entails the use of ABCs. Since
current-driven homogenization does not consider the physical boundary
of a sample, derivation of the ABCs is outside of its theoretical
framework.  Besides, the use of the ABCs would defeat the very purpose
of homogenization because all the applications of metamaterials
discussed so far in the literature rely heavily on the existence of
local constitutive parameters.

Hence there exists a second step in which the nonlocal tensor
$\Sigma(\omega,{\bf k})$ is used to derive purely local tensors
$\epsilon$ and $\mu$ (here we restrict attention to media with a
center-symmetric lattice cells and do not introduce or discuss
magneto-electric coupling parameters). This second step is based on
the proposition that, at high frequencies, magnetization of matter is
physically and mathematically indistinguishable from weak nonlocality
of the dielectric
response~\cite{agranovich_book_66,landau_ess_84,agranovich_04_1,agranovich_06_1,agranovich_06_2,agranovich_09_1}.
We will give an exact prescription for completing this step, too.

We now turn to the mathematical details needed to complete the two
steps mentioned above. We work in the frequency domain and the
time-dependence factor $\exp(-i \omega t)$ is suppressed. The
dependence of various physical quantities on $\omega$ is assumed but
not indicated explicitly except in a few cases, such as in the
notation $\Sigma(\omega,{\bf k})$, where both arguments $\omega$ and
${\bf k}$ are customarily included.  The free-space wave number $k_0$
and wavelength $\lambda_0$ are defined by
\begin{equation*}
\label{k0_def}
k_0 = \omega/c \ , \ \ \lambda_0 = 2\pi/ k_0 \ .
\end{equation*}
Finally, the Gaussian system of units is used throughout.

\subsection{Step One: calculation of the nonlocal permittivity tensor
  $\Sigma(\omega,{\bf k})$}
\label{subsec:step_one}

Consider an infinite, periodic, intrinsically-non\-mag\-ne\-tic
composite characterized by the permittivity function
$\tilde{\epsilon}({\bf r})$. Here the tilde symbol has been used to
indicate that $\tilde{\epsilon}({\bf r})$ is the true parameter of the
composite varying on a fine spatial scale, as opposed to the
spatially-uniform {\em effective medium parameters} $\epsilon$ and
$\mu$. We assume for simplicity that the composite is orthorhombic so
that
\begin{equation}
\label{periodicity}
\tilde{\epsilon}(x + h_x, y + h_y, z + h_z) = \tilde{\epsilon}(x,y,z) \ ,
\end{equation}
\noindent
where $h_x, h_y$ and $h_z$ are the lattice periods. Note that
$\tilde{\epsilon}({\bf r})$ is a macroscopic quantity and that we
consider the composite exclusively within the framework of macroscopic
electrodynamics.

In the current-driven homogenization theory, it is assumed that the
system is excited by an ``impressed'' or external electric current
${\bf J}_{\rm ext}({\bf r})$ in the form of an infinite plane wave,
viz,
\begin{equation}
\label{J_ext}
 {\bf J}_{\rm ext}({\bf r}) =  \frac{\omega}{4\pi i} {\bf J} e^{i {\bf k}\cdot {\bf r}} \ .
\end{equation}
\noindent
Here ${\bf J}$ is the amplitude, ${\bf k}$ is an arbitrary wave vector
which defines the ``forced'' Bloch-periodicity, and the $\omega/4\pi
i$ factor has been introduced for convenience. Note that ${\bf J}_{\rm
  ext}({\bf r})$ is not subject to constitutive relations and is not
equivalent to the current induced in the medium by the electric and
magnetic fields. Maxwell's equations for the system just described
have the following form:
\begin{subequations}
\label{Maxwell}
\begin{align}
\label{McH}
& \nabla \times {\bf H}({\bf r}) = - ik_0 \left[ \tilde{\epsilon}({\bf r}) {\bf
  E}({\bf r}) + {\bf J} e^{i {\bf k} \cdot {\bf r}} \right] \ , \\
\label{McE}
& \nabla \times {\bf E}({\bf r}) =   ik_0 {\bf H}({\bf r}) \ .
\end{align}
\end{subequations}
\noindent
In some generalizations~\cite{fietz_10_1}, a similar wave of magnetic
current is included in (\ref{McE}). However, inclusion of electric
current only will prove sufficient for our purposes.

Obviously, the solution to (\ref{Maxwell}) has the property of
``forced'' Bloch-periodicity~\cite{fn3}. This can be expressed
mathematically as
\begin{equation}
\label{Forced_Bloch}
{\bf E}({\bf r}) = e^{i {\bf k} \cdot {\bf r}} {\bf F}({\bf r}) \ ,
\end{equation}
\noindent
where ${\bf F}({\bf r})$ satisfies the periodicity condition
(\ref{periodicity}), and similarly for all other fields. The averaging
procedure is then defined as ``low-pass filtering'' of the fields
(e.g., Ref.~\onlinecite{silveirinha_09_1}). The averaged quantities
are defined according to
\begin{eqnarray}
\label{averaging}
{\bf E}_{\rm av} = \frac{1}{V} \int_{\mathbb C} e^{-i {\bf k} \cdot
  {\bf r}} {\bf E}({\bf r}) d^3 r
= \frac{1}{V} \int_{\mathbb C} {\bf F}({\bf r}) d^3 r \ .
\end{eqnarray}
\noindent
Here $V = h_x h_y h_z = \int_{\mathbb C} d^3 r$ and ${\mathbb C}$ denotes the
unit cell. Similar definitions can be given for averages of all other
fields, including the field of displacement ${\bf D}({\bf r}) =
\tilde{\epsilon}({\bf r}) {\bf E}({\bf r})$.

The nonlocal permittivity tensor is then defined as the linear
coefficient between ${\bf D}_{\rm av}$ and ${\bf E}_{\rm av}$, viz,
\begin{equation}
\label{Sigma_def}
{\bf D}_{\rm av} = \Sigma(\omega,{\bf k}) {\bf E}_{\rm av} \ .
\end{equation}
\noindent
If all Cartesian components of ${\bf E}_{\rm av}$ and ${\bf D}_{\rm
  av}$ are known, (\ref{Sigma_def}) contains three linear equations
for the tensor elements of $\Sigma(\omega,{\bf k})$. By considering
three different polarizations of ${\bf J}$, we can construct a set of
nine linear equations.  However, in non-gyrotropic media, the tensor
$\Sigma(\omega,{\bf k})$ is symmetric~\cite{agranovich_book_66} and
has, therefore, only six independent elements. We can force the set to
be formally well-determined by requiring that ${\bf k} \cdot {\bf
  J}=0$.

In this regard, it is useful to note that the averaged fields satisfy
${\bf k}$-space Maxwell's equations with a spatially-uniform
source~\cite{fn4}:
\begin{equation}
\label{Maxwell_av}
{\bf k} \times {\bf H}_{\rm av} = - k_0 \left( {\bf D}_{\rm av} +
  {\bf J} \right)  \ , \ \
 {\bf k} \times {\bf E}_{\rm av} =   k_0 {\bf H}_{\rm av} \ .
\end{equation}
\noindent
Consequently, ${\bf k}\cdot \left( {\bf D}_{\rm av} + {\bf J} \right)
= 0$. If ${\bf k}\cdot {\bf J} \neq 0$ [the current wave in
(\ref{J_ext}) is not transverse], we also have ${\bf k}\cdot {\bf
  D}_{\rm av} \neq 0$. This means, of course, that, in addition to the
external current (\ref{J_ext}), we have included into consideration an
external wave of charge density $\rho_{\rm ext}({\bf r}) = ({\bf k}
\cdot {\bf J} / \omega ) \exp(i {\bf k} \cdot {\bf r})$. However, in
the homogenized sample, we expect $\nabla \cdot {\bf D} = 0$ to hold.
In this paper, we use only a transverse external current wave but note
that more general excitation schemes have been
considered~\cite{fietz_10_1}.

Let us further specialize to the case of a two-component composite in
which the function $\tilde{\epsilon}({\bf r})$ can take two discrete
values $\epsilon_a$ and $\epsilon_b$. We will write ${\mathbb C} =
{\mathbb C}_a \cup {\mathbb C}_b$
and $\tilde{\epsilon}({\bf r}) = \epsilon_a$ if ${\bf r} \in {\mathbb C}_a$,
$\tilde{\epsilon}({\bf r}) = \epsilon_b$ if ${\bf r} \in {\mathbb C}_b$. In
this case, ${\bf E}_{\rm av}={\bf Q}_a + {\bf Q}_b$, ${\bf D}_{\rm av}
= \epsilon_a {\bf Q}_a + \epsilon_b {\bf Q}_b$, where
\begin{equation*}
\label{Q_def}
{\bf Q}_a = \int_{{\mathbb C}_a} {\bf F} ({\bf r})d^3r \ , \ \ {\bf Q}_b =
\int_{{\mathbb C}_b} {\bf F} ({\bf r})d^3r \ .
\end{equation*}
\noindent
Therefore, equation (\ref{Sigma_def}) takes the form
\begin{equation}
\label{Sigma_def_twocomp}
\left( {\bf Q}_a \epsilon_a + {\bf Q}_b \epsilon_b \right)
= \Sigma(\omega,{\bf k}) \left( {\bf Q}_a + {\bf Q}_b \right) \ .
\end{equation}
\noindent
From the linearity of (\ref{Maxwell}), we have ${\bf Q}_a = \tau_a
{\bf J}$, ${\bf Q}_b = \tau_b {\bf J}$, where $\tau_a$ and $\tau_b$
are two tensors. If $\tau_a + \tau_b$ is invertible, we can solve
(\ref{Sigma_def_twocomp}) to obtain
\begin{equation}
\label{Sigma_tau}
\Sigma(\omega,{\bf k}) = \left( \tau_a \epsilon_a + \tau_b \epsilon_b
\right)  \left( \tau_a + \tau_b \right)^{-1} \ .
\end{equation}
The above equation implies that introduction of the external current
(\ref{J_ext}) is not required to define the function
$\Sigma(\omega,{\bf k})$ mathematically. In fact, this statement is
general and applies to any periodic structure in any number of
dimensions, as long as the intrinsic constitutive laws are linear. In
Sec.~\ref{sec:Bloch}, we will demonstrate the same point from
Bloch-wave analysis. In Sec.~\ref{subsec:disc_nonl}, we will show that
$\Sigma(\omega,{\bf k})$ does not characterize the medium completely
but can only be used to find the law of dispersion.

\subsection{Step Two: calculation of local parameters}
\label{subsec:step_two}

The proposition that magnetization (nontrivial magnetic permeability)
of matter is indistinguishable from nonlocality of the dielectric
response is based on the equivalence of expressions for the induced
current that are obtained in both models for infinite plane waves.
Here we recount these arguments insomuch as they are needed for
deriving the main results of this paper.

Consider two electromagnetically-homogeneous media. The first medium
is characterized by a nonlocal permittivity tensor $\Sigma(\omega,{\bf
  k})$ and $\mu=1$. In fact, the auxiliary field ${\bf H}$ is not
introduced for this medium, so that $\mu$ is, strictly speaking, not
defined. The macroscopic electrodynamics is then built using the
fields ${\bf E}$, ${\bf B}$ and ${\bf D}$ with the account of
spatially-nonlocal relationship between ${\bf D}$ and ${\bf E}$.  The
induced current in such a medium is given by
\begin{equation}
\label{Jd1}
{\bf J}_{\rm ind}^{(1)} = -\frac{i\omega}{4\pi} \left[
  \Sigma(\omega,{\bf k}) - 1 \right] {\bf E} \ .
\end{equation}
\noindent
Here we assume, as is done in all relevant
references~\cite{agranovich_book_66,landau_ess_84,agranovich_04_1,agranovich_06_1,agranovich_06_2,agranovich_09_1},
that ${\bf E}$ is an infinite plane wave with the wave vector ${\bf
  k}$.

The second medium is characterized by purely local tensors $\epsilon$
and $\mu$ and the induced current in this medium is given by
\begin{equation}
\label{Jd2a}
{\bf J}_{\rm ind}^{(2)} = -\frac{i\omega}{4\pi} \left(
  \epsilon - 1 \right) {\bf E} + c\nabla \times {\bf M} \ ,
\end{equation}
\noindent
where ${\bf M}$ is the vector of magnetization. Using the definition
of ${\bf M}$ and macroscopic Maxwell's equations, we can also write
\begin{equation}
\label{Jd2b}
{\bf J}_{\rm ind}^{(2)} = -\frac{i\omega}{4\pi} \left[
  (\epsilon - 1) - \frac{1}{k_0^2} {\bf k} \times \left(1 - \mu^{-1}
  \right) {\bf k} \times \right] {\bf E} \ .
\end{equation}
\noindent
This expression can be compared to (\ref{Jd1}). In general, of course,
there is no equivalence between (\ref{Jd1}) and (\ref{Jd2b}). But in
the so-called weak nonlocality regime [defined more precisely after
Eq.~(\ref{beta_Sigma}) below], $\Sigma(\omega,{\bf k})$ is well
approximated by its second-order expansion in powers of ${\bf k}$.  In
this case, one can look for the condition under which (\ref{Jd1}) and
(\ref{Jd2b}) agree to second order in ${\bf k}$. In non-gyrotropic
media, the expansion of $\Sigma(\omega,{\bf k})$ has the form
\begin{equation}
\label{Sigma_exp}
\Sigma(\omega,{\bf k}) = \Sigma(\omega,0) - \frac{1}{k_0^2} {\bf k}
\times \beta {\bf k} \times + \ldots \ \ ,
\end{equation}
\noindent
where $\beta$ is a tensor.

We are interested in the condition under which the expression in the
square brackets in (\ref{Jd2b}) is equivalent to $[\Sigma(\omega,{\bf
  k}) - 1]$ computed to second order in ${\bf k}$. It is easy to see
that this condition is $\epsilon = \Sigma(\omega,0)$ and $\mu = (1 -
\beta)^{-1}$ where the last equation implies tensor inverse.  In
isotropic media $\mu$ and $\beta$ are reduced to scalars. In the case
of cubic symmetry, when the tensors $\beta$ and $\mu$ are diagonal in
the rectangular reference frame $XYZ$, we have $\mu_{\alpha\alpha} =
(1 - \beta_{\alpha\alpha})^{-1}$, where $\alpha=x,y,z$. Note that all
three principal values of $\beta$ can now be different. The conclusion
that is typically drawn from this analysis~\cite{agranovich_09_1} is
that the introduction of local parameter $\mu$ is physically
indistinguishable from the account of the second-order term in
expansion (\ref{Sigma_exp}).

If the function $\Sigma(\omega,{\bf k})$ is known (it is computed
directly in Step One of the current-driven homogenization
prescription), the tensor $\beta$ can be easily computed from
(\ref{Sigma_exp}). In the case of cubic symmetry, the relevant
formulas are
\begin{subequations}
\label{beta_Sigma}
\begin{align}
\label{beta_xx}
\beta_{xx} = \frac{k_0^2}{2}\frac{\partial^2 \Sigma_{yy}}{\partial k_z^2} =
\frac{k_0^2}{2}\frac{\partial^2 \Sigma_{zz}}{\partial k_y^2} =
-k_0^2 \frac{\partial^2 \Sigma_{yz}}{\partial k_y\partial k_z} \ , \\
\label{beta_yy}
\beta_{yy} = \frac{k_0^2}{2}\frac{\partial^2 \Sigma_{xx}}{\partial k_z^2} =
\frac{k_0^2}{2}\frac{\partial^2 \Sigma_{zz}}{\partial k_x^2} =
-k_0^2 \frac{\partial^2 \Sigma_{xz}}{\partial k_x\partial k_z} \ , \\
\label{beta_zz}
\beta_{zz} = \frac{k_0^2}{2}\frac{\partial^2 \Sigma_{xx}}{\partial k_y^2} =
\frac{k_0^2}{2}\frac{\partial^2 \Sigma_{yy}}{\partial k_z^2} =
-k_0^2 \frac{\partial^2 \Sigma_{xy}}{\partial k_x\partial k_y} \ .
\end{align}
\end{subequations}
\noindent
All derivatives in the above equations must be evaluated at ${\bf k}=0$.

We can now formulate the condition of weak nonlocality more precisely.
Let's assume that we have applied the prescription and computed the
local parameters $\epsilon$ and $\mu$ at a given frequency. These
parameters can now be used to compute the natural wave vector of the
medium, ${\bf q}$, using the dispersion relation [e.g., for a uniaxial
crystal, see (\ref{q_hmg}) below]. We then evaluate the nonlocal
permittivity $\Sigma(\omega,{\bf k})$ at ${\bf k} = {\bf q}$.  The
nonlocality is weak if the expansion (\ref{Sigma_exp}) computed to
second order accurately approximates $\Sigma(\omega,{\bf q})$. Thus,
in the weak nonlocality regime, higher-order terms in the expansion
(\ref{Sigma_exp}) can be neglected.

At this point, we can make two important observations. First, the
above discussion applies only to infinite media. In any finite
magnetic medium, additional surface currents exist. These currents are
not included in (\ref{Jd2a}). Consequently, the equivalence of
currents is, in principle, not complete: it does not apply to the
surface currents. As a result, introduction of a nontrivial magnetic
permeability and a dynamic correction to the permittivity~\cite{fn6},
as described above, can yield a first nonvanishing correction to the
dispersion relation but not to the impedance of the medium. A related
point is that, in finite samples, ${\bf J}_{\rm ind}^{(2)}$ is not
reduced to a quadratic form (in Cartesian components of ${\bf k}$)
even in the case of natural magnetics.

The second observation is more subtle. The local parameters that
satisfy the requirement of current equivalence are not unique when the
current is evaluated on-shell, that is, for ${\bf k} = {\bf q}$. There
exists an infinite set of such parameters, related to each other by
the transformation (\ref{renorm}) (stated below), all of which yield
exactly the same law of dispersion and the same induced current
(\ref{Jd2b}). However, in current-driven homogenization, the variable
${\bf k}$ in (\ref{Jd2b}) is viewed as a free parameter. If we follow
this ideology and require the current equivalence to hold for all
values of ${\bf k}$, we would obtain an unambiguous ``prescription''
for computing the local effective parameters. But the only
physically-realizable case is ${\bf k} = {\bf q}$. Therefore, it is
not clear why the pair of effective parameters predicted by
current-driven homogenization is ``better'' than any other pair
obtained by the transformation (\ref{renorm}). This point will be
illustrated numerically in Sec.~\ref{subsec:Ex_A}.

\section{Exact solution in one-dimensional layered medium}
\label{sec:exact}

\begin{figure}
  \psfig{file=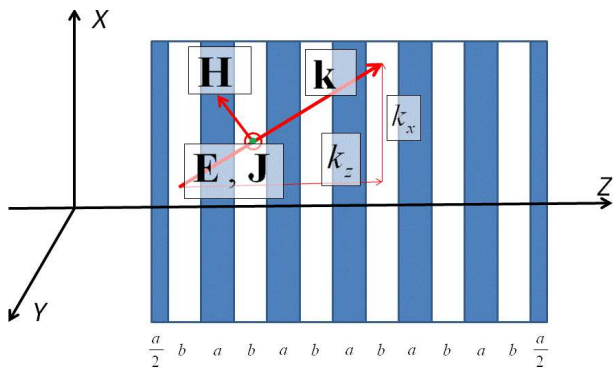,width=8.2cm,bbllx=40bp,bblly=130bp,bburx=940bp,bbury=660bp,clip=}
\caption{\label{geom} Geometry of wave propagation in the case of
  s-polarization. Here ${\bf k}=(k_x,0,k_z)$ is the wave vector of the
  external current wave [Eq.~(\ref{J_ext})]. A finite symmetric slab
  containing $N=6$ unit cells is shown. Each cell consists of three
  layers of the widths $(a/2,b,a/2)$. Equivalently, we can view the
  unit cells as consisting of two layers of the widths $(a,b)$,
  provided that one half of the first $a$-type layer has been cut off
  and moved from the left face of the slab to its right face. Note
  that the sample shown in the figure has a center of symmetry.}
\end{figure}

The geometry considered is illustrated in Fig.~\ref{geom}. A
one-dimensional periodic medium consists of alternating intrinsically
nonmagnetic layers of widths $a$ and $b$ and scalar permittivities
$\epsilon_a$ and $\epsilon_b$, respectively. The period of the system
is given by $h=a+b$. The layered medium described here can be
considered as a special case of the three-dimensional orthorhombic
lattice obtained in the limit $h_x=h_y=0$, $h_z=h>0$. Note that the
medium shown in Fig.~\ref{geom} is finite and terminated by half-width
$a$-type layers.  However, in the current-driven homogenization
theory, the medium is assumed to be infinite.

We will consider only the special case of s-polarization, when the
wave vector ${\bf k}$ lies in the plane $XZ$ of the rectangular frame
shown in Fig.~\ref{geom} and the amplitude ${\bf J}$ of the external
current (\ref{J_ext}) is collinear with the $Y$-axis, so that ${\bf k}
= (k_x,0,k_z)$ and ${\bf J} = (0,J_y,0)$. According to
(\ref{beta_Sigma}), this is sufficient to uniquely define the
following elements of the effective permittivity and permeability
tensors: $\epsilon_{xx} = \epsilon_{yy}$, $\mu_{xx} = \mu_{yy}$ and
$\mu_{zz}$.

We will need to introduce the following notations:
\begin{subequations}
\label{wv_defs}
\begin{align}
\label{k_a_b_def}
& k_a^2    = k_0^2 \epsilon_a     \ , \ \ k_b^2    = k_0^2 \epsilon_b
\ , \\
\label{kappa_a_b_def}
& \kappa_a = \sqrt{k_a^2 - k_x^2} \ , \ \ \kappa_b = \sqrt{k_b^2 - k_x^2} \ , \\
\label{phi_a_b_def}
& \phi_a   = \kappa_a a           \ , \ \ \phi_b   = \kappa_b b \ , \\
\label{theta_a_b_def}
& \theta_a = k_z a                \ , \ \ \theta_b = k_z b      \ , \\
& p_a      = a/h                  \ , \ \  p_b     = b/h        \ ,
\end{align}
\end{subequations}
and also the standard homogenization result for a periodic layered medium:

\begin{subequations}
\label{eps_mu_hmg}
\begin{align}
\label{eps_hmg}
&\epsilon_{\parallel} =
   p_a \epsilon_a + p_b\epsilon_b \ , \ \ \epsilon_\perp =
              \frac{1}{p_a/\epsilon_a + p_b/\epsilon_b} \ , \\
\label{mu_hmg}
&\mu_\parallel=\mu_\perp = 1 \ .
\end{align}
\end{subequations}
\noindent
Here the quantities indexed by ``$\parallel$'' and ``$\perp$'' give
the standard homogenization results for the elements of the
permittivity and permeability tensors that correspond to the direction
of the electric field parallel and perpendicular to the layers,
respectively. Throughout the paper, the branches of all square roots
are defined by the condition $0 \leq {\rm arg}(\sqrt{z}) < \pi$.

We wish to solve Eq.~(\ref{Maxwell}) in which $\tilde{\epsilon}({\bf
  r})$ is equal to $\epsilon_a$ in the $a$-type layers and to
$\epsilon_b$ in the $b$-type layers. Without loss of generality, we
can consider the unit cell $0<z\leq h=a+b$, which contains two layers:
the first layer ($a$-type) is contained between the planes $z=0$ and
$z=a$ and the second ($b$-type) layer is contained between the planes
$z=a$ and $z=h$. We can seek the solution in each homogeneous region
excluding its boundaries as a particular solution to the inhomogeneous
equation plus the general solution to the homogeneous equation, viz,
\begin{subequations}
\label{Fields_pg}
\begin{align}
& E_y(x,z)=e^{i {\bf k} \cdot {\bf r}} \left[ {\mathscr E}_p(z)
  \hspace*{1.5mm} +
  {\mathscr E}_g(z) \hspace*{1.5mm} \right] \ , \\
& H_x(x,z)=e^{i {\bf k} \cdot {\bf r}} \left[ {\mathscr H}_p(z) +
  {\mathscr H}_g(z) \right] \ .
\end{align}
\end{subequations}
\noindent
Here the subscripts ``$p$'' and ``$g$'' denote the particular and the
general solution, respectively, and the overall exponential factor
$\exp(i {\bf k} \cdot {\bf r})$ is written out explicitly.

The particular solution is given by
\begin{equation}
\label{Sol_part}
{\mathscr E}_p(z) = J_y f(z) \ , \ \  {\mathscr
  H}_p(z) = -\frac{k_z}{k_0} J_y f(z) \ ,
\end{equation}
\noindent
where
\begin{equation*}
\label{f_part}
f(z) =
\left\{
\begin{array}{ll}
{\displaystyle \frac{k_0^2}{k^2 - k_a^2}} \equiv f_a \ , & 0<z<a \\
{\displaystyle \frac{k_0^2}{k^2 - k_b^2}} \equiv f_b \ , & a<z<h
\end{array} \right. \ . \\
\end{equation*}
\noindent
We emphasize that (\ref{Sol_part}) is a particular solution in the
open intervals $0<z<a$ and $a<z<h$. To satisfy boundary conditions at
the interfaces, we must add to (\ref{Sol_part}) the general solution
to the corresponding homogeneous problem. The latter can be easily
stated:

\begin{subequations}
\label{Sol_gen}
\begin{align}
\label{Sol_gen_E}
& {\mathscr E}_g(z) = \hspace*{9mm} J_y \Delta e^{-i k_z z} F_{\mathscr E}(z) \ , \\
\label{Sol_gen_H}
& {\mathscr H}_g(z) = -\frac{k_z}{k_0} J_y \Delta e^{-i k_z z} F_{\mathscr H}(z) \ ,
\end{align}
\end{subequations}
where
\begin{equation*}
\label{Delta_def}
\Delta = f_b - f_a = \frac{k_0^2}{k^2 - k_b^2} - \frac{k_0^2}{k^2 - k_a^2}
\end{equation*}
\noindent
and
\begin{subequations}
\label{F_eh_def}
\begin{align}
\label{F_e_def}
& F_{\mathscr E}(z) = \hspace*{0cm}  \\
& \left\{
\begin{array}{ll}
\hspace*{9mm} A_a e^{i\kappa_a z} \hspace*{6mm} + B_a e^{-i\kappa_a z} \hspace*{7mm} , & 0<z<a \\
e^{i\theta_a} [-A_b e^{i\kappa_b (z-a)} - B_b e^{-i\kappa_b (z-a)}] \ , & a<z<h
\end{array} \right. \ ; \nonumber \\
\label{F_h_def}
& F_{\mathscr H}(z) = \hspace*{0cm} \\
& \left\{
\begin{array}{ll}
\hspace*{5.5mm} {\displaystyle \frac{\kappa_a}{k_z}} [ \hspace*{2.5mm} A_a e^{ i
  \kappa_a z} \hspace*{5.5mm} - B_a e^{-i \kappa_a z}
\hspace*{5.5mm} ] \ , & 0<z<a  \\
{\displaystyle \frac{\kappa_b}{k_z}} e^{i\theta_a} [-A_b e^{ i
  \kappa_b (z-a)} + B_b e^{-i \kappa_b (z-a)}] \ , & a<z<h
\end{array}
\right. \ . \nonumber
\end{align}
\end{subequations}
\noindent
In these expressions, various $z$-independent factors have been
introduced for convenience and $A_a,B_a,A_b,B_b$ is a set of
coefficients to be determined from the boundary conditions.  The
latter require continuity of all tangential field components at the
interfaces $z=0,a,h$ and can be stated as follows:
\begin{subequations}
\label{BC}
\begin{align}
& F_{\mathscr E}(a-0) - F_{\mathscr E}(a+0) = e^{i\theta_a} \ , \\
& F_{\mathscr H}(a-0) - F_{\mathscr H}(a+0) = e^{i\theta_a} \ , \\
& e^{i\theta_a} F_{\mathscr E}(0)   - e^{-i\theta_b} F_{\mathscr E}(h) = e^{i\theta_a} \ , \\
& e^{i\theta_a} F_{\mathscr H}(0)   - e^{-i\theta_b} F_{\mathscr H}(h) = e^{i\theta_a} \ .
\end{align}
\end{subequations}
\noindent
This results in a set of four equations for the unknown coefficients
$A_a,B_a,A_b,B_b$, which are stated in Appendix~\ref{app:A}.

It may seem confusing that the right-hand side in (\ref{BC}) does not
go to zero when $J_y \rightarrow 0$; in fact, (\ref{BC}) does not
contain $J_y$ at all. However, the electromagnetic fields $E_x(x,z)$
and $H_y(x,z)$ computed according to (\ref{Fields_pg})-(\ref{Sol_gen})
are proportional to $J_y$. Note that the most general solution to
(\ref{Maxwell}) is a superposition of the solution derived here (whose
amplitude is proportional to $J_y$) and the natural Bloch mode of the
medium with an arbitrary amplitude. To remove the nonuniqueness, one
can either consider the boundary of the medium and thus abandon the
infinite medium model, or, alternatively, apply the additional
boundary condition requiring ``forced'' Bloch-periodicity
(\ref{Forced_Bloch}).  The latter approach is used in current-driven
homogenization and in the derivations of this section.

The solution to (\ref{BC}) is given by
\begin{equation}
\label{ABCD_det}
A_a = \frac{{\mathscr A}_a}{2 \mathscr D} \ , \ \
B_a = \frac{{\mathscr B}_a}{2 \mathscr D} \ , \ \
A_b = \frac{{\mathscr A}_b}{2 \mathscr D} \ , \ \
B_b = \frac{{\mathscr B}_b}{2 \mathscr D} \ ,
\end{equation}
\noindent
where
\begin{equation}
\label{D_def}
{\mathscr D} = \cos (k_z h) - \cos (q_z h)
\end{equation}
\noindent
and the closed-form expressions for ${\mathscr A}_a$, ${\mathscr
  B}_a$, ${\mathscr A}_b$ and ${\mathscr B}_b$ are given in
Appendix~\ref{app:A}. In (\ref{D_def}), $q_z$ is the $z$-projection of
the natural Bloch wave vector ${\bf q}$ computed under the assumption
that its $X$-projection is equal to $k_x$ (that is, $q_x = k_x$). The
factor $\cos(q_z h)$ is defined by the equation
\begin{eqnarray}
\label{q_z_B}
\cos(q_z h) && = \cos(\phi_a)\cos(\phi_b)  \nonumber \\
&& - \frac{1}{2}\left(\frac{\kappa_a}{\kappa_b}  +
  \frac{\kappa_b}{\kappa_a} \right) \sin(\phi_a) \sin(\phi_b) \ .
\end{eqnarray}
\noindent
Evidently, if $k_z = \pm q_z + 2 \pi n/h$, where $n$ is an arbitrary
integer, the matrix in (\ref{BC}) is singular.

We now simplify the expression for the electric field $E_y(x,z)$.
After some rearrangement, we can write
\begin{eqnarray*}
E_y(x,z)  = \hspace*{5cm} \nonumber \\
 \frac{J_y k_0^4  e^{i {\bf k} \cdot {\bf r}}}{(k^2 - k_a^2)(k^2 -
  k_b^2)} \frac{1}{\mathscr D}  \left\{
\begin{array}{ll}
F_a(z) , & 0<z<a  \\
F_b(z) , & a<z<h
\end{array} \right. \ ,
\end{eqnarray*}
\noindent
where
\begin{subequations}
\label{F_a_b}
\begin{align}
F_a(z) = {\mathscr D} & \frac{k^2 - k_b^2}{k_0^2} + \frac{1}{2} e^{-ik_z z}(\epsilon_b - \epsilon_a) \nonumber \\
& \times  \left[
  {\mathscr A}_a e^{i\kappa_a z} + {\mathscr B}_a e^{-i\kappa_a z}
\right] \ , \\
F_b(z) = {\mathscr D} & \frac{k^2 - k_a^2}{k_0^2} - \frac{1}{2} e^{ik_z (a-z)}(\epsilon_b - \epsilon_a) \nonumber \\
& \times \left[
  {\mathscr A}_b e^{i\kappa_b (z-a)} + {\mathscr B}_b e^{-i\kappa_b (z-a)}
\right] \ .
\end{align}
\end{subequations}
\noindent
The $yy$-component of the nonlocal permittivity tensor
$\Sigma(\omega,{\bf k})$ is computed by using
(\ref{Sigma_def_twocomp}) or (\ref{Sigma_tau}), which results in
\begin{equation}
\label{eps_nl_Q}
\Sigma_{yy}(\omega,{\bf k}) = \frac{Q_a \epsilon_a + Q_b
  \epsilon_b}{Q_a + Q_b} \ ,
\end{equation}
\noindent
where
\begin{equation}
\label{Q_def_1D}
Q_a = \int_0^a F_a(z) dz \ , \ \ Q_b = \int_a^h F_b(z)dz \ .
\end{equation}
\noindent
The integrals in (\ref{Q_def_1D}) are easily computed analytically;
these intermediate results are omitted. Note that $Q_a$ and $Q_b$
depend implicitly on both $\omega$ and ${\bf k}$, which are considered
as mathematically-independent variables in the current-driven
homogenization theory; this dependence is indicated explicitly in the
notation $\Sigma(\omega,{\bf k})$.

Equations (\ref{F_a_b})-(\ref{Q_def_1D}) together with the expressions
for the expansion coefficients given in Appendix~\ref{app:A}
constitute a closed-form solution for $\Sigma_{yy}(\omega,{\bf k})$.
This solution contains only elementary functions, has no branch
ambiguities [see the note after Eq.~(\ref{wv_defs})] and can be easily
programmed. Note that the quantities $Q_a$ and $Q_b$ defined in
(\ref{Q_def_1D}) have no singularities when viewed as functions of
${\bf k}$. However, $\Sigma_{yy}(\omega,{\bf k})$ has singularities at
the roots of the equation $Q_a + Q_b = 0$. This completes Step One of
the current-driven homogenization prescription, at least for the case
of s-polarization.

We now proceed with Step Two. For the local effective permittivity, we
have
\begin{equation*}
\epsilon_{yy} = \Sigma_{yy}(\omega,0) =  \left. \frac{Q_a \epsilon_a + Q_b
  \epsilon_b}{Q_a + Q_b} \right\vert_{{\bf k}=0} \ .
\end{equation*}
\noindent
We note that this expression contains the dynamic correction to the
permittivity~\cite{fn6}. From symmetry, we also have $\epsilon_{xx} =
\epsilon_{yy}$. The remaining nontrivial component of the permittivity
tensor is $\epsilon_{zz}$; this element cannot be computed by
considering only s-polarization of the external current.

To compute the elements of the permeability tensor, we use the first
equality in (\ref{beta_xx}) and the second equality in
(\ref{beta_zz}). More specifically, we have
\begin{equation*}
\label{beta_xx_zz}
\beta_{xx} = \frac{k_0^2}{2}
\left. \frac{\partial^2 \Sigma_{yy}}{\partial k_z^2} \right\vert_{{\bf
    k}=0} \ , \ \ \beta_{zz} = \frac{k_0^2}{2}
\left. \frac{\partial^2 \Sigma_{yy}}{\partial k_x^2}
\right\vert_{{\bf k}=0} \ .
\end{equation*}
Using (\ref{eps_nl_Q}), we can obtain the following formulas for
$\beta_{xx}$ and $\beta_{zz}$ in terms of $Q_a$ and $Q_b$:
\begin{subequations}
\label{beta_comp}
\begin{align}
\beta_{xx} = \left. k_0^2(\epsilon_b - \epsilon_a) \frac{Q_a
  Q_b^{(zz)} - Q_b Q_a^{(zz)}}{(Q_a + Q_b)^2}
\right\vert_{{\bf k}=0} \ , \\
\beta_{zz} = \left. k_0^2(\epsilon_b - \epsilon_a) \frac{Q_a
  Q_b^{(xx)} - Q_b Q_a^{(xx)}}{(Q_a + Q_b)^2}
\right\vert_{{\bf k}=0} \ .
\end{align}
\end{subequations}
\noindent
In these expressions, $Q_a^{(xx)}$ denotes the second derivative of
$Q_a$ with respect to $k_x$ evaluated at ${\bf k}=0$, etc. In deriving
(\ref{beta_comp}), we have used the fact that $Q_a^{(x)} = Q_a^{(z)} =
Q_b^{(x)} = Q_b^{(z)} = 0$. Note that equation (\ref{beta_comp}) is
invariant under the permutation of indexes $a \leftrightarrow b$.

The elements of the effective permeability tensor are expressed in
terms of $\beta_{xx}$, $\beta_{zz}$ as
\begin{equation}
\label{mu_eff_beta}
\mu_{xx} = \frac{1}{1 - \beta_{xx}} \ , \ \ \mu_{zz} = \frac{1}{1 - \beta_{zz}} \ .
\end{equation}
\noindent
From symmetry, we also have $\mu_{yy}=\mu_{xx}$. Closed-form
expressions for $\epsilon_{xx} = \epsilon_{yy}$, $\mu_{xx}=\mu_{yy}$
and $\mu_{zz}$ are given in Appendix~\ref{app:B}. These expressions
contain only elementary trigonometric functions but are fairly
cumbersome. However, the small-$h$ asymptotic approximations of these expressions
have the following simple form:
\begin{subequations}
\label{asympt}
\begin{align}
\label{eps_yy}
\epsilon_{xx} &= \epsilon_{yy} = \epsilon_\parallel +
\frac{\left(\epsilon_a - \epsilon_b \right)^2}{12} \nonumber \\
& \times \left( p_a p_b
\right)^2 \left( k_0
  h\right)^2  + O(h^4) \ , \\
\label{mu_xx}
 \mu_{xx} &= \mu_{yy} = 1 + \frac{\left( \epsilon_a - \epsilon_b\right)^2}{240} \nonumber \\
& \times \left( p_a p_b \right)^2 (1 + 2 p_a p_b) (k_0h)^4 + O(h^6) \ , \\
\label{mu_zz}
 \mu_{zz} &= 1 - \frac{\left( \epsilon_a - \epsilon_b\right)^2}{720} \nonumber \\
& \times \left( p_a p_b \right)^2 (1 + 2 p_a p_b) (k_0h)^4 + O(h^6) \ .
\end{align}
\end{subequations}
\noindent
Thus, the first nonvanishing corrections to the effective permeability
tensor are obtained to fourth order in $h$. Moreover, the corrections
to $\mu_{xx}$ and $\mu_{zz}$ differ by the constant factor $-3$.
Consequently, current-driven homogenization, when applied to the 1D
periodic structure considered in this section, guarantees that at
least one of the principal values of the permeability tensor has a
negative imaginary part for sufficiently small values of $h$, provided
that ${\rm Im}(\epsilon_a - \epsilon_b)^2 \neq 0$.

\section{Transmission and reflection by homogeneous and layered slabs}
\label{sec:TR}

In what follows, we will need to refer to the formulas for the
transmission ($T$) and reflection ($R$) coefficients of homogeneous
and layered slabs. These formulas are well known and are adduced here
mainly for reference. However, they also reveal some important
features that will help us analyze the numerical results of the next
section. We still work in the geometry of Fig.~\ref{geom} and, in
analogy to (\ref{kappa_a_b_def}), denote the $z$-projection of the
incident wave vector by $\kappa_0$, so that
\begin{equation*}
\label{k_iz}
\kappa_0 = \sqrt{k_0^2 - k_x^2} \ .
\end{equation*}
\noindent
Note that, for $k_x > k_0$, the incident wave is evanescent. All
formulas given below are parameterized by $k_x$.

Any slab with a one-dimensional distribution of electromagnetic
parameters is completely characterized by its characteristic matrix
$M$. Suppose the slab occupies the region $0<z<L$. Then the tangential
components of the electric and magnetic field at the left and right
faces of the slab are related by
\begin{equation*}
\label{M_def}
\begin{bmatrix}
E_L    \\
H_L
\end{bmatrix}
=
\begin{bmatrix}
M_{11} & M_{12} \\
M_{21} & M_{22} \\
\end{bmatrix}
\begin{bmatrix}
E_0    \\
H_0
\end{bmatrix}
\ .
\end{equation*}
The general property of all characteristic matrices is 
%unitarity, from which it follows that 
${\rm det}(M)=1$. If the slab has a center of symmetry (as is the case
in this paper), the left and right incidence directions are equivalent
which can be mathematically stated as $M_{11} = M_{22}$. Under these
conditions, the transfer matrix can be written as~\cite{feng_10_1}
\begin{equation*}
\label{M_symm}
M =
\begin{bmatrix}
\hspace*{7mm} \cos \theta  & (-i/{\mathscr Z}) \sin\theta \\
-i {\mathscr Z} \sin \theta           & \hspace*{12mm} \cos\theta \\
\end{bmatrix} \ ,
\end{equation*}
where $\theta$ and ${\mathscr Z}$ are the optical depth and the
generalized impedance of the slab~\cite{fn2}.

The transmission and reflection coefficients can be expressed in terms
of $\theta$ and ${\mathscr Z}$ as
\begin{subequations}
\label{T_R_theta_Zet}
\begin{eqnarray}
\label{T_theta_Zet}
T = \frac{1}{\cos\theta - i X_{+}\left({\mathscr Z}_0,{\mathscr
      Z}\right) \sin\theta} \ , \\
\label{R_theta_Zet}
R = \frac{-i X_{-} \left( {\mathscr Z}_0, {\mathscr Z} \right)
  \sin\theta}{\cos\theta - i X_{+} \left( {\mathscr Z}_0, {\mathscr
      Z} \right) \sin\theta } \ ,
\end{eqnarray}
\end{subequations}
\noindent
where
\begin{equation*}
\label{X_pm_def}
X_\pm({\mathscr Z}_1,{\mathscr Z}_2) =
\frac{1}{2}\left(\frac{{\mathscr Z}_1}{{\mathscr Z}_2} \pm
                 \frac{{\mathscr Z}_2}{{\mathscr Z}_1} \right)
\end{equation*}
\noindent
and ${\mathscr Z}_0 = \kappa_0/k_0 $ is the generalized impedance of
free space (we assume that the slab is embedded in a vacuum or air). The
quantities $T$ and $R$ defined in (\ref{T_R_theta_Zet}) relate the
amplitudes of the transmitted and reflected tangential field (electric
in the case of s-polarization or magnetic field in the case of
p-polarization) measured at the planes $z=L$ (for $T$) or $z=0$ (for
$R$) to the amplitude of the incident wave at $z=0$. The specific
expressions for $\theta$ and ${\mathscr Z}$ depend on polarization and
it should be kept in mind that, at normal incidence, $T_s = T_p$ but
$R_s = -R_p$, where the subscripts indicate the particular mathematical
expression applicable to a given polarization state.

Specific expressions for $\theta$ and ${\mathscr Z}$ for homogeneous
and layered slabs are given below.\\

\paragraph{Homogeneous Anisotropic Slab.}
Consider a slab characterized by purely local diagonal tensors
$\epsilon = {\rm diag}(\epsilon_\perp, \epsilon_\perp,
\epsilon_\parallel)$ and $\mu = {\rm diag} (\mu_\perp, \mu_\perp,
\mu_\parallel)$. Then
\begin{equation}
\label{theta_Zet_hmg}
\theta = q_z L \ , \ \ {\mathscr Z} = q_z / (k_0 \eta_\parallel) \ ,
\end{equation}
\noindent
where
\begin{equation}
\label{q_hmg}
q_z = \sqrt{k_0^2 \epsilon_\parallel \mu_\parallel - k_x^2
  \left(\eta_\parallel / \eta_\perp \right) }
\end{equation}
\noindent
and $\eta$ refers to $\mu$ for s-polarization and to $\epsilon$ for
p-polarization.\\

\paragraph{Layered Slab.}
Consider a layered slab of total width $L=Nh$ containing $N$
unit cells arranged as shown in Fig.~\ref{geom}. Each cell
consists of three consecutive layers of the widths $(a/2,b,a/2)$,
where $a+b=h$, and the permittivities $\epsilon_a$ and $\epsilon_b$,
respectively. Then
\begin{subequations}
\label{q_Zet_Layered}
\begin{align}
\label{q_Layered}
& \theta = q_z L = N q_z h \ , \\
\label{Zet_Layered}
& {\mathscr Z^2} = {\mathscr Z}_a^2 \nonumber \\
& \times
\frac{\sin\phi_a \cos\phi_b - X_- \sin\phi_b + X_+ \cos\phi_a \cos\phi_b}
     {\sin\phi_a \cos\phi_b + X_- \sin\phi_b + X_+ \cos\phi_a
       \cos\phi_b} \ .
\end{align}
\end{subequations}
Here $\phi_a,\phi_b$ are defined in (\ref{phi_a_b_def}), $X_\pm =
X_\pm ({\mathscr Z}_a, {\mathscr Z}_b)$, ${\mathscr Z}_a$ and
${\mathscr Z}_b$ are the generalized impedances of each layer and
$q_z$ is the natural Bloch wave number of the medium defined by the
following equation:
\begin{eqnarray}
\label{q_z_B_gen}
\cos(q_z h) = \cos(\phi_a)\cos(\phi_b) - X_+ \sin(\phi_a) \sin(\phi_b)
\ .
\end{eqnarray}
\noindent
Note that (\ref{q_z_B}) is a special case of (\ref{q_z_B_gen}) (for
s-polarization and nonmagnetic layers). Also Eq.~(\ref{q_z_B_gen})
defines $\cos(q_zh)$ but not $\sin(q_z h)$. The latter quantity can be
computed by using one of the formulas
\begin{align*}
& \sin (q_z h) \nonumber \\
& = \frac{\mathscr Z}{{\mathscr Z}_a} \left(\sin\phi_a\cos\phi_b + X_-
  \sin\phi_b + X_+\cos\phi_a \sin\phi_b \right) \\
& = \frac{{\mathscr Z}_a}{\mathscr Z} \left(\sin\phi_a\cos\phi_b - X_-
  \sin\phi_b + X_+\cos\phi_a \sin\phi_b \right) \ ,
\end{align*}
\noindent
where ${\mathscr Z}$ is determined by taking an arbitrary branch of
the square root of (\ref{Zet_Layered}); the resultant transfer matrix
is invariant with respect to this choice.

The formulas given above illustrate several important points. First,
the transmission of a thick, highly transparent slab is very sensitive
to small errors in $q_z$. This is because the trigonometric functions
such as $\cos\theta = \cos(q_z L)$ incur a substantial phase shift
when $q_z$ is changed by $\sim \pi/L$.  If $L\rightarrow \infty$ (and
losses can still be ignored), any homogenization theory is expected to
be unstable numerically because a small error in medium parameters can
propagate to become a significant error in $T$. This instability,
however, is of little practical importance because, in most cases, the
illumination is not monochromatic. What we are discussing here are,
essentially, the resonances of a Fabry-Perot etalon. In most
applications to optical imaging and microscopy, illumination is more
broadband than a line of a single-mode laser and the interference
effects are unobservable. Under these conditions, the expressions
(\ref{T_R_theta_Zet}) can be regularized by Gaussian integration with
respect to $k_0$ or $L$. However, in this paper, we only consider
strictly monochromatic light.

Second, an error in the generalized impedance will also result in an
error in both $T$ and $R$ and, in some cases, this error can also be
dramatic. An illustrative example is the case $X_+\rightarrow -1$,
which is the operation regime of the so-called perfect
lens~\cite{pendry_00_1}. Indeed, we can re-write (\ref{T_theta_Zet})
as
\begin{equation}
\label{T_Perf_Lens}
T = \frac{2}{(1-X_+)\exp(i\theta) + (1+X_+) \exp(-i\theta)} \ .
\end{equation}
\noindent
If $X_+$ is exactly equal to $-1$, (\ref{T_Perf_Lens}) predicts
$T=\exp(-i\theta)$. For evanescent waves and a macroscopically-thick
slab, the factor $\exp(i\theta)$ is exponentially small. Therefore, if
we make a small error~\cite{fn5} in $X_+$, say, $X_+=-1 + \delta$,
such that $\vert \exp(i\theta)\vert \ll \vert \delta \vert \ll 1$,
Eq.~(\ref{T_Perf_Lens}) would predict $T=(2/\delta) \exp(i\theta)$,
which is dramatically different from the former result. We note that
there are other similar situations and that the condition $X_+\approx
- 1$ is not special in this respect.

Therefore, a homogenization theory must predict correctly both the
optical depth and the impedance of a medium. But can local effective
medium parameters be found in such a way as to predict $\theta$ and
${\mathscr Z}$ correctly and simultaneously? Standard EMTs do allow
this asymptotically in the limit $h\rightarrow 0$. In contrast,
extended EMTs that consider an infinite medium cannot predict
correctly the impedance of the medium.  The current-driven
homogenization theory is of this variety: it predicts correctly the
first nonvanishing correction to $\theta$ (compared to the standard
homogenization result) but does not provide any meaningful corrections
or approximations to ${\mathscr Z}$.  Moreover, this correction to
$\theta$ yields a valid approximation only in a limited range of $h$,
as will be shown below.  In a more general case, it is not possible to
find local parameters that predict correctly (for all angles of
incidence) even $\theta$ alone.  Therefore, the claims that the
current-driven homogenization theory is rigorous and completely
general~\cite{silveirinha_07_1,silveirinha_09_1,alu_11_2} are
exaggerated.

We finally note that the above analysis could be extended to
three-dimensional orthorhombic lattices by integrating out
higher-order harmonics in the $xy$-plane, i.e. by low-pass filtering.

\section{Numerical examples}
\label{sec:num}

In this section, we consider several examples of computing the local
effective parameters of one-dimensional layered media according to the
current-driven homogenization prescription. We note that the maximum
possible value of the numerical factor $\left( p_a p_b \right)^2 (1 +
2 p_a p_b)$ in (\ref{asympt}) is $3/32$ and it is achieved when
$p_a=p_b=1/2$. These volume fractions are used in all the numerical
examples shown below. The $b$-type medium is assumed to be air or
vacuum with $\epsilon_b=1$. For the $a$-type medium, three different
examples will be considered. In Example A, the $a$-type medium is a
lossy dielectric considered at a fixed frequency and varying values of
$h$ and $k_x$. Example B is similar to example A, but the $a$-type
medium is a conductor. In Example C, the $a$-type medium is an
idealized Drudean metal considered at a fixed value of $h$, $k_x=0$
and varying $\lambda_0$. Thus, in Example C we account for the
frequency dispersion in metal.

The results of current-driven homogenization will be compared to the
standard homogenization result (\ref{eps_mu_hmg}) and to the results of
S-parameter
retrieval~\cite{koschny_03_1,chen_04_1,feng_10_1,menzel_08_1}. As is
well known, retrieving effective parameters of a slab from the
transmission and reflection coefficients at normal or near-normal
incidence is an ill-posed inverse problem. We have used several
different methods of regularizing the inverse solution, some of which
have been proposed in the literature~\cite{chen_04_1} and others have
been devised by us; see Appendix~\ref{app:C} for full details. All
these various modalities of the retrieval technique yield
approximately the same result when $h/\lambda_0 \ll 1$ but deviate
strongly for larger values of this ratio. In the figures below, the
retrieval results are shown only for the range of $h/\lambda_0$ within
which the technique is numerically stable.

\subsection{Example A}
\label{subsec:Ex_A}

We start with the case where the $a$-type medium is a lossy dielectric
characterized by $\epsilon_a = 4.0 + 0.1 i$ at a given wavelength
$\lambda_0$. In Example A, we assume that $h$ and $k_x$ can vary while
$\lambda_0$ is fixed. Then the physically-measurable quantities of
interest are functions of the dimensionless variables $h/\lambda_0$
and $k_x/k_0$, and the actual value of $\lambda_0$ is unimportant. The
sample consists of $N=50$ symmetric unit cells of the type
$(a/2,b,a/2)$ arranged as shown in Fig.~\ref{geom}. As noted above,
we take $a=b=h/2$ in all numerical experiments.

\begin{figure}
\psfig{file=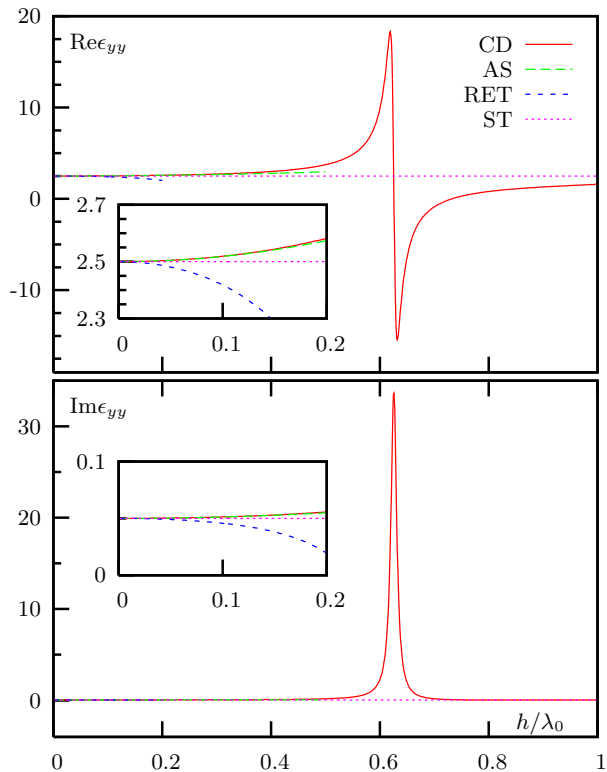,width=8.5cm,bbllx=40bp,bblly=475bp,bburx=295bp,bbury=790bp,clip=}
\caption{\label{EX_A_eyy} Example A. Real (top) and imaginary (bottom) parts of
  $\epsilon_{yy}$ as functions of $h/\lambda_0$. The various curves
  shown are obtained as follows: CD - by current-driven homogenization
  [formulas given in Appendix~\ref{app:B}]; AS - the small-$h$
  asymptotes of the former [Eq.~(\ref{asympt})]; RET - by S-parameter
  retrieval [see Appendix~\ref{app:C}]; and ST - the standard
  homogenization result $\epsilon_\parallel$ [defined in
  Eq.~(\ref{eps_hmg})]. Insets show the details of all curves for
  small $h$.}
\end{figure}

\begin{figure}
\psfig{file=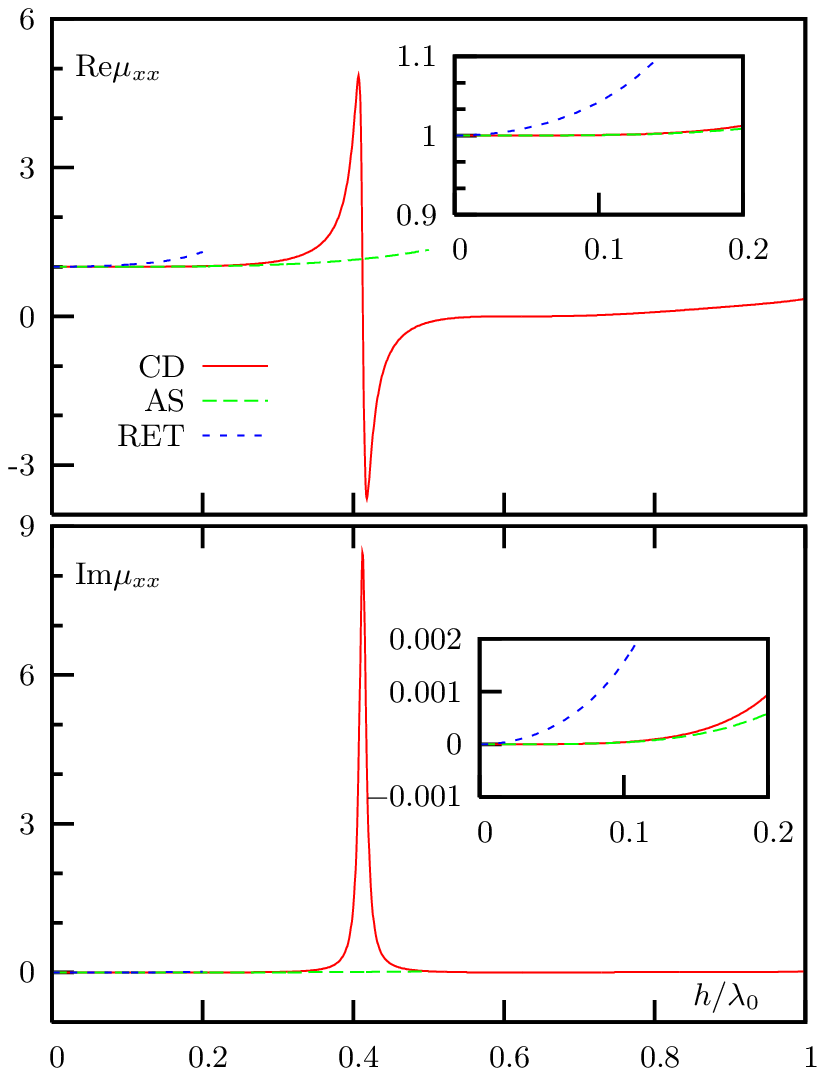,width=8.5cm,bbllx=40bp,bblly=475bp,bburx=295bp,bbury=790bp,clip=}
\caption{\label{EX_A_mxx} Example A. Same as in Fig.~\ref{EX_A_eyy} but for
  $\mu_{xx}$. The standard homogenization result $\mu_{xx}=1$ is not
  shown.}
\end{figure}

\begin{figure}
\psfig{file=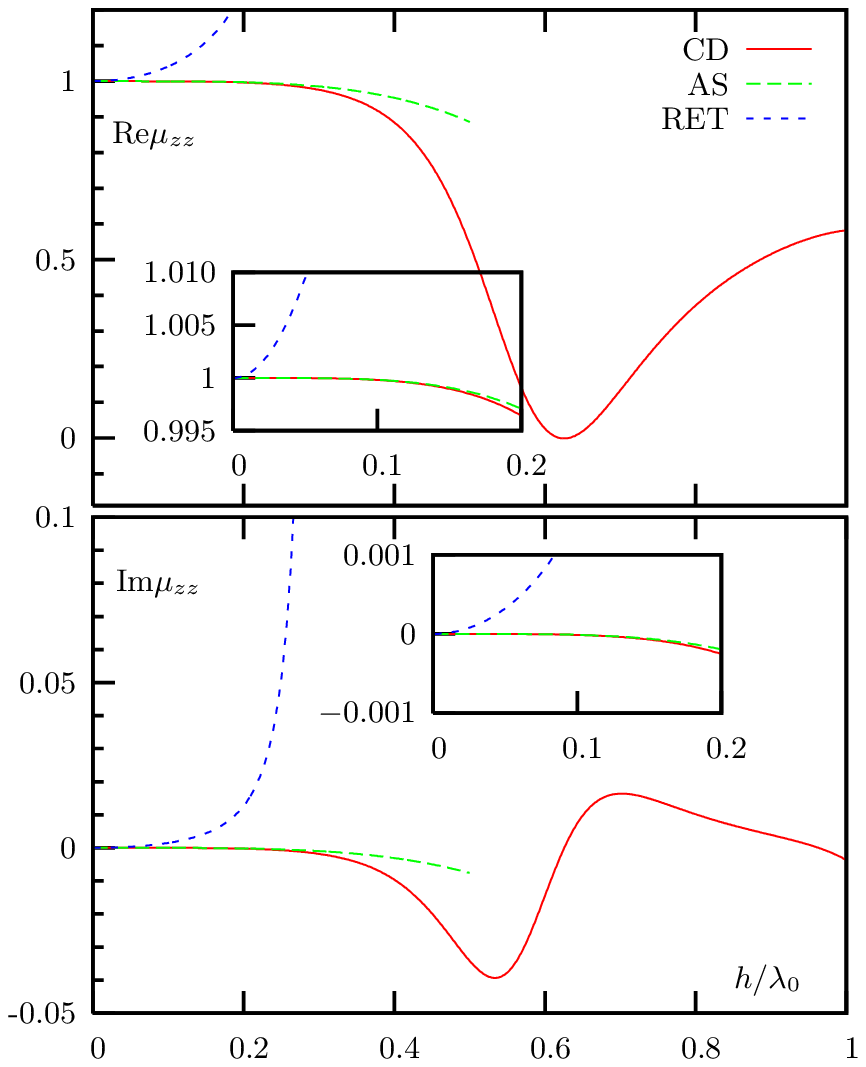,width=8.5cm,bbllx=40bp,bblly=475bp,bburx=295bp,bbury=790bp,clip=}
\caption{\label{EX_A_mzz} Example A. Same as in Fig.~\ref{EX_A_eyy}
  but for $\mu_{zz}$. The standard homogenization result $\mu_{zz}=1$
  is not shown.}
\end{figure}

In Figs.~\ref{EX_A_eyy}-\ref{EX_A_mzz}, we illustrate the predictions
of current-driven homogenization for all the components of the
permittivity and permeability tensors that can be obtained in
s-polarization. The results are compared to the predictions of the
S-parameter retrieval method. It can be seen that current-driven
homogenization produces the standard homogenization result when
$h\rightarrow 0$. This much could be inferred from considering the
asymptotic expansions (\ref{asympt}). In fact, for $h/\lambda_0
\lesssim 0.2$, all curves displayed in the figures are close to the
standard homogenization result. We note that this is, approximately,
the same range of $h$ in which S-parameter retrieval is numerically
stable. The interpretation of this fact is obvious: for sufficiently
small ratios of $h/\lambda_0$, the transmission and reflection
properties of the sample are well fitted by purely local effective
permittivity $\epsilon$ and $\mu=1$.

Both current-driven homogenization and S-parameter retrieval provide
corrections to the standard homogenization result. As long as these
corrections are small, they can yield a homogenization result that
appears to be ``reasonable''. Yet, all the dramatic features of
current-driven homogenization occur for $h/\lambda_0 > 0.2$, where the
S-parameter retrieval is unstable (that is, the retrieved result
strongly depends on the particular implementation of the retrieval
technique).  In particular, the resonance in $\mu_{xx}$ occurs at
$h/\lambda_0 \approx 0.4$. Is this result physically reasonable?  We
will address this question now by considering the transmission and
reflection coefficients of a finite slab.

\begin{figure}
\psfig{file=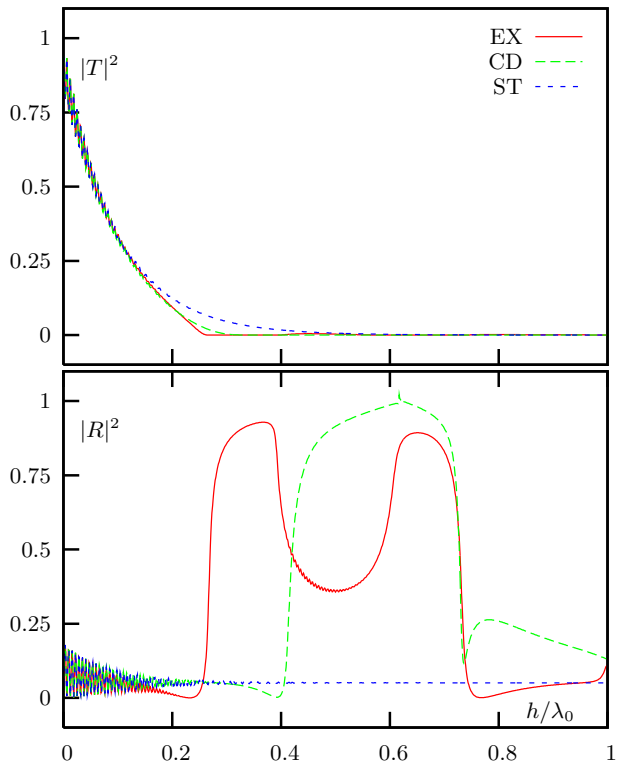,width=8.5cm,bbllx=40bp,bblly=475bp,bburx=295bp,bbury=790bp,clip=}
\caption{\label{EX_A_T2_R2_norm}
  Example A. Absolute values squared of the transmission (top) and
  reflection (bottom) coefficients at normal incidence as functions of
  $h/\lambda_0$. The various curves shown are obtained as follows: EX
  - the exact result [Eqs.~(\ref{T_R_theta_Zet})]; CD - equivalent
  homogeneous slab with current-driven effective parameters; and ST -
  same as above but for effective parameters obtained by standard
  homogenization.}
\end{figure}

In Fig.~\ref{EX_A_T2_R2_norm}, we plot $\vert T\vert^2$ and $\vert
R\vert^2$ at normal incidence as functions of $h/\lambda_0$. It can be
seen that the different methods used to compute the transmittance and
reflectance yield very similar result for $h/\lambda_0 \lesssim 0.2$
but very different results for $h/\lambda_0 >0.2$. Overall, when both
$T$ and $R$ are considered, current-driven homogenization does not
provide a meaningful correction to the standard homogenization result
(\ref{eps_mu_hmg}).  In other words, at small $h/\lambda_0$, both methods
predict approximately the same result while at larger values of
$h/\lambda_0$ both methods simultaneously fail. This is clearly
visible in the case of $R$ but is also true for $T$, which is very
small when $h/\lambda_0 >0.4$. In the latter case, both current-driven
and standard homogenization generate relative errors in $\vert T
\vert^2$ of many orders of magnitude, as could be verified by
utilizing logarithmic vertical scale (data not shown).  Of course,
this result is expected for standard homogenization, which is an
asymptotic theory. However, the current-driven homogenization theory
was claimed to have predictive power beyond the limit of small $h$
and, in particular, in the region of the parameter space where it
predicts nontrivial magnetic effects. This claim appears not to be
supported by the data of Fig.~\ref{EX_A_T2_R2_norm}.

\begin{figure}
\psfig{file=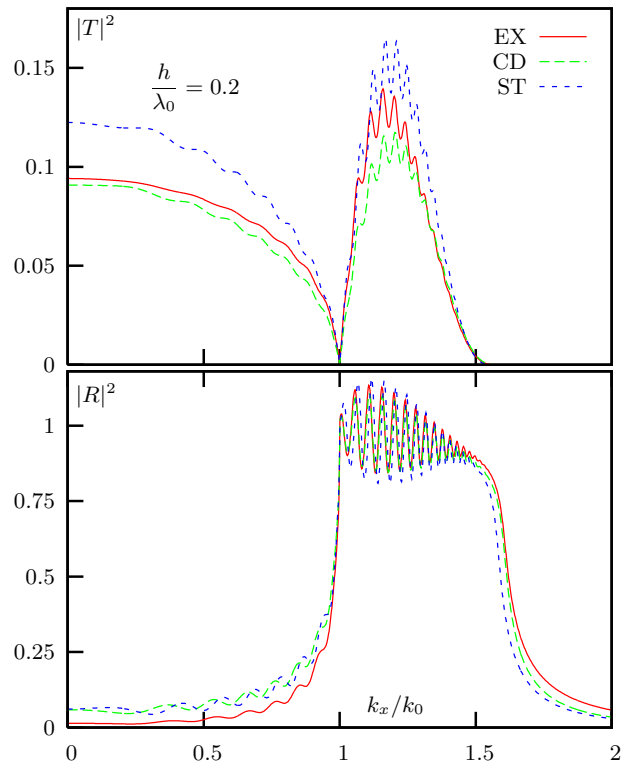,width=8.5cm,bbllx=40bp,bblly=475bp,bburx=295bp,bbury=790bp,clip=}
\caption{\label{EX_A_T2_R2_h=0.2_kx}
  Example A. Absolute values squared of the transmission (top) and
  reflection (bottom) coefficients computed as functions of $k_x/k_0$
  for $h/\lambda_0=0.2$. Same curve labels as in
  Fig.~\ref{EX_A_T2_R2_norm} have been used.}
\end{figure}

Nevertheless, if we focus on $T$ alone, current-driven homogenization
provides a slightly more accurate result compared to standard
homogenization when $h/\lambda_0$ is in a small vicinity of $0.2$. Let
us, therefore, consider in more detail transmission and reflection by
the slab at $h/\lambda_0=0.2$. The effective parameters obtained at
this value of $h/\lambda_0$ by current-driven homogenization are
listed in Table~\ref{tab:1} and the dependence of $\vert T \vert^2$
and $\vert R \vert^2$ on the angle of incidence is illustrated in
Fig.~\ref{EX_A_T2_R2_h=0.2_kx}. In the case of $T$, current-driven
homogenization provides a noticeable improvement over the standard
homogenization result when $k_x<k_0$ (in fact, the standard formula
predicts the phase of $T$ incorrectly in this range of $k_x$) but not
when $k_x>k_0$, i.e., not when the incident wave is evanescent. In the
case of $\vert R \vert^2$, no improvement is observed. We note that
the values of $\vert R \vert^2$ in Fig.~\ref{EX_A_T2_R2_h=0.2_kx} can
exceed unity for $k_x>k_0$, when both the incident and the reflected
waves are evanescent.

\begin{figure}
\psfig{file=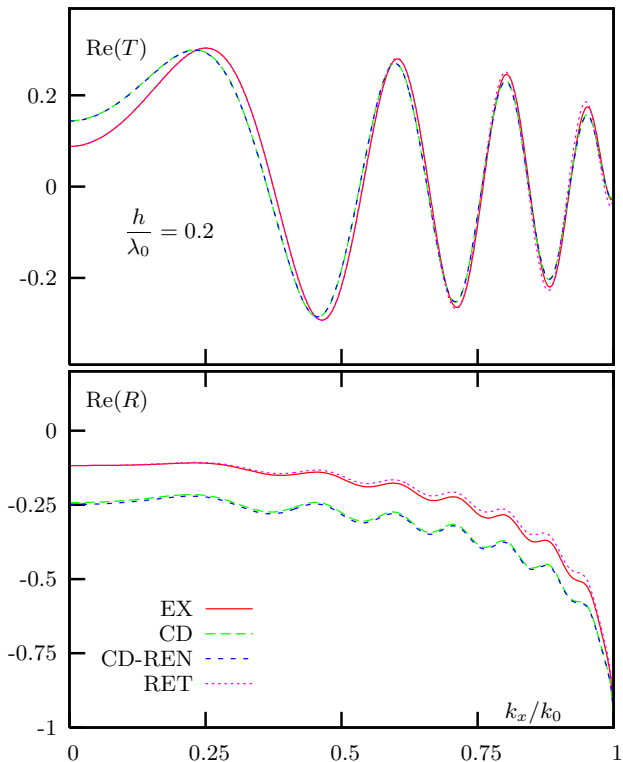,width=8.5cm,bbllx=40bp,bblly=475bp,bburx=295bp,bbury=790bp,clip=}
\caption{\label{EX_A_Re_T_Re_R_h=0.2_kx} Example A. Same as in
  Fig.~\ref{EX_A_T2_R2_h=0.2_kx} but for the real parts of $T$ and
  $R$. Additional curve labels: CD-REN - current-driven homogenization
  with renormalization (\ref{renorm}); and RET - S-parameter
  retrieval. See Table~\ref{tab:1} for numerical values of the
  effective parameters used for each curve. Note that CD and CD-REN
  curves are visually indistinguishable; EX and RET curves are
  indistinguishable in the upper plot but slightly different in the
  bottom plot.}
\end{figure}

We will discuss below the reason why current-driven homogenization
predicts $\vert T \vert^2$ more accurately than the standard
homogenization result at $h/\lambda_0 = 0.2$, but it is useful to note
right away that this has nothing to do with an accurate prediction of
$\mu$. In fact, the values of $\mu$ computed by current-driven
homogenization are not optimal. To illustrate this point, consider the
data of Fig.~\ref{EX_A_Re_T_Re_R_h=0.2_kx}. Here we plot the real
parts of $T$ and $R$ and introduce two additional curves.  The first
of these curves (labeled CD-REN) was obtained by taking the
current-driven effective parameters listed in Table~\ref{tab:1} and
renormalizing them according to the formula
\begin{equation}
\label{renorm}
\epsilon \rightarrow \xi \epsilon \ , \ \ \mu \rightarrow
\mu/\xi \ .
\end{equation}
with the renormalization factor $\xi=\mu_{xx}$. In (\ref{renorm}),
$\epsilon$ nd $\mu$ are tensors while $\xi$ is a scalar.
Renormalization (\ref{renorm}) does not affect the equivalence of the
induced currents (\ref{Jd1}) and (\ref{Jd2b}) when each formula is
evaluated on-shell, that is, for ${\bf k}={\bf q}$. The renormalized
parameters are also given in Table~\ref{tab:1}. It can be seen that
the renormalized effective parameters have dramatically different
values of $\mu - 1$. Yet, $T$ and $R$ computed by using both sets of
parameters are virtually indistinguishable.

Moreover, the effective parameters obtained by current-driven
homogenization are in no way optimal if the goal of homogenization is
to fit the transmission and reflection data as closely as possible.
The latter aim is, in fact, achieved by the S-parameter retrieval
procedure. In Fig.~\ref{EX_A_Re_T_Re_R_h=0.2_kx} we show an additional
curve (labeled RET), which was computed using the effective parameters
obtained by S-parameter retrieval. More specifically, $\epsilon_{yy}$
and $\mu_{xx}$ have been computed by Method 2 and $\mu_{zz}$ was
computed by Method 3, where the various methods of S-parameter
retrieval are described in Appendix~\ref{app:C}. The particular choice
of methods is explained as follows: Method 2 is more stable
numerically but, unlike Method 3, it does not allow one to compute
$\mu_{zz}$. Returning to Fig.~\ref{EX_A_Re_T_Re_R_h=0.2_kx}, we
observe that the curve labeled RET provides a much better fit to both
$T$ and $R$ in a wide range of incidence angles. This is in spite of
the fact that the effective parameters labeled as RET in
Table~\ref{tab:1} are very different from those labeled as either CD
or CD-REN.

At this point, we note that the magnetic effects predicted by
current-driven homogenization at $h/\lambda_0 = 0.2$ are tiny; $\vert
\mu - 1 \vert$ does not exceed $\sim 0.01$ and the condition of weak
nonlocality is very well satisfied. Yet the relative errors in $T$ and
$R$ produced by current-driven homogenization are significant -- they
are at least of the same order of magnitude as $\vert \mu - 1 \vert$
or greater. In particular, the relative errors in ${\rm Re}(R)$ or
$\vert R \vert^2$ at normal incidence exceed $100\%$. To distinguish
the two effects, it would suffice to measure the reflection
coefficient at normal incidence.

\begin{table}
\begin{tabular}{|r|r|r|}
\hline
Effective            &                   &                   \\
Parameters           & $h/\lambda_0=0.2$ & $h/\lambda_0=0.3$ \\
\hline
                     & \multicolumn{2}{|c|}{ST} \\
\hline
$\epsilon_\parallel$ & \multicolumn{2}{|c|}{$2.50 + i 0.05$}\\
$\mu_\parallel$      & \multicolumn{2}{|c|}{$1 + i 0$}\\
$\mu_\perp$          & \multicolumn{2}{|c|}{$1 + i 0$}\\
\hline
                     & \multicolumn{2}{|c|}{CD} \\
\hline
$\epsilon_{yy} - \epsilon_\parallel$    & $ (820. + i 56.6)
10^{-4} $ & $(21.4 + i 6.05) 10^{-2}$ \\
$\mu_{xx}-1$                            & $ (126. + i 9.45) 10^{-4} $
& $(11.5 + i 1.11) 10^{-2}$\\
$\mu_{zz}-1$                            & $-(35.9 + i 2.55) 10^{-4} $
& $-(24.0 + i 1.84) 10^{-3}$\\
$n=\epsilon_{yy}\mu_{xx}$               & $ (261.  + i 5.88) 10^{-2}$
& $(30.3 + i 1.03) 10^{-1}$ \\
\hline
                & \multicolumn{2}{|c|}{CD-REN} \\
\hline
$\epsilon_{yy} - \epsilon_\parallel$    & $ (114. + i 8.80)
10^{-3}$ & $ (52.5 + i 5.32) 10^{-2}$ \\
$\mu_{xx}-1$                            & $  0$ & 0\\
$\mu_{zz}-1$                            & $-(16.0 + i 1.17) 10^{-3}$ &
$ -(12.5 + i 1.04)10^{-2}$
\\
$n=\epsilon_{yy}\mu_{xx}$               & $ (261. + i 5.88) 10^{-2}$
& $(30.3 + i 1.03) 10^{-1}$\\
\hline
                & \multicolumn{2}{|c|}{RET} \\
\hline
$\epsilon_{yy} - \epsilon_\parallel$    & $-(48.4 + i 3.07)
10^{-2}$ & N/A \\
$\mu_{xx}-1$                            & $ (30.3 + i 1.74) 10^{-2}$ &
N/A \\
$\mu_{zz}-1$                            & $ (25.9 + i 1.27) 10^{-2}$ &
N/A \\
$n=\epsilon_{yy}\mu_{xx}$               & $ (263.  + i 6.03) 10^{-2}$
& N/A \\
\hline
\end{tabular}
\caption{\label{tab:1} Effective parameters for Example A obtained by various methods
  at $h/\lambda_0=0.2$ and $h/\lambda_0=0.3$. ST - the standard
  homogenization result; CD - current-driven homogenization; CD-REN -
  current-driven homogenization with renormalization (\ref{renorm}); RET - retrieved
  parameters. The numbers have been rounded off to three
  significant figures. However, all plots shown in
  this paper utilize either double or quadruple-precision computations.}
\end{table}

\begin{figure}
\psfig{file=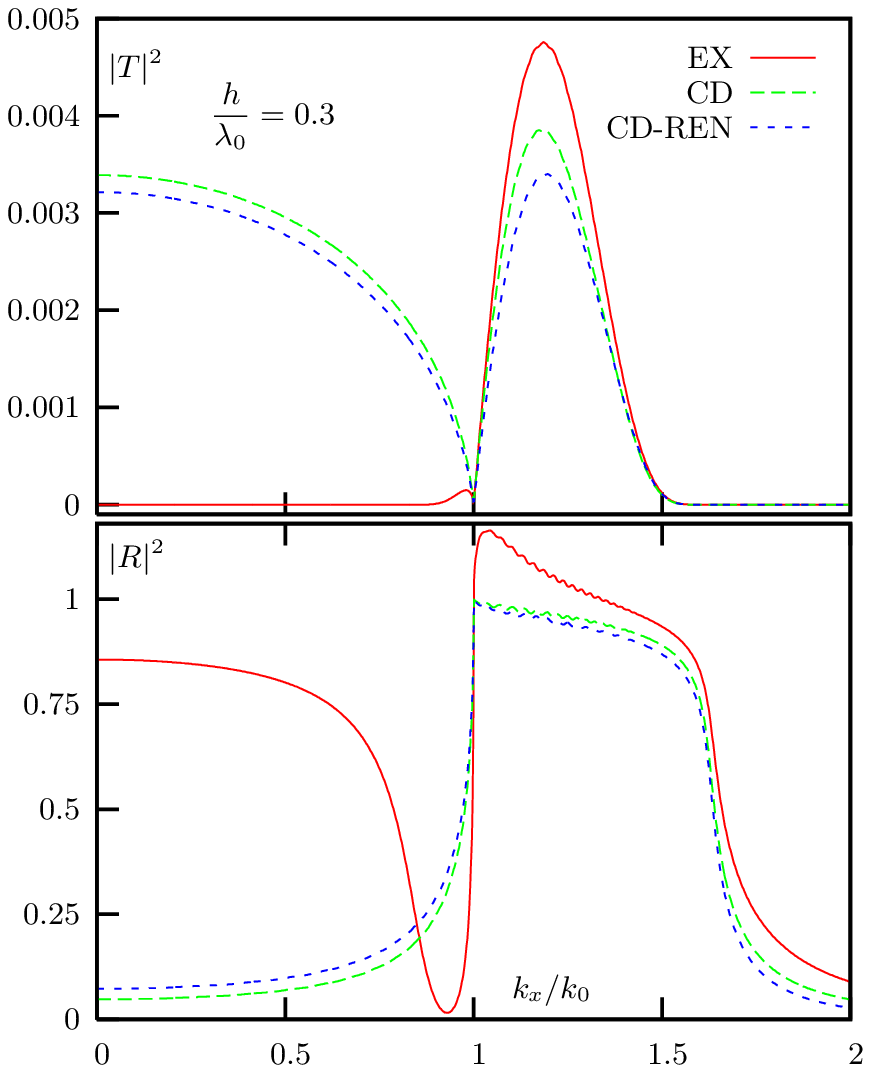,width=8.5cm,bbllx=40bp,bblly=475bp,bburx=295bp,bbury=790bp,clip=}
\caption{\label{EX_A_T2_R2_h=0.3_kx}
  Example A. Same as in Fig.~\ref{EX_A_T2_R2_h=0.2_kx} but for
  $h/\lambda_0 = 0.3$.}
\end{figure}

\begin{figure}
\psfig{file=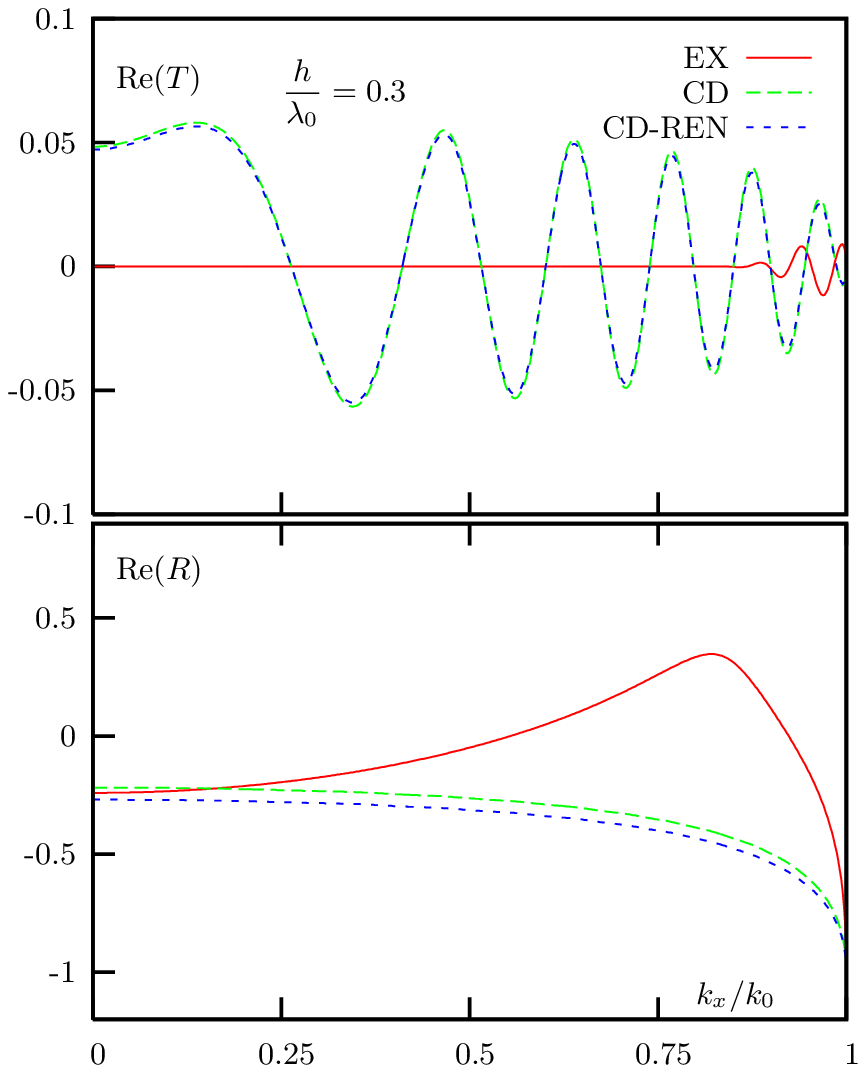,width=8.5cm,bbllx=40bp,bblly=475bp,bburx=295bp,bbury=790bp,clip=}
\caption{\label{EX_A_Re_T_Re_R_h=0.3_kx}
  Example A. Same as in Fig.~\ref{EX_A_Re_T_Re_R_h=0.2_kx} but for
  $h/\lambda_0 = 0.3$.}
\end{figure}

Now let us turn to the case $h/\lambda_0=0.3$, which is illustrated in
Figs.~\ref{EX_A_T2_R2_h=0.3_kx},\ref{EX_A_Re_T_Re_R_h=0.3_kx}.
S-parameter retrieval is unstable at this point and standard
homogenization is inapplicable (the corresponding curves are not shown).
Therefore, if current-driven homogenization could produce reasonable
predictions at $h/\lambda_0=0.3$, it would constitute a valid and
useful approximation. However, the data of
Figs.~\ref{EX_A_T2_R2_h=0.3_kx},\ref{EX_A_Re_T_Re_R_h=0.3_kx} do not
support this hypothesis. The relative errors in $R$ and $T$ for
current-driven homogenization are at this point dramatic and can be
many orders of magnitude. For $k_x<k_0$, the relative errors are many
orders of magnitude in $T$ and at least an order of magnitude in $R$.
Both the phase and the amplitude of $T$ and $R$ are predicted with
significant errors in the whole range of $k_x$ considered. We note
that, in the bottom plot of Fig.~\ref{EX_A_Re_T_Re_R_h=0.3_kx}, the
real part of $R$ is reproduced correctly at normal incidence by
current-driven homogenization. However, as can be seen from the data
of Fig.~\ref{EX_A_T2_R2_h=0.3_kx}, $\vert R \vert^2$ at normal
incidence is predicted with a large error. Consequently, ${\rm Im}R$
is also predicted with a large error (data not shown).

All this is in spite of the fact that current-driven homogenization
does not yet predict any dramatic magnetic effects at
$h/\lambda_0=0.3$. Indeed, for this value of $h/\lambda_0$,
current-driven homogenization predicts that $\vert \mu - 1 \vert$ does
not exceed $\approx 0.1$.  Moreover, in can be seen from the data of
Figs.~\ref{EX_A_T2_R2_h=0.3_kx},\ref{EX_A_Re_T_Re_R_h=0.3_kx} that
renormalization of effective parameters according to (\ref{renorm})
does not worsen or, indeed, noticeably modify the predictions of
current-driven homogenization.

At even larger values of $h/\lambda_0$, predictions of current-driven homogenization
are widely inaccurate. In particular,
current-driven homogenization cannot be relied upon at
$h/\lambda_0=0.4$, when $\mu_{xx}$ experiences a dramatic resonance.
This is evident from the data of Fig.~\ref{EX_A_T2_R2_norm} and there
is no need to support this conclusion with additional graphics. A
question, however, remains: why did we observe a moderate
improvement in $T$ at $h/\lambda_0=0.2$? The
reason is that current-driven homogenization provides a first
nonvanishing correction to the Bloch wave number $q_z$ but not to the
impedance ${\mathscr Z}$. Both quantities enter the formulas for $T$
and $R$ (\ref{T_R_theta_Zet}). As was discussed in Sec.~\ref{sec:TR},
under some circumstances, $T$ can be more sensitive to errors in $q_z$
than to errors in ${\mathscr Z}$.  This does not mean that errors in
${\mathscr Z}$ are insignificant. As could be seen in
Figs.~\ref{EX_A_Re_T_Re_R_h=0.2_kx}, an error in the impedance that
current-driven homogenization entails translates into an error in $T$
and $R$, which is at least of the same order of magnitude or greater
than $\vert \mu - 1 \vert$.

\begin{figure}
\psfig{file=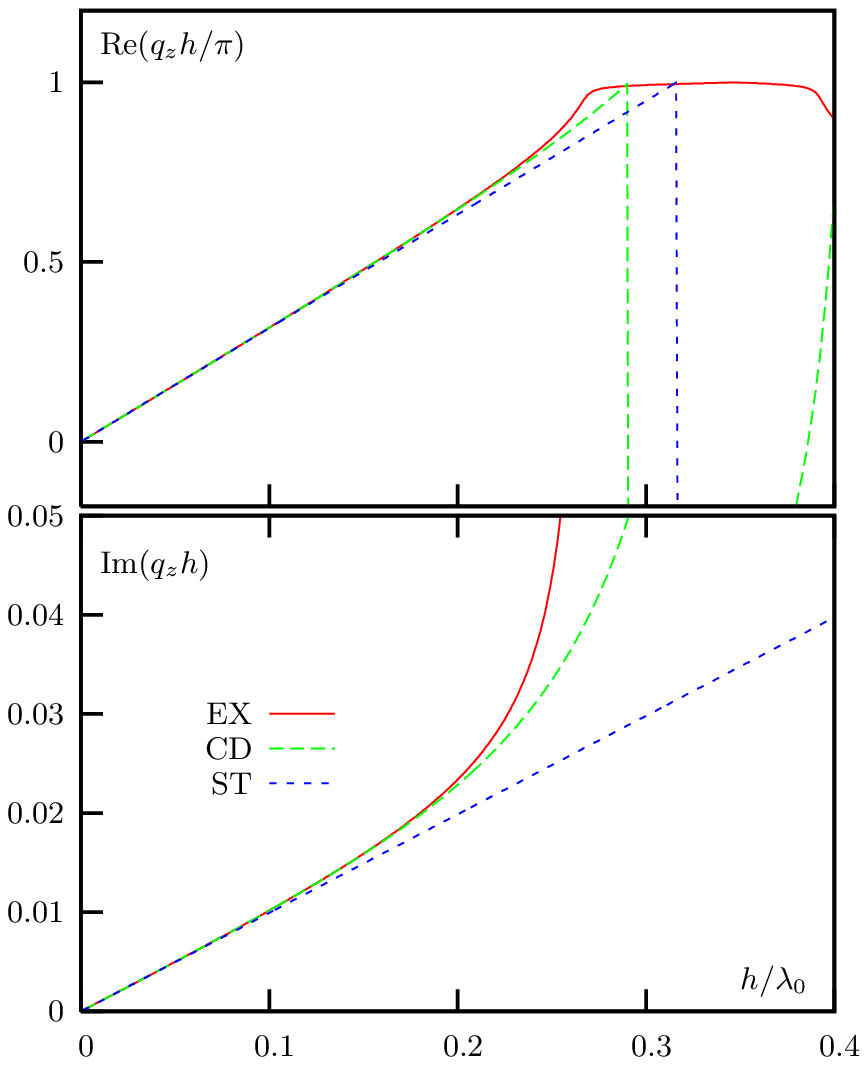,width=8.5cm,bbllx=40bp,bblly=475bp,bburx=295bp,bbury=790bp,clip=}
\caption{\label{EX_A_Re_qh_Im_qh_norm}
  Example A. Real (top) and imaginary (bottom) parts of the unit cell
  optical depth parameter, $q_zh$, at normal incidence as a function
  of $h/\lambda_0$.}
\end{figure}

\begin{figure}
\psfig{file=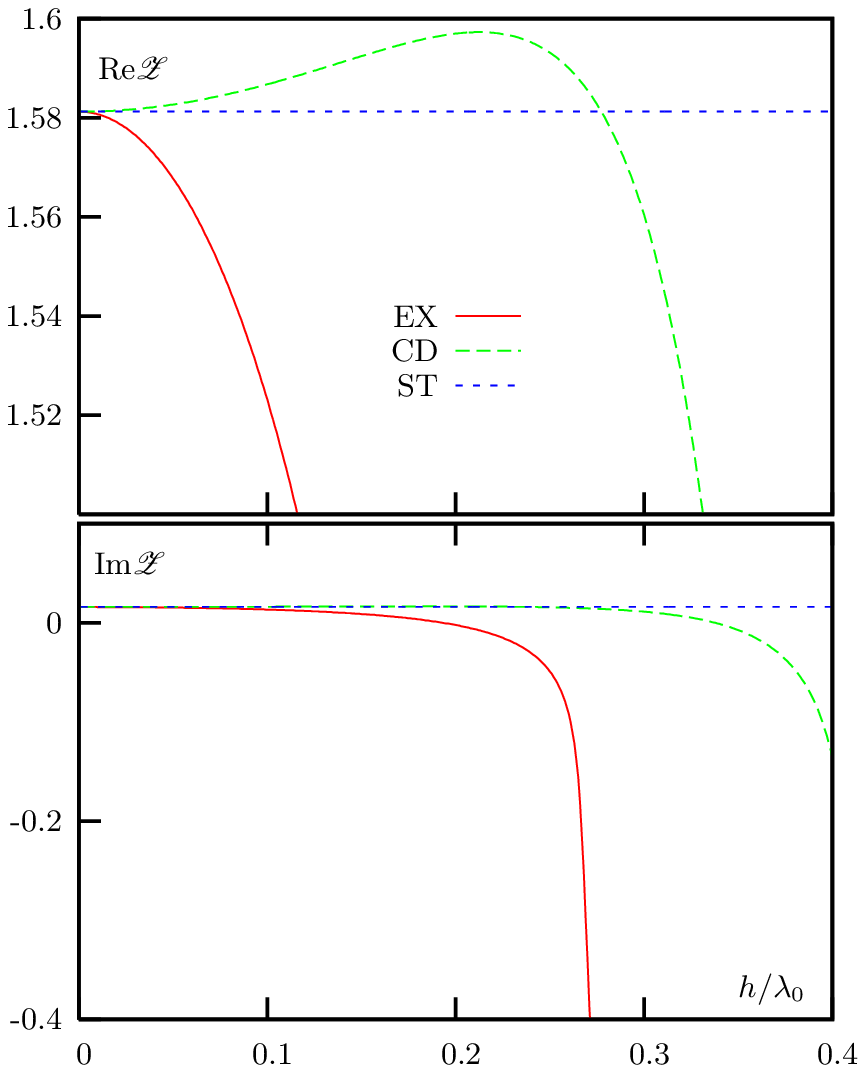,width=8.5cm,bbllx=40bp,bblly=475bp,bburx=295bp,bbury=790bp,clip=}
\caption{\label{EX_A_Re_zet_Im_zet_norm}
  Example A. Real (top) and imaginary (bottom) parts of the
  generalized impedance ${\mathscr Z}$, at normal incidence as a
  function of $h/\lambda_0$.}
\end{figure}

The above point is illustrated in
Figs.~\ref{EX_A_Re_qh_Im_qh_norm},\ref{EX_A_Re_zet_Im_zet_norm}. Here
we plot $q_z h$ and ${\mathscr Z}$ at normal incidence as functions of
$h/\lambda_0$; predictions of current-driven and standard
homogenization are compared to the exact result given in
(\ref{q_Zet_Layered}),(\ref{q_z_B_gen}). It can be seen that, at
$h/\lambda_0 \approx 0.2$, current-driven homogenization provides a
slightly more accurate result for $q_z$. In the case of ${\mathscr
  Z}$, current-driven homogenization does not provides a better
approximation at any value of $h/\lambda_0$.

\subsection{Example B}
\label{subsec:Ex_B}

\begin{figure}
\psfig{file=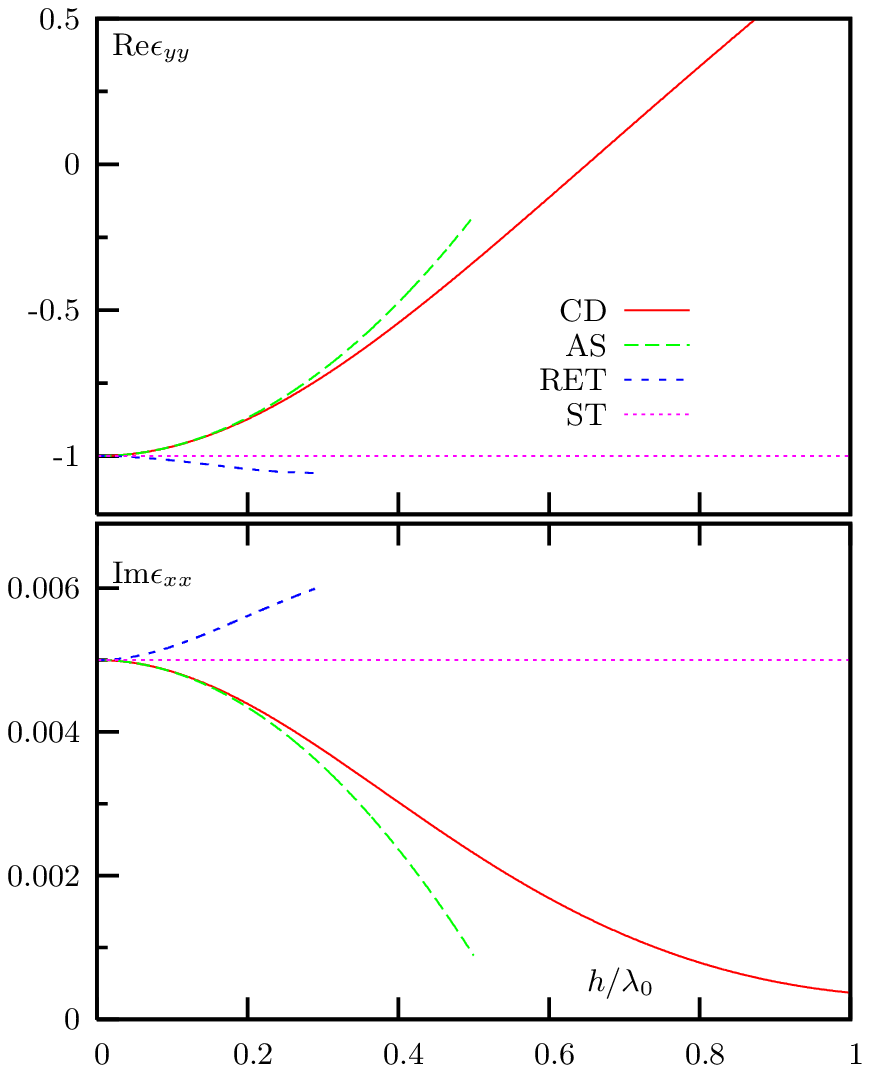,width=8.5cm,bbllx=40bp,bblly=475bp,bburx=295bp,bbury=790bp,clip=}
\caption{\label{EX_B_eyy} Example B. Real (top) and imaginary (bottom) parts of
  $\epsilon_{yy}$ as functions of $h/\lambda_0$. Same curve labels as
  in Fig.~\ref{EX_A_eyy} have been used.}
\end{figure}

\begin{figure}
\psfig{file=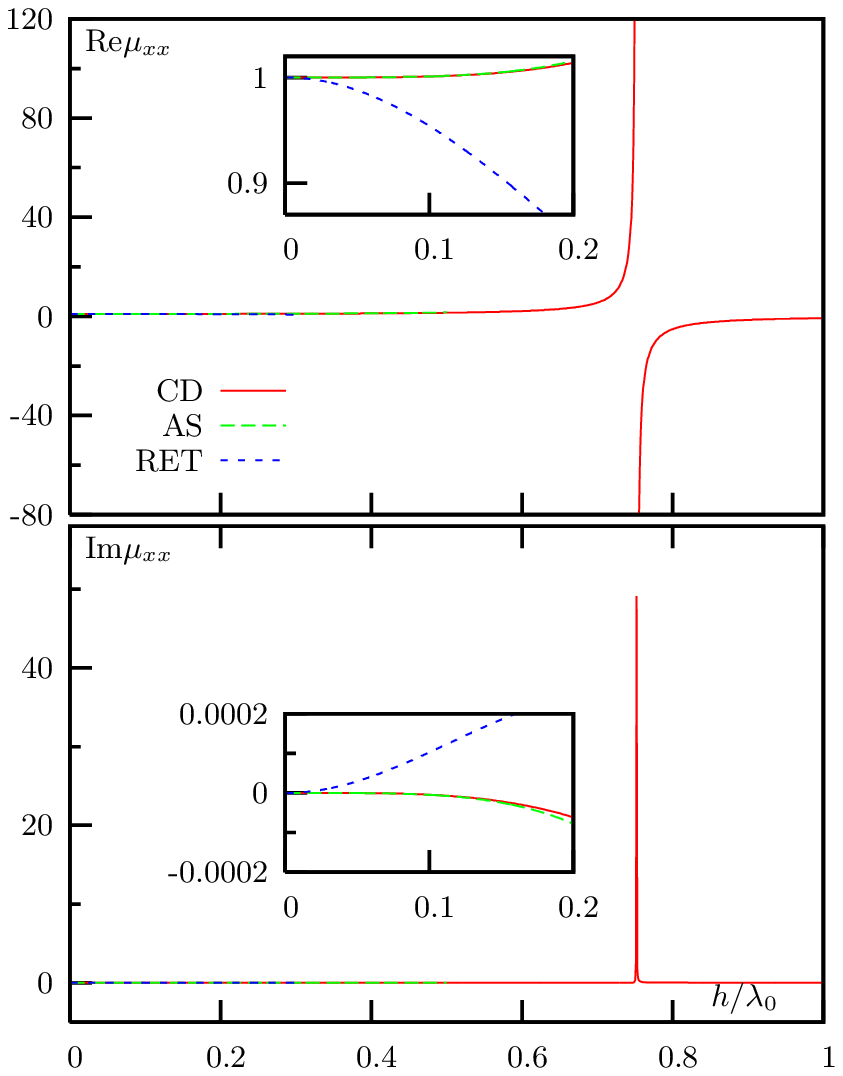,width=8.5cm,bbllx=40bp,bblly=475bp,bburx=295bp,bbury=790bp,clip=}
\caption{\label{EX_B_mxx} Example B. Same as in Fig.~\ref{EX_B_eyy}
  but for $\mu_{xx}$. The standard homogenization result $\mu_{xx}=1$
  is not shown.}
\end{figure}

\begin{figure}
\psfig{file=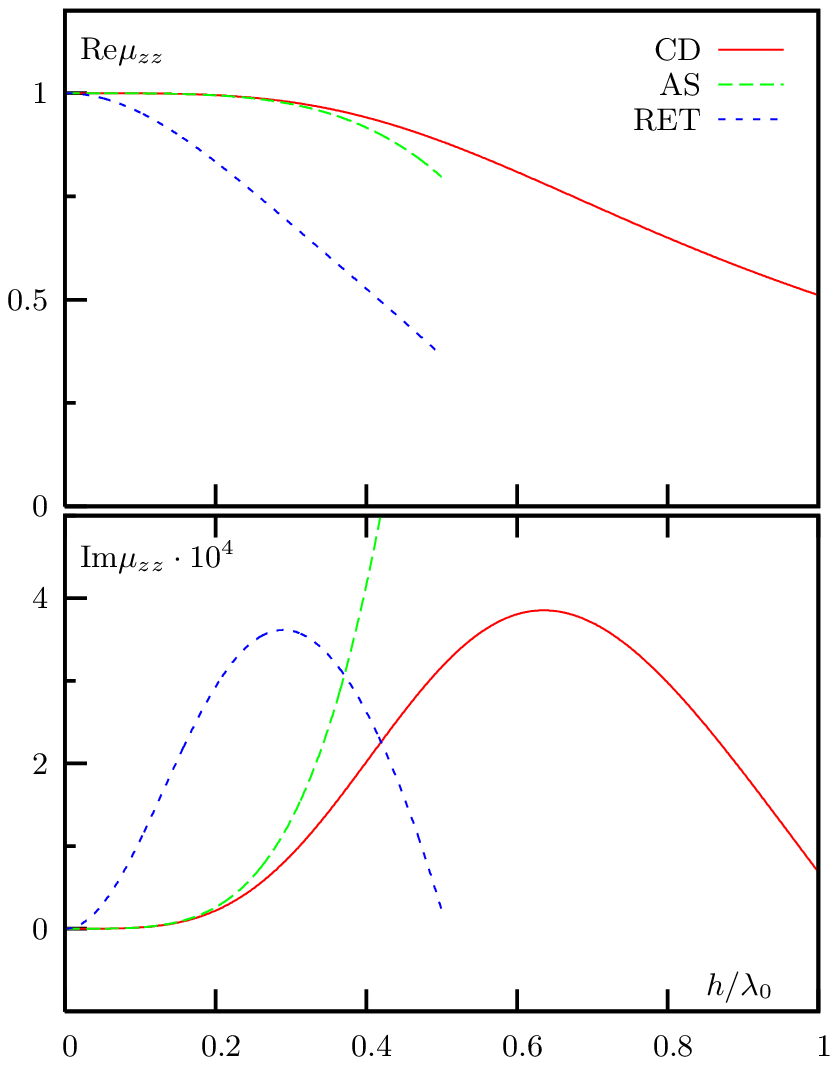,width=8.5cm,bbllx=40bp,bblly=475bp,bburx=295bp,bbury=790bp,clip=}
\caption{\label{EX_B_mzz} Example B. Same as in Fig.~\ref{EX_B_eyy}
  but for $\mu_{zz}$. The standard homogenization result $\mu_{zz}=1$
  is not shown.}
\end{figure}

We now turn to the case when the $a$-type medium is a
high-conductivity metal with $\epsilon_a=-3+0.01i$ at a given fixed
wavelenghth $\lambda_0$. The sample consists of $5$ symmetric unit
cells of the type $(a/2,b,a/2)$, where $a=b$ as before.  Effective
parameters for Example B are plotted as functions of $h/\lambda_0$ in
Figs.~\ref{EX_B_eyy}-\ref{EX_B_mzz}.  Current-driven homogenization
predicts that $\mu_{xx}$ experiences a resonance near the point
$h/\lambda_0 = 0.75$ while $\mu_{zz}$ exhibits no dramatic effects.

\begin{figure}
\psfig{file=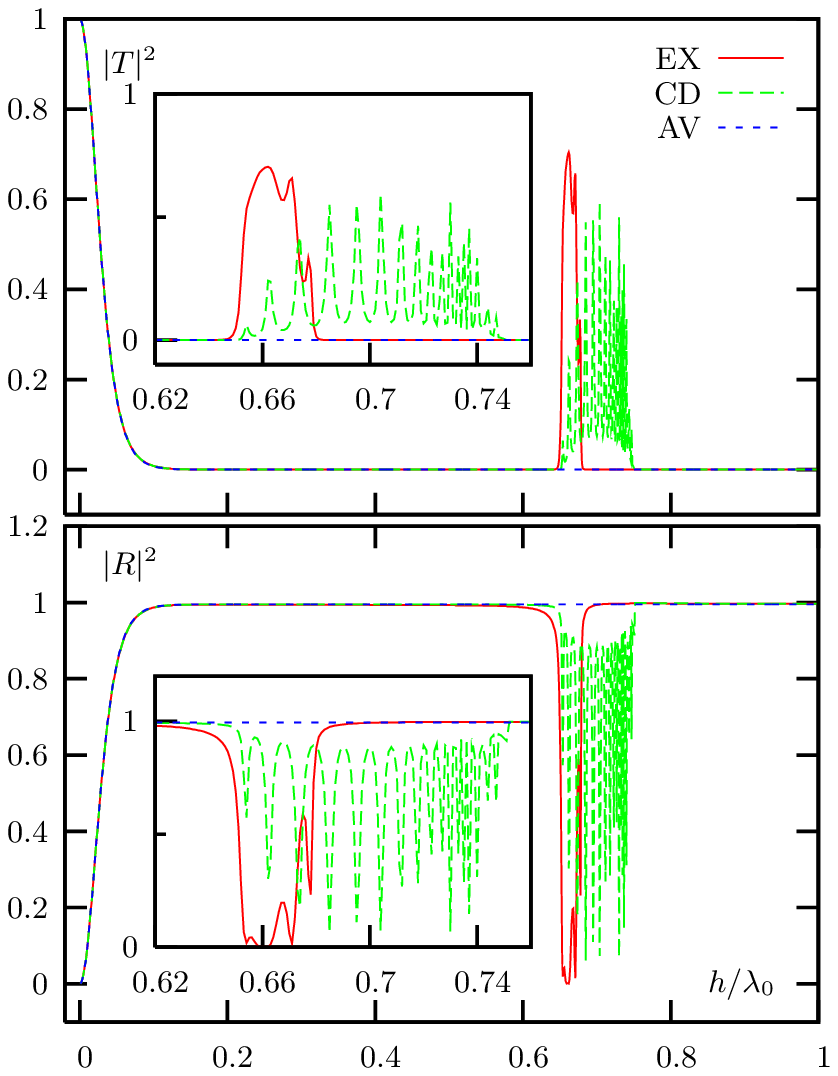,width=8.5cm,bbllx=40bp,bblly=475bp,bburx=295bp,bbury=790bp,clip=}
\caption{\label{EX_B_T2_R2_norm} Same as in
  Fig.~\ref{EX_A_T2_R2_norm} but for Example B.}
\end{figure}

\begin{figure}
\psfig{file=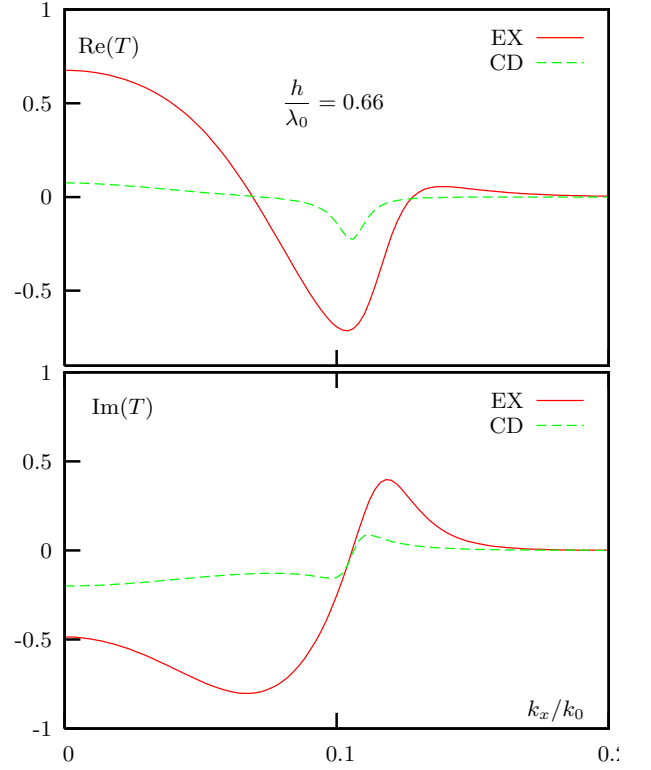,width=8.5cm,bbllx=40bp,bblly=475bp,bburx=295bp,bbury=790bp,clip=}
\caption{\label{EX_B_Re_T_Im_T_h=0.66_kx}
  Example B. Real (top) and imaginary (bottom) parts of $T$ as
  functions of $k_x/k_0$ for $h/\lambda_0 =0.66$. EX - exact result,
  CD - current-driven homogenization. Only the range of $k_x/k_0$ is
  shown for which $T$ computed by either method is not negligibly
  small.}
\end{figure}

\begin{figure}
\psfig{file=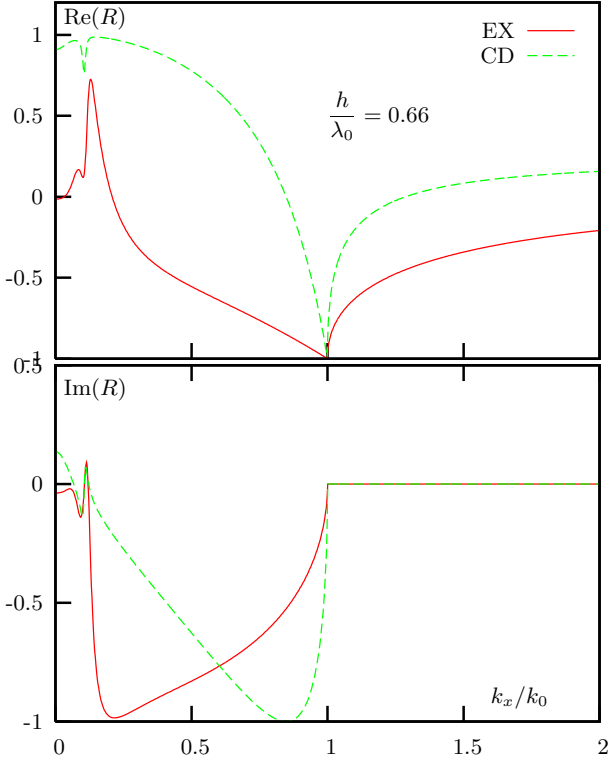,width=8.5cm,bbllx=40bp,bblly=475bp,bburx=295bp,bbury=790bp,clip=}
\caption{\label{EX_B_Re_R_Im_R_h=0.66_kx} Example B. Same as in
  Fig.~\ref{EX_B_Re_T_Im_T_h=0.66_kx} but for $R$.}
\end{figure}

In Fig.~\ref{EX_B_T2_R2_norm}, we display the predictions of various
theories for $\vert T \vert^2$ and $\vert R \vert^2$. The conclusion
that can be made is that current-driven homogenization does not
provide a meaningful correction or a noticeable improvement of
precision compared to standard homogenization in the whole range of
$h/\lambda_0$ considered. In fact, there are fairly significant
intervals of $h/\lambda_0$ (clearly visible in the insets) in which
standard homogenization predicts correctly $\vert T \vert^2 \approx
0$ and $\vert R \vert^2 \approx 1$ while current driven homogenization
is widely off the mark. Perhaps, current-driven homogenization can be
credited with predicting a transparency window which exists in reality
for relatively large values of $h/\lambda_0$ and which is not
predicted for obvious reasons by standard homogenization formulas.
Unfortunately, the transparency window is predicted for wrong values
of $h/\lambda_0$ and, in the true transparency window, both $T$ and
$R$ are predicted with the wrong phase and amplitude. This is
illustrated in Figs.~\ref{EX_B_Re_T_Im_T_h=0.66_kx} and
\ref{EX_B_Re_R_Im_R_h=0.66_kx} where we plot real and imaginary parts
of both $T$ and $R$ as functions of $k_x/k_0$. It can be seen that
there is no correspondence between current-driven homogenization and
exact result.

The discrepancy is even more pronounced for the values of
$h/\lambda_0$ such that the exact transmission coefficient is close to
zero but current-driven homogenization predicts significant
transmission.

\begin{figure}
\psfig{file=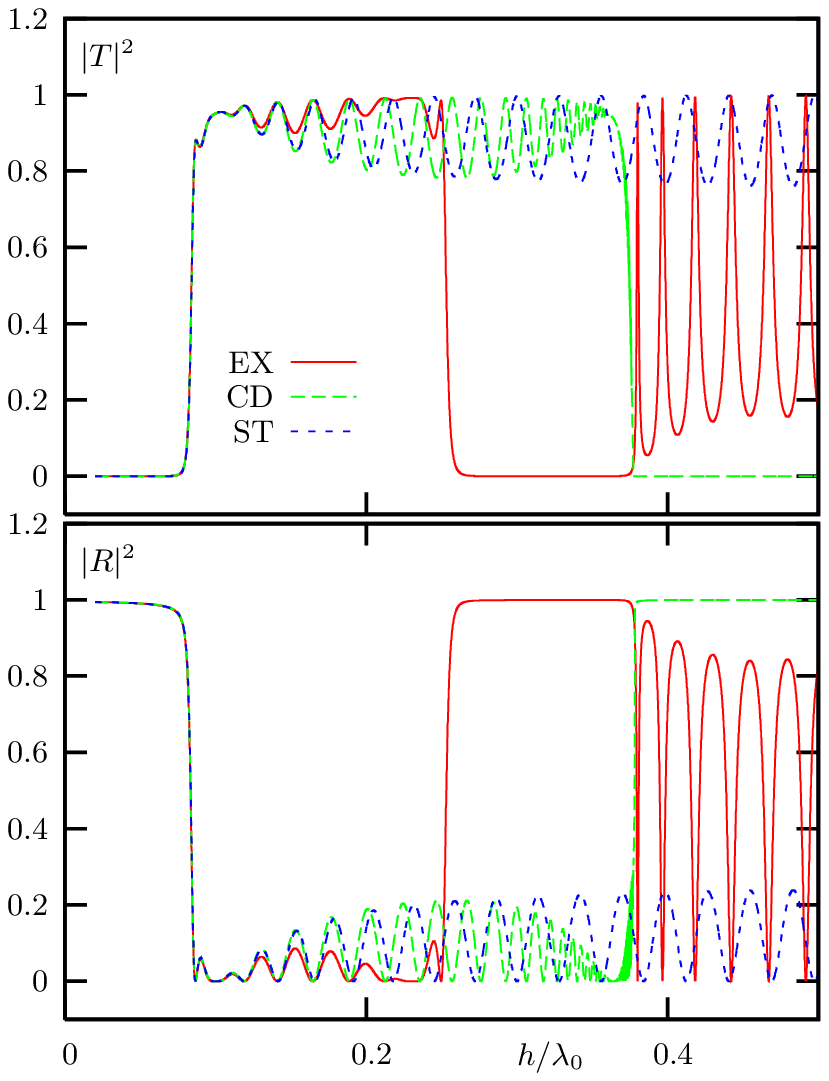,width=8.5cm,bbllx=40bp,bblly=475bp,bburx=295bp,bbury=790bp,clip=}
\caption{\label{EX_C_T2_R2_norm} Example C. Same as in
  Fig.~\ref{EX_A_T2_R2_norm} but for Example C. Note that, in Example
  C, $h$ is fixed while $\lambda_0$ varies.}
\end{figure}

\subsection{Example C}

In this example, we consider spectral dependencies of $\vert R
\vert^2$ and $\vert T\vert^2$ at normal incidence with the account of
frequency dispersion in the constituents of the composite. The
$a$-type medium is an idealized Drudean metal described by the
permittivity function
\begin{equation}
\label{Drude}
\epsilon_a = \epsilon_0 - \frac{\omega_p^2}{\omega(\omega + i \gamma)}
\ .
\end{equation}
\noindent
and the $b$-type medium is vacuum or air. The sample consists of
$N=10$ symmetric unit cells of the type $(a/2,b,a/2)$, where, again,
$a=b$.  We have chosen the parameters in (\ref{Drude}) to represent
the experimental values for silver: $\epsilon_0=5$ and
$\omega_p/\gamma=500$. The lattice period $h$ is fixed so that
$h/\lambda_p = 0.2$, where $\lambda_p = 2\pi c / \omega_p$ is the
wavelength at the plasma frequency $\omega_p$.  In the case of silver,
$\lambda_p = 136{\rm nm}$ so that $h\approx 27{\rm nm}$. The
free-space wavelength $\lambda_0$ is varied. In this case, all
physical quantities of interest can be expressed as functions of the
dimensionless variables $h/\lambda_0$, $\lambda_0/\lambda_p$ and
$k_x/k_0$ (we will take $k_x=0$ in this example).

We do not display the effective parameters obtained by different
methods as nothing qualitatively new compared to the previously
considered examples emerges in Example C. Note that the current-driven
permeability $\mu_{xx}$ experiences a sharp resonance at $h/\lambda_0
\approx 0.38$ while $\epsilon_{yy}$ and $\mu_{zz}$ do not exhibit any
dramatic features. In Fig.~\ref{EX_C_T2_R2_norm}, we plot $\vert T
\vert^2$ and $\vert R \vert^2$ for $\lambda_p/\lambda_0$ varying from
$0$ to $2.5$. The corresponding parameter $h/\lambda_0$ varies from
$0$ to $0.5$. Again, the data clearly demonstrate that current-driven
homogenization does not provide a meaningful correction to the
standard result.

\section{Bloch-wave analysis of the current-driven homogenization theory}
\label{sec:Bloch}

In this section, we consider the current-driven homogenization theory
from a more general point of view. We assume that the medium is
intrinsically-nonmagnetic, three-dimensional and periodic, and that
its true permittivity function $\tilde{\epsilon}({\bf r})$ satisfies
the periodicity condition (\ref{periodicity}). Any such function can
be expanded into a Fourier series
\begin{equation}
\label{eps_def}
\tilde{\epsilon}({\bf r}) = \sum_{\bf g} \epsilon_{\bf g} e^{i{\bf g} \cdot
{\bf r}} \ ,
\end{equation}
\noindent
where
\begin{equation*}
\label{g_def}
{\bf g}=2\pi \left( \frac{\hat{\bf x} n_x}{h_x} + \frac{\hat{\bf y}
    n_y}{h_y} + \frac{\hat{\bf z} n_z}{h_z} \right) \ .
\end{equation*}
\noindent
are the reciprocal lattice vectors, which can be viewed as
three-dimensional summation indices, and $n_x$, $n_y$ and $n_z$ are
arbitrary integers. We can seek the solution to (\ref{Maxwell}) in the
form of a Bloch wave
\begin{equation*}
\label{E_Bloch}
{\bf E}({\bf r}) = \sum_{\bf g} {\bf
  E}_{\bf g} e^{i ({\bf k} + {\bf g}) \cdot {\bf r}} \ .
\end{equation*}
\noindent
The displacement ${\bf D}({\bf r}) = \tilde{\epsilon}({\bf r}){\bf
  E}({\bf r})$ can be similarly expanded. Given the periodicity of
$\tilde{\epsilon}({\bf r})$ expressed in (\ref{eps_def}), we can
find the relation between the expansion coefficients ${\bf D}_{\bf
  g}$ and ${\bf E}_{\bf g}$:
\begin{equation}
\label{Dg_Eg}
{\bf D}_{\bf g} = \sum_{\bf p} \epsilon_{\bf p} {\bf E}_{{\bf
    g} - {\bf p}} \ .
\end{equation}
\noindent
Upon substitution of the expansions into (\ref{Maxwell}), we find the
following system of equations for ${\bf E}_{\bf g}$:
\begin{eqnarray}
\label{main}
({\bf k} + {\bf g}) &\times& ({\bf k} + {\bf g})\times {\bf E}_{\bf
  g} \nonumber \\
&+& k_0^2 \left[ \sum_{\bf p} \epsilon_{\bf p} {\bf
  E}_{{\bf g} - {\bf p}} + {\bf J} \delta_{{\bf g}0} \right] = 0 \ .
\end{eqnarray}
\noindent
Equations of this kind are well known in the theory of photonic
crystals~\cite{joannopoulos_book_95,sakoda_book_05}, except that here
we have included the free term ${\bf J} \delta_{{\bf g}0}$.  However,
the following analysis (proposed by us earlier~\cite{markel_12_1}) is
rarely used.

Let us write (\ref{main}) for the special cases ${\bf g}=0$ and ${\bf
  g}\neq 0$ separately. We note that ${\bf E}_0 = {\bf E}_{\rm av}$,
where the low-pass filtered averages are defined in (\ref{averaging}),
and $\epsilon_0$ is the usual arithmetic average of
$\tilde{\epsilon}({\bf r})$ (without low-pass filtering). We thus
obtain:
\begin{subequations}
\label{main-spec_cases}
\begin{align}
\label{main_g=0}
& \underline{{\bf g}=0 \ :} \\
& {\bf k} \times {\bf k} \times {\bf E}_{\rm av} + k_0^2
\left[ \epsilon_0 {\bf E}_{\rm av} + \sum_{{\bf p} \neq 0}
  \epsilon_{\bf p} {\bf E}_{-{\bf p}} + {\bf J} \right] =  0 \ ,
\nonumber  \\
\label{main_g=/=0}
& \underline{{\bf g} \neq 0 \ :} \\
& ({\bf k} + {\bf g})\times ({\bf k} + {\bf g})\times {\bf E}_{\bf
  g} + k_0^2 \left[ \sum_{{\bf p} \neq {\bf g}} \epsilon_{\bf p}
{\bf E}_{{\bf g} - {\bf p}} + \epsilon_{\bf g} {\bf
  E}_{\rm av} \right] = 0 \ . \nonumber
\end{align}
\end{subequations}
\noindent
Note that only the first of these two equations is affected by our
choice to include the external current in (\ref{Maxwell}). We now
utilize the linearity of equation (\ref{main_g=/=0}), from which we
can write:
\begin{equation}
\label{Sigma_def_g}
\sum_{{\bf p} \neq 0} \epsilon_{\bf p} {\bf E}_{-{\bf p}} =
\left[\Sigma(\omega,{\bf k}) -
\epsilon_0 \right] {\bf E}_{\rm av} \ ,
\end{equation}
\noindent
where $\Sigma(\omega,{\bf k})$ is a tensor to be determined by solving
(\ref{main_g=/=0}). Here the factor $\epsilon_0$ (the average
permittivity of the composite) has been introduced for convenience and
does not result in any loss of generality. Then the ${\bf g}=0$
equation (\ref{main_g=0}) takes the following form:
\begin{equation}
\label{g=0_Sigma}
{\bf k} \times {\bf k} \times {\bf E}_{\rm av} + k_0^2
\left[ \Sigma(\omega,{\bf k}) {\bf E}_{\rm av} + {\bf J} \right] = 0 \ .
\end{equation}
This is, essentially, the same equation as (\ref{Maxwell_av}).
Consequently, $\Sigma(\omega,{\bf k})$ is the same tensor as the one
appearing in the current-driven homogenization theory.

We note that inclusion into Maxwell's equations of the external
current (\ref{J_ext}) is not needed to compute $\Sigma(\omega,{\bf
  k})$, which is completely defined by the infinite set of equations
(\ref{main_g=/=0}). We can refer to this set as to the cell problem.
In what follows, we assume that the cell problem can be solved by
means of linear algebra and that the tensor $\Sigma(\omega,{\bf k})$
can be computed.

Since $\Sigma(\omega,{\bf k})$ is defined completely by solving the
cell problem, it is useful to consider what would happen if we set ${\bf
  J}=0$ in (\ref{g=0_Sigma}). Obviously, this would result in an
eigenproblem
\begin{equation}
\label{g=0_Sigma_J=0}
\left[ \left({\bf k} \times {\bf k} \times \right) + k_0^2
\Sigma(\omega,{\bf k}) \right] {\bf E}_{\rm av} = 0 \ .
\end{equation}
\noindent
The above equation has nontrivial solutions only when ${\bf k} = {\bf
  q}$, where the Bloch wave vector ${\bf q}$ is determined from the equation
\begin{equation}
\label{g=0_Sigma_det=0}
{\rm det}\left[ \left({\bf k} \times {\bf k} \times \right) + k_0^2
\Sigma(\omega,{\bf k}) \right] = 0 \ .
\end{equation}
\noindent
The solution to this equation, viewed as a function of frequency,
yields the dispersion equation of the medium, ${\bf q} = {\bf
  q}(\omega)$. The dispersion relation is physically measurable and
the same is true for the on-shell tensor $\Sigma(\omega,{\bf q})$.
For example, in the simplest case of transverse waves, the dispersion
equation takes the form $q^2 = k_0^2 \Sigma(\omega,{\bf q})$ and the
quantity $\Sigma(\omega,{\bf q})$ can be referred to as the
propagation constant (index of refraction squared) of the Bloch
mode~\cite{menzel_10_2,alaee_13_1}.

We can seek approximate solutions to (\ref{g=0_Sigma_det=0}) by using
the limit ${\bf k}\rightarrow 0$. As was discussed in
Sec.~\ref{subsec:step_two}, the tensor $\Sigma(\omega,{\bf k})$ can be
formally expanded in a non-gyrotropic medium according to
(\ref{Sigma_exp}). We substitute this expansion into
(\ref{g=0_Sigma_det=0}) and obtain
\begin{equation}
\label{g=0_Sigma_det=0_wnl}
{\rm det}\left[ \left({\bf k} \times (1 - \beta) {\bf k} \times \right) + k_0^2
\Sigma(\omega, 0) \right] = 0 \ .
\end{equation}
\noindent
Equation (\ref{g=0_Sigma_det=0_wnl}) is a valid approximation to the
dispersion equation in the weak nonlocality regime and its solution
yields the first nonvanishing solution to ${\bf q}$ (compared to the
limit $h\rightarrow 0$). Also, (\ref{g=0_Sigma_det=0_wnl}) coincides
with the dispersion equation in a homogeneous medium with local
parameters $\epsilon = \Sigma(\omega, 0)$ and $\mu = (1-\beta)^{-1}$.
If we make this identification, we would arrive at the same
homogenization result as in the current-driven homogenization theory,
except that we did not need to introduce the external current.
However, this identification is not mathematically justified due to
the reasons already discussed by us in Sec.~\ref{subsec:step_two}.
Here we reiterate these arguments in the somewhat new light of the
Bloch-wave analysis.

Firstly and most importantly, it can be easily seen that
multiplication of (\ref{g=0_Sigma_det=0_wnl}) by a scalar $\xi$ does
not alter the dispersion equation or the value of ${\bf q}$ but doing
so does alter the impedance ${\mathscr Z}$. Therefore, the above
procedure is not expected to yield a meaningful correction to
${\mathscr Z}$. This was illustrated above in
Fig.~\ref{EX_A_Re_zet_Im_zet_norm}.  In fact, S-parameter retrieval
predicts a much more accurate ${\mathscr Z}$ while keeping
approximately the same dispersion relation. This was illustrated in
Fig.~\ref{EX_A_Re_T_Re_R_h=0.2_kx}. And in general, it could not be
reasonably expected that a theory that considers infinite media and
disregards the physical boundary would predict the impedance
correctly.

Second, the procedure described above is clearly inapplicable outside
of the weak nonlocality regime and, in particular, when $\Vert \beta
\Vert \sim 1$. In this region of parameters, introduction of the local
permittivity and permeability tensors does not result in a correct
dispersion relation, even approximately. But this is exactly the
region of parameters where current-driven homogenization predicts
magnetic resonances. Consequently, this prediction is mathematically
unjustified.

\section{Discussion}
\label{sec:disc}

\subsection{Current-driven excitation model and the theory of
  nonlocality}
\label{subsec:disc_nonl}

The current-driven homogenization theory is deeply rooted in the
theory of natural electromagnetic nonlocality (spatial
dispersion)~\cite{agranovich_book_66,landau_ess_84}. The latter is, of
course, a very successful theory, which has predicted and described
theoretically such diverse phenomena as optical activity, additional
waves, and anisotropy of crystals with cubic symmetry, etc. However,
current-driven homogenization and, more generally, current-driven
excitation model take certain analogies too far or apply them
unscrupulously.

The basic idea behind the current-driven excitation model can be
traced to Ref.~\onlinecite{agranovich_book_66}.  We translate the
relevant text from the Russian edition of this book (Moscow, Nauka,
1965, p.~34), using only a slight change of notations:

\begin{quote}
  ``Generally, the arguments of the tensor $\Sigma(\omega,{\bf k})$
  are mathematically-independent. This fact follows already from the
  definition (1.6) [equivalent to Eqs.~\ref{D_E_nonl},\ref{Sigma_f}
  below (authors' comment)] but can be at times not entirely obvious.
  This is so because, in optics, one encounters very frequently wave
  propagation in the absence of sources in the medium itself, in which
  case ${\bf k}$ is a function of $\omega$; for example, for normal
  homogeneous plane waves, ${\bf k} = (\omega/c) \tilde{n}(\omega,
  \hat{\bf s}) \hat{\bf s}$. But if ${\bf k} = {\bf k}(\omega)$, then
  the spatial dispersion appears to be indistinguishable from the
  frequency dispersion. This observation raises a question [about the
  physical nature of spatial nonlocality (authors' comment)] and the
  answer to this question is the following. The tensor
  $\Sigma(\omega,{\bf k})$ is introduced for fields of the most
  general form, obtained when the sources ${\bf J}_{\rm ext}({\bf r})$
  and $\rho_{\rm ext}({\bf r})$ spatially overlap with the medium.
  Under these conditions, it is possible to create a field ${\bf E}$
  with arbitrary and mathematically independent $\omega$ and ${\bf k}$
  (the Fourier components $\tilde{\bf E}(\omega, {\bf k})$ is
  ultimately expressed in terms of ${\bf J}_{\rm ext}(\omega, {\bf
    k})$ and $\rho_{\rm ext}(\omega, {\bf k})$; see \$2.1). From this,
  it follows immediately that all problems involving wave propagation
  can be solved if $\Sigma(\omega, {\bf k})$ is known.''
\end{quote}

The last sentence in the above is only partially correct. In the case
of natural nonlocality, $\Sigma(\omega,{\bf k})$ is the
spatial Fourier transform of the influence function $\sigma(\omega;
{\bf r},{\bf r}^\prime)$, which appears in the nonlocal relation
between ${\bf D}(\omega,{\bf r})$ and ${\bf E}(\omega,{\bf r})$, viz,:
\begin{equation}
\label{D_E_nonl}
{\bf D}(\omega, {\bf r}) =
\int_V \sigma(\omega; {\bf r},{\bf r}^\prime)
{\bf E}(\omega, {\bf r}^\prime) d^3 r^\prime \ .
\end{equation}
\noindent
If both points ${\bf r}$ and ${\bf r}^\prime$ are sufficiently far
from the boundary of the medium, we can write $\sigma(\omega;{\bf
  r},{\bf r}^\prime) = f(\omega,{\bf r}-{\bf r}^\prime)$. Then
$\Sigma(\omega,{\bf k})$ is defined as the spatial Fourier transform
of $f(\omega,{\bf r})$:
\begin{equation}
\label{Sigma_f}
\Sigma(\omega,{\bf k}) = \int f(\omega,{\bf r}) e^{i {\bf k}
  \cdot {\bf r}} d^3 r \ .
\end{equation}
But this is insufficient to solve the boundary-value problem for any
finite shape. To that end, we would need to know how
$\sigma(\omega;{\bf r},{\bf r}^\prime)$ behaves when at least one of
the points ${\bf r}$ and ${\bf r}^\prime$ is close to the boundary.
One can consider an approximation of the type
\begin{eqnarray}
\label{sigma_approx}
\sigma(\omega;{\bf r},{\bf r}^\prime) &=& S({\bf r}) f(\omega, {\bf r} - {\bf
  r}^\prime) S({\bf r}^\prime) \nonumber \\
&+& [1-S({\bf r})] \delta({\bf r}- {\bf
  r}^\prime) [1-S({\bf r}^\prime)] \ ,
\end{eqnarray}
\noindent
where $S({\bf r})$ is the shape function: it is equal to unity inside
the medium and to zero outside (in vacuum). If (\ref{sigma_approx})
holds, then the statement under consideration is correct: Maxwell's
equations can be written in a closed form using only the Fourier
transform of $f(\omega, {\bf r})$ and the Fourier transform of the
shape function. Therefore, if $\Sigma(\omega,{\bf k})$ is known, then
Maxwell's equations can be solved in a finite sample, at least in
principle. We note that the familiar relation ${\bf D}(\omega,{\bf k})
= \Sigma(\omega,{\bf k}) {\bf E}(\omega,{\bf k})$ does not hold in
this case and Maxwell's equations, written in the ${\bf k}$-domain,
contain an integral transform and cannot be solved by algebraic
manipulation. This difficulty is known and explained in \$~10 of
Ref.~\onlinecite{agranovich_book_66} but appears to be scarcely
appreciated in the modern literature. But regardless of this
difficulty, there is no reason to believe that (\ref{sigma_approx}) is
generally true. This approximation can be applicable, perhaps, if the
nonlocal interaction between two points is transmitted only along the
line of sight and if the body is convex. However, the first of these
assumptions is difficult to justify.

The same analysis applies to current-driven homogenization of periodic
composites. The knowledge of the function $\Sigma(\omega,{\bf k})$, as
defined by (\ref{averaging}) or by (\ref{Sigma_def_g}), allows one to
find the law of dispersion but is insufficient to solve any boundary
value problem.  Therefore, this function is not an intrinsic physical
characteristic of a composite.  It is, rather, an auxiliary
mathematical function, which appears when a certain ansatz is
substituted into Maxwell's equations written for an infinite periodic
medium.

Now the difference between the classical theory of nonlocality and the
current-driven homogenization theory becomes apparent. In the former
case, the real-space influence function $\sigma(\omega; {\bf r},{\bf
  r}^\prime)$ is derived from first principles (e.g., from a
microscopic theory) or introduced phenomenologically and then it
completely characterizes the electromagnetic properties of a
macroscopic object of any shape in the sense that it renders Maxwell's
equations closed.  As discussed above, under some limited conditions,
the dependence of $\sigma(\omega; {\bf r},{\bf r}^\prime)$ on the two
variables ${\bf r}$ and ${\bf r}^\prime$ can be simplified, e.g., as
in (\ref{sigma_approx}), and then there exists a one-to-one
correspondence between $\sigma(\omega; {\bf r},{\bf r}^\prime)$ and
$\Sigma(\omega,{\bf k})$. But in the case of current-driven
homogenization theory, the low-pass filtering (\ref{averaging}) does
not follow from any first principle. Moreover, if (\ref{averaging}) is
accepted as the fundamental definition of $\Sigma(\omega,{\bf k})$,
there is no way to establish a one-to-one correspondence between the
latter and the real-space function $\sigma(\omega; {\bf r},{\bf
  r}^\prime)$. As a result, the knowledge of $\Sigma(\omega,{\bf k})$,
thus defined, is insufficient to solve a boundary-value problem in any
finite sample.

Another obvious distinction between the two theories is that, for
natural nonlocality, the influence range (the characteristic value of
$\vert {\bf r} - {\bf r}^\prime \vert$ for which $\sigma(\omega; {\bf
  r},{\bf r}^\prime)$ is not negligibly small) is of the order of the
atomic scale. Therefore, in the optical range, the condition of weak
nonlocality is satisfied with extremely good precision and all the
effects of nonlocality are, essentially, small perturbations. In the
case of current-driven homogenization, the influence range is $h$ and
the effects claimed as a result of current-driven homogenization
(e.g., $\mu \approx -1$) are dramatic and nonperturbative.

So far, we have discussed the physical and mathematical meaning of the
function $\Sigma(\omega,{\bf k})$ as it is used both in the theory of
natural nonlocality and in current-driven homogenization. The next
important point to consider is the unjustified assumption of the
current-driven homogenization theory (and, more generally, of the
current-driven excitation model) that the external current of the form
(\ref{J_ext}) can, under some unspecified conditions, be created in
the medium. This assumption also grows conceptually from the above
quote. In reality, ``wave propagation in the absence of sources in the
medium itself'' is encountered in optics (and, more generally, in
macroscopic electrodynamics) not just ``very frequently,'' but always,
without any known exceptions. Of course a medium can be optically
active and emit some kind of radiation from its volume.  However, in
all such cases, the current inside the medium is subject to (linear or
nonlinear) constitutive relations and cannot be created or controlled
by an experimentalist at will. 

Finally, another relevant misconception, which has been widely
popularized in recent
years~\cite{agranovich_04_1,agranovich_06_1,agranovich_06_2,agranovich_09_1},
is the proposition of equivalence of weak nonlocality of the
dielectric response and nontrivial magnetic permeability. We have
discussed this point in this article in much detail and have
demonstrated that the equivalence exists only for the dispersion
relation but not for the impedance of the medium. It can be argued
that this is exactly what was meant by Landau and
Lifshitz~\cite{landau_ess_84} since none of the relevant chapters
consider boundary conditions in any form. The current-driven
homogenization theory has taken this statement of equivalence out of
its proper context and applied it to the problem of homogenization
wherein the boundary conditions play the central role.

It can be concluded that the classical theory of spatial dispersion is
concerned primarily with certain physical effects such as rotation of
the plane of polarization or appearance of additional waves, which are
not present in the purely local regime but can be described as
perturbations if small nonlocal corrections to the permittivity tensor
are taken into account. The corrections are either introduced
phenomenologically or computed using a microscopic theory. The theory
of spatial dispersion was never meant to be used for rigorous solution
of boundary value problems and, therefore, the discussion of the
dispersion equation sufficed in the vast majority of cases. For this
reason, certain remarks appearing in the classical texts on the
subject, such as the now famous remark of Landau and Lifshitz
regarding the equivalence of nonlocality and magnetism, apply only to
the dispersion relation. In the case of the current-driven
homogenization theory, all these limitations have been disregarded.

\subsection{Homogenization by spatial Fourier transform}
\label{subsec:disc_princ}

Throughout the paper, we have emphasized the critical importance of
taking the boundary effects into account in electromagnetic
homogenization, particularly in the case of metamaterials whose
lattice cell size typically constitutes an appreciable fraction of the
vacuum wavelength. Consequently, theories relying entirely on the bulk
behavior of waves cannot be accurate; they may be capable of finding
the effective index but not the impedance. We note that the role of
boundary conditions has been elucidated and emphasized in the
literature previously~\cite{simovski_09_1,felbacq_00_1}; however, in
this work, we have presented a detailed case study using an exactly
solvable model.

Generally speaking, Fourier-based homogenization theories should be
applied with extreme care because Fourier analysis makes it difficult
to account for material interfaces, which break the discrete
translational invariance of a periodic composite.

It is feasible to devise a theory in which Fourier analysis is applied
in the medium and in the empty space separately. In this case,
however, only natural Bloch modes will exist in the material, and no
other values of ${\bf k}$ will appear. The intuitive perception that a
small localized source (e.g., a nano-antenna) embedded in a composite
(e.g., in one of the empty voids) would generate within the material
the whole spectrum of waves with all possible real-valued ${\bf k}$'s
is correct only in a very narrow technical sense. In fact, within any
area away from the small source, the waves with various ${\bf k}$ will
interfere to produce the natural Bloch modes of the periodic
structure. Only the latter are physically measurable.

\subsection{Current-driven homogenization}
\label{subsec:disc_cd}

As an illustration of the general principles stated above, we have
critically analyzed the current-driven homogenization theory of
Refs.~\onlinecite{silveirinha_07_1,alu_11_2}, which does not account
for the boundaries of the medium but derives the effective parameters
from the behavior of waves in the bulk. In addition, this model relies
on the use of physically-unrealizable sources inside the medium, with
no justification as to why the results thus obtained should be
experimentally relevant.

The numerical results of Sec.~\ref{sec:num} are therefore not
surprising.  They demonstrate that current driven homogenization does
not yield accurate results in the range of parameters where it
predicts nontrivial magnetic effects. In particular, the errors of the
transmission and reflection coefficients $T$ and $R$ are of the same
order of magnitude or, in some cases, much larger than the deviation
of the magnetic permeability from unity, $\Vert \mu -1 \Vert$, where
$\mu$ is the local permeability tensor predicted by current-driven
homogenization.

In most cases considered, current-driven homogenization does not
provide a noticeable improvement in accuracy compared to the standard
homogenization result (\ref{eps_mu_hmg}). In the cases when such
improvement can be observed, e.g., in Fig.~\ref{EX_A_T2_R2_norm}, this
is due to a correction in the effective permittivity $\epsilon$ rather
than to an accurate prediction of $\mu$. We note in passing that the
correction to the magnetic permeability produced by the current-driven
model is asymptotically $\mathcal{O}(kh)^4$, which is different from
the $\mathcal{O}(kh)^2$ asymptote that follows from S-parameter
retrieval.

\section{Summary}
\label{sec:sum}

This paper has three main conclusions. 

First, careful consideration of boundary conditions is required in all
effective medium theories (EMTs).

Second, all EMTs have an applicability range, and wherever a
homogenization result is obtained, it is important to verify that the
parameters of the composite are within this range or, otherwise,
validate the result with direct simulation.

Third, there are many EMTs that yield the standard homogenization
result in the limit $h\rightarrow 0$ but different $h$-dependent
corrections to the former. Validating that these corrections are
physically meaningful requires consideration of finite samples and
cannot be done by investigating an infinite periodic composite (this
conclusion is closely related to the first one).

\section*{Acknowledgments}

This research was supported by the US National Science Foundation
under Grant DMS1216970.

\bibliographystyle{apsrev}

\bibliography{abbrev,book,master,local}

\begin{thebibliography}{58}
\expandafter\ifx\csname natexlab\endcsname\relax\def\natexlab#1{#1}\fi
\expandafter\ifx\csname bibnamefont\endcsname\relax
  \def\bibnamefont#1{#1}\fi
\expandafter\ifx\csname bibfnamefont\endcsname\relax
  \def\bibfnamefont#1{#1}\fi
\expandafter\ifx\csname citenamefont\endcsname\relax
  \def\citenamefont#1{#1}\fi
\expandafter\ifx\csname url\endcsname\relax
  \def\url#1{\texttt{#1}}\fi
\expandafter\ifx\csname urlprefix\endcsname\relax\def\urlprefix{URL }\fi
\providecommand{\bibinfo}[2]{#2}
\providecommand{\eprint}[2][]{\url{#2}}

\bibitem[{\citenamefont{Simovski}(2009)}]{simovski_09_1}
\bibinfo{author}{\bibfnamefont{C.~R.} \bibnamefont{Simovski}},
  \bibinfo{journal}{Opt. Spectrosc.} \textbf{\bibinfo{volume}{107}},
  \bibinfo{pages}{766} (\bibinfo{year}{2009}).

\bibitem[{\citenamefont{Simovski}(2011)}]{simovski_11_1}
\bibinfo{author}{\bibfnamefont{C.~R.} \bibnamefont{Simovski}},
  \bibinfo{journal}{J. Opt.} \textbf{\bibinfo{volume}{13}},
  \bibinfo{pages}{103001} (\bibinfo{year}{2011}).

\bibitem[{\citenamefont{Felbacq and
  Bouchitte}(2005{\natexlab{a}})}]{felbacq_05_1}
\bibinfo{author}{\bibfnamefont{D.}~\bibnamefont{Felbacq}} \bibnamefont{and}
  \bibinfo{author}{\bibfnamefont{G.}~\bibnamefont{Bouchitte}},
  \bibinfo{journal}{New J. Phys.} \textbf{\bibinfo{volume}{7}},
  \bibinfo{pages}{159} (\bibinfo{year}{2005}{\natexlab{a}}).

\bibitem[{\citenamefont{Felbacq and
  Bouchitte}(2005{\natexlab{b}})}]{felbacq_05_2}
\bibinfo{author}{\bibfnamefont{D.}~\bibnamefont{Felbacq}} \bibnamefont{and}
  \bibinfo{author}{\bibfnamefont{G.}~\bibnamefont{Bouchitte}},
  \bibinfo{journal}{Phys. Rev. Lett.} \textbf{\bibinfo{volume}{94}},
  \bibinfo{pages}{183902} (\bibinfo{year}{2005}{\natexlab{b}}).

\bibitem[{\citenamefont{Bohren}(1986)}]{bohren_86_1}
\bibinfo{author}{\bibfnamefont{C.~F.} \bibnamefont{Bohren}},
  \bibinfo{journal}{J. Atmospheric Sci.} \textbf{\bibinfo{volume}{43}},
  \bibinfo{pages}{468} (\bibinfo{year}{1986}).

\bibitem[{\citenamefont{Bohren}(2009)}]{bohren_09_1}
\bibinfo{author}{\bibfnamefont{C.~F.} \bibnamefont{Bohren}},
  \bibinfo{journal}{J. Nanophotonics} \textbf{\bibinfo{volume}{3}},
  \bibinfo{pages}{039501} (\bibinfo{year}{2009}).

\bibitem[{\citenamefont{Wellander and Kristensson}(2003)}]{wellander_03_1}
\bibinfo{author}{\bibfnamefont{N.}~\bibnamefont{Wellander}} \bibnamefont{and}
  \bibinfo{author}{\bibfnamefont{G.}~\bibnamefont{Kristensson}},
  \bibinfo{journal}{SIAM J. Appl. Math.} \textbf{\bibinfo{volume}{64}},
  \bibinfo{pages}{170} (\bibinfo{year}{2003}).

\bibitem[{\citenamefont{Markel and Schotland}(2012)}]{markel_12_1}
\bibinfo{author}{\bibfnamefont{V.~A.} \bibnamefont{Markel}} \bibnamefont{and}
  \bibinfo{author}{\bibfnamefont{J.~C.} \bibnamefont{Schotland}},
  \bibinfo{journal}{Phys. Rev. E} \textbf{\bibinfo{volume}{85}},
  \bibinfo{pages}{066603} (\bibinfo{year}{2012}).

\bibitem[{\citenamefont{Pendry}(2001)}]{pendry_01_1}
\bibinfo{author}{\bibfnamefont{J.}~\bibnamefont{Pendry}},
  \bibinfo{journal}{Physics World}  (\bibinfo{year}{2001}).

\bibitem[{\citenamefont{Lewin}(1947)}]{lewin_47_1}
\bibinfo{author}{\bibfnamefont{L.}~\bibnamefont{Lewin}}, \bibinfo{journal}{J.
  Inst. Elec. Eng.} \textbf{\bibinfo{volume}{94}}, \bibinfo{pages}{65}
  (\bibinfo{year}{1947}).

\bibitem[{\citenamefont{Khizhnyak}(1957{\natexlab{a}})}]{khizhnyak_57_1}
\bibinfo{author}{\bibfnamefont{N.~A.} \bibnamefont{Khizhnyak}},
  \bibinfo{journal}{Sov. Phys. Tech. Phys.} \textbf{\bibinfo{volume}{27}},
  \bibinfo{pages}{2006} (\bibinfo{year}{1957}{\natexlab{a}}).

\bibitem[{\citenamefont{Khizhnyak}(1957{\natexlab{b}})}]{khizhnyak_57_2}
\bibinfo{author}{\bibfnamefont{N.~A.} \bibnamefont{Khizhnyak}},
  \bibinfo{journal}{Sov. Phys. Tech. Phys.} \textbf{\bibinfo{volume}{27}},
  \bibinfo{pages}{2014} (\bibinfo{year}{1957}{\natexlab{b}}).

\bibitem[{\citenamefont{Khizhnyak}(1959)}]{khizhnyak_59_1}
\bibinfo{author}{\bibfnamefont{N.~A.} \bibnamefont{Khizhnyak}},
  \bibinfo{journal}{Sov. Phys. Tech. Phys.} \textbf{\bibinfo{volume}{29}},
  \bibinfo{pages}{604} (\bibinfo{year}{1959}).

\bibitem[{\citenamefont{Niklasson et~al.}(1981)\citenamefont{Niklasson,
  Granqvist, and Hunderi}}]{niklasson_81_1}
\bibinfo{author}{\bibfnamefont{G.~A.} \bibnamefont{Niklasson}},
  \bibinfo{author}{\bibfnamefont{C.~G.} \bibnamefont{Granqvist}},
  \bibnamefont{and} \bibinfo{author}{\bibfnamefont{O.}~\bibnamefont{Hunderi}},
  \bibinfo{journal}{Appl. Opt.} \textbf{\bibinfo{volume}{20}},
  \bibinfo{pages}{26} (\bibinfo{year}{1981}).

\bibitem[{\citenamefont{Doyle}(1989)}]{doyle_89_1}
\bibinfo{author}{\bibfnamefont{W.~T.} \bibnamefont{Doyle}},
  \bibinfo{journal}{Phys. Rev. B} \textbf{\bibinfo{volume}{39}},
  \bibinfo{pages}{9852} (\bibinfo{year}{1989}).

\bibitem[{\citenamefont{Waterman and Pedersen}(1986)}]{waterman_86_1}
\bibinfo{author}{\bibfnamefont{P.~C.} \bibnamefont{Waterman}} \bibnamefont{and}
  \bibinfo{author}{\bibfnamefont{N.~E.} \bibnamefont{Pedersen}},
  \bibinfo{journal}{J. Appl. Phys.} \textbf{\bibinfo{volume}{59}},
  \bibinfo{pages}{2609} (\bibinfo{year}{1986}).

\bibitem[{\citenamefont{Cherednichenko and
  Guenneau}(2007)}]{cherednichenko_07_1}
\bibinfo{author}{\bibfnamefont{K.~D.} \bibnamefont{Cherednichenko}}
  \bibnamefont{and} \bibinfo{author}{\bibfnamefont{S.}~\bibnamefont{Guenneau}},
  \bibinfo{journal}{Waves in Random and Complex Media}
  \textbf{\bibinfo{volume}{17}}, \bibinfo{pages}{627} (\bibinfo{year}{2007}).

\bibitem[{\citenamefont{Guenneau et~al.}(2007)\citenamefont{Guenneau, Zolla,
  and Nicolet}}]{guenneau_07_1}
\bibinfo{author}{\bibfnamefont{S.}~\bibnamefont{Guenneau}},
  \bibinfo{author}{\bibfnamefont{F.}~\bibnamefont{Zolla}}, \bibnamefont{and}
  \bibinfo{author}{\bibfnamefont{A.}~\bibnamefont{Nicolet}},
  \bibinfo{journal}{Waves in Random and Complex Media}
  \textbf{\bibinfo{volume}{17}}, \bibinfo{pages}{653} (\bibinfo{year}{2007}).

\bibitem[{\citenamefont{Guenneau and Zolla}(2011)}]{guenneau_11_1}
\bibinfo{author}{\bibfnamefont{S.}~\bibnamefont{Guenneau}} \bibnamefont{and}
  \bibinfo{author}{\bibfnamefont{F.}~\bibnamefont{Zolla}},
  \bibinfo{journal}{Prog. Electromagnetic Res.} \textbf{\bibinfo{volume}{27}},
  \bibinfo{pages}{91} (\bibinfo{year}{2011}).

\bibitem[{\citenamefont{Craster et~al.}(2011)\citenamefont{Craster, Kaplunov,
  Nolde, and Guenneau}}]{craster_11_1}
\bibinfo{author}{\bibfnamefont{R.~V.} \bibnamefont{Craster}},
  \bibinfo{author}{\bibfnamefont{J.}~\bibnamefont{Kaplunov}},
  \bibinfo{author}{\bibfnamefont{E.}~\bibnamefont{Nolde}}, \bibnamefont{and}
  \bibinfo{author}{\bibfnamefont{S.}~\bibnamefont{Guenneau}},
  \bibinfo{journal}{J. Opt. Soc. Am. A} \textbf{\bibinfo{volume}{28}},
  \bibinfo{pages}{1032} (\bibinfo{year}{2011}).

\bibitem[{\citenamefont{Tsukerman}(2011{\natexlab{a}})}]{tsukerman_11_1}
\bibinfo{author}{\bibfnamefont{I.}~\bibnamefont{Tsukerman}},
  \bibinfo{journal}{J. Opt. Soc. Am. B} \textbf{\bibinfo{volume}{28}},
  \bibinfo{pages}{577} (\bibinfo{year}{2011}{\natexlab{a}}).

\bibitem[{\citenamefont{Pors et~al.}(2011)\citenamefont{Pors, Tsukerman, and
  Bozhevolnyi}}]{pors_11_1}
\bibinfo{author}{\bibfnamefont{A.}~\bibnamefont{Pors}},
  \bibinfo{author}{\bibfnamefont{I.}~\bibnamefont{Tsukerman}},
  \bibnamefont{and} \bibinfo{author}{\bibfnamefont{S.~I.}
  \bibnamefont{Bozhevolnyi}}, \bibinfo{journal}{Phys. Rev. E}
  \textbf{\bibinfo{volume}{84}}, \bibinfo{pages}{016609}
  (\bibinfo{year}{2011}).

\bibitem[{\citenamefont{Tsukerman}(2011{\natexlab{b}})}]{tsukerman_11_2}
\bibinfo{author}{\bibfnamefont{I.}~\bibnamefont{Tsukerman}},
  \bibinfo{journal}{J. Opt. Soc. Am. B} \textbf{\bibinfo{volume}{28}},
  \bibinfo{pages}{2956} (\bibinfo{year}{2011}{\natexlab{b}}).

\bibitem[{\citenamefont{Silveirinha}(2007)}]{silveirinha_07_1}
\bibinfo{author}{\bibfnamefont{M.~G.} \bibnamefont{Silveirinha}},
  \bibinfo{journal}{Phys. Rev. B} \textbf{\bibinfo{volume}{75}},
  \bibinfo{pages}{115104} (\bibinfo{year}{2007}).

\bibitem[{\citenamefont{Alu}(2011)}]{alu_11_2}
\bibinfo{author}{\bibfnamefont{A.}~\bibnamefont{Alu}}, \bibinfo{journal}{Phys.
  Rev. B} \textbf{\bibinfo{volume}{84}}, \bibinfo{pages}{075153}
  (\bibinfo{year}{2011}).

\bibitem[{\citenamefont{Silveirinha}(2009)}]{silveirinha_09_1}
\bibinfo{author}{\bibfnamefont{M.~G.} \bibnamefont{Silveirinha}},
  \bibinfo{journal}{Phys. Rev. B} \textbf{\bibinfo{volume}{80}},
  \bibinfo{pages}{235120} (\bibinfo{year}{2009}).

\bibitem[{\citenamefont{Costa et~al.}(2009)\citenamefont{Costa, Silveirinha,
  and Maslovski}}]{costa_09_1}
\bibinfo{author}{\bibfnamefont{J.~T.} \bibnamefont{Costa}},
  \bibinfo{author}{\bibfnamefont{M.~G.} \bibnamefont{Silveirinha}},
  \bibnamefont{and} \bibinfo{author}{\bibfnamefont{S.~I.}
  \bibnamefont{Maslovski}}, \bibinfo{journal}{Phys. Rev. B}
  \textbf{\bibinfo{volume}{80}}, \bibinfo{pages}{235124}
  (\bibinfo{year}{2009}).

\bibitem[{\citenamefont{Fietz and Shvets}(2010{\natexlab{a}})}]{fietz_10_1}
\bibinfo{author}{\bibfnamefont{C.}~\bibnamefont{Fietz}} \bibnamefont{and}
  \bibinfo{author}{\bibfnamefont{G.}~\bibnamefont{Shvets}},
  \bibinfo{journal}{Physica B} \textbf{\bibinfo{volume}{405}},
  \bibinfo{pages}{2930} (\bibinfo{year}{2010}{\natexlab{a}}).

\bibitem[{\citenamefont{Fietz and Shvets}(2010{\natexlab{b}})}]{fietz_10_2}
\bibinfo{author}{\bibfnamefont{C.}~\bibnamefont{Fietz}} \bibnamefont{and}
  \bibinfo{author}{\bibfnamefont{G.}~\bibnamefont{Shvets}},
  \bibinfo{journal}{Phys. Rev. B} \textbf{\bibinfo{volume}{82}},
  \bibinfo{pages}{205128} (\bibinfo{year}{2010}{\natexlab{b}}).

\bibitem[{\citenamefont{Fietz and Shvets}(2010{\natexlab{c}})}]{fietz_10_3}
\bibinfo{author}{\bibfnamefont{C.}~\bibnamefont{Fietz}} \bibnamefont{and}
  \bibinfo{author}{\bibfnamefont{G.}~\bibnamefont{Shvets}}, in
  \emph{\bibinfo{booktitle}{{Metamaterials: Fundamentals and Appplications
  III}}} (\bibinfo{publisher}{SPIE}, \bibinfo{year}{2010}{\natexlab{c}}), pp.
  \bibinfo{pages}{77540V--1--8}.

\bibitem[{\citenamefont{Costa et~al.}(2011)\citenamefont{Costa, Silveirinha,
  and Alu}}]{costa_11_1}
\bibinfo{author}{\bibfnamefont{J.~T.} \bibnamefont{Costa}},
  \bibinfo{author}{\bibfnamefont{M.~G.} \bibnamefont{Silveirinha}},
  \bibnamefont{and} \bibinfo{author}{\bibfnamefont{A.}~\bibnamefont{Alu}},
  \bibinfo{journal}{Phys. Rev. B} \textbf{\bibinfo{volume}{83}},
  \bibinfo{pages}{165120} (\bibinfo{year}{2011}).

\bibitem[{\citenamefont{Silveirinha}(2011)}]{silveirinha_11_1}
\bibinfo{author}{\bibfnamefont{M.~G.} \bibnamefont{Silveirinha}},
  \bibinfo{journal}{Phys. Rev. B} \textbf{\bibinfo{volume}{83}},
  \bibinfo{pages}{165104} (\bibinfo{year}{2011}).

\bibitem[{\citenamefont{Alu et~al.}(2011)\citenamefont{Alu, Yaghjian, Shore,
  and Silveirinha}}]{alu_11_3}
\bibinfo{author}{\bibfnamefont{A.}~\bibnamefont{Alu}},
  \bibinfo{author}{\bibfnamefont{A.~D.} \bibnamefont{Yaghjian}},
  \bibinfo{author}{\bibfnamefont{R.~A.} \bibnamefont{Shore}}, \bibnamefont{and}
  \bibinfo{author}{\bibfnamefont{M.~G.} \bibnamefont{Silveirinha}},
  \bibinfo{journal}{Phys. Rev. B} \textbf{\bibinfo{volume}{84}},
  \bibinfo{pages}{054305} (\bibinfo{year}{2011}).

\bibitem[{\citenamefont{Fietz and Soukoulis}(2012)}]{fietz_12_1}
\bibinfo{author}{\bibfnamefont{C.}~\bibnamefont{Fietz}} \bibnamefont{and}
  \bibinfo{author}{\bibfnamefont{C.~M.} \bibnamefont{Soukoulis}},
  \bibinfo{journal}{Phys. Rev. B} \textbf{\bibinfo{volume}{86}},
  \bibinfo{pages}{085146} (\bibinfo{year}{2012}).

\bibitem[{\citenamefont{Chebykin et~al.}(2011)\citenamefont{Chebykin, Orlov,
  Vozianova, Maslovski, Kivshar, and Belov}}]{chebykin_11_1}
\bibinfo{author}{\bibfnamefont{A.~V.} \bibnamefont{Chebykin}},
  \bibinfo{author}{\bibfnamefont{A.~A.} \bibnamefont{Orlov}},
  \bibinfo{author}{\bibfnamefont{A.~V.} \bibnamefont{Vozianova}},
  \bibinfo{author}{\bibfnamefont{S.~I.} \bibnamefont{Maslovski}},
  \bibinfo{author}{\bibfnamefont{Y.~S.} \bibnamefont{Kivshar}},
  \bibnamefont{and} \bibinfo{author}{\bibfnamefont{P.~A.} \bibnamefont{Belov}},
  \bibinfo{journal}{Phys. Rev. B} \textbf{\bibinfo{volume}{84}},
  \bibinfo{pages}{115438} (\bibinfo{year}{2011}).

\bibitem[{\citenamefont{Chebykin et~al.}(2012)\citenamefont{Chebykin, Orlov,
  Simovski, Kivshar, and Belov}}]{chebykin_12_1}
\bibinfo{author}{\bibfnamefont{A.~V.} \bibnamefont{Chebykin}},
  \bibinfo{author}{\bibfnamefont{A.~A.} \bibnamefont{Orlov}},
  \bibinfo{author}{\bibfnamefont{C.~R.} \bibnamefont{Simovski}},
  \bibinfo{author}{\bibfnamefont{Y.~S.} \bibnamefont{Kivshar}},
  \bibnamefont{and} \bibinfo{author}{\bibfnamefont{P.~A.} \bibnamefont{Belov}},
  \bibinfo{journal}{Phys. Rev. B} \textbf{\bibinfo{volume}{86}},
  \bibinfo{pages}{115420} (\bibinfo{year}{2012}).

\bibitem[{\citenamefont{Markel}(2010)}]{markel_10_2}
\bibinfo{author}{\bibfnamefont{V.~A.} \bibnamefont{Markel}},
  \bibinfo{journal}{J. Phys.: Condens. Matter} \textbf{\bibinfo{volume}{22}},
  \bibinfo{pages}{485401} (\bibinfo{year}{2010}).

\bibitem[{\citenamefont{Agranovich and Ginzburg}(1966)}]{agranovich_book_66}
\bibinfo{author}{\bibfnamefont{V.}~\bibnamefont{Agranovich}} \bibnamefont{and}
  \bibinfo{author}{\bibfnamefont{V.}~\bibnamefont{Ginzburg}},
  \emph{\bibinfo{title}{Spatial Dispersion in Crystal Optics and the Theory of
  Excitons}} (\bibinfo{publisher}{Wiley-Interscience}, \bibinfo{address}{New
  York}, \bibinfo{year}{1966}).

\bibitem[{\citenamefont{Landau and Lifshitz}(1984)}]{landau_ess_84}
\bibinfo{author}{\bibfnamefont{L.~D.} \bibnamefont{Landau}} \bibnamefont{and}
  \bibinfo{author}{\bibfnamefont{L.~P.} \bibnamefont{Lifshitz}},
  \emph{\bibinfo{title}{{Electrodynamics of Continuous Media}}}
  (\bibinfo{publisher}{Pergamon Press}, \bibinfo{address}{Oxford},
  \bibinfo{year}{1984}).

\bibitem[{\citenamefont{Agranovich et~al.}(2004)\citenamefont{Agranovich, Shen,
  Baughman, and Zakhidov}}]{agranovich_04_1}
\bibinfo{author}{\bibfnamefont{V.~M.} \bibnamefont{Agranovich}},
  \bibinfo{author}{\bibfnamefont{Y.~R.} \bibnamefont{Shen}},
  \bibinfo{author}{\bibfnamefont{R.~H.} \bibnamefont{Baughman}},
  \bibnamefont{and} \bibinfo{author}{\bibfnamefont{A.~A.}
  \bibnamefont{Zakhidov}}, \bibinfo{journal}{Phys. Rev. B}
  \textbf{\bibinfo{volume}{69}}, \bibinfo{pages}{165112}
  (\bibinfo{year}{2004}).

\bibitem[{\citenamefont{Agranovich et~al.}(2006)\citenamefont{Agranovich,
  Gartstein, and Zakhidov}}]{agranovich_06_1}
\bibinfo{author}{\bibfnamefont{V.~M.} \bibnamefont{Agranovich}},
  \bibinfo{author}{\bibfnamefont{Y.~N.} \bibnamefont{Gartstein}},
  \bibnamefont{and} \bibinfo{author}{\bibfnamefont{A.~A.}
  \bibnamefont{Zakhidov}}, \bibinfo{journal}{Phys. Rev. B}
  \textbf{\bibinfo{volume}{73}}, \bibinfo{pages}{045114}
  (\bibinfo{year}{2006}).

\bibitem[{\citenamefont{Agranovich and Gartstein}(2006)}]{agranovich_06_2}
\bibinfo{author}{\bibfnamefont{V.~M.} \bibnamefont{Agranovich}}
  \bibnamefont{and} \bibinfo{author}{\bibfnamefont{Y.~N.}
  \bibnamefont{Gartstein}}, \bibinfo{journal}{Phys. Usp.}
  \textbf{\bibinfo{volume}{49}}, \bibinfo{pages}{1029} (\bibinfo{year}{2006}).

\bibitem[{\citenamefont{Agranovich and Gartstein}(2009)}]{agranovich_09_1}
\bibinfo{author}{\bibfnamefont{V.~M.} \bibnamefont{Agranovich}}
  \bibnamefont{and} \bibinfo{author}{\bibfnamefont{Y.~N.}
  \bibnamefont{Gartstein}}, \bibinfo{journal}{Metamaterials}
  \textbf{\bibinfo{volume}{3}}, \bibinfo{pages}{1} (\bibinfo{year}{2009}).

\bibitem[{fn3()}]{fn3}
\bibinfo{note}{Strictly speaking, an additional Bloch wave with the ''natural''
  wave vector ${\bf q}$ can be added to (\ref{Forced_Bloch}). See the remark
  after Eq.~(\ref{BC}).}

\bibitem[{fn4()}]{fn4}
\bibinfo{note}{Note that Eqs.~(\ref{Maxwell_av}) are not closed and can not be
  solved without resorting to additional equations. A linear relationship
  between ${\bf D}_{\rm av}$ and ${\bf E}_{\rm av}$ can only be established by
  solving Eqs.~(\ref{Maxwell}).}

\bibitem[{fn6()}]{fn6}
\bibinfo{note}{The standard homogenization resul is $\epsilon=\Sigma(0,0)$
  while the result of current-driven homogenization is
  $\epsilon=\Sigma(\omega,0)$. We refer to the difference $\Sigma(\omega,0) -
  \Sigma(0,0)$ as to the dynamic correction to the permittivity.}

\bibitem[{\citenamefont{Feng}(2010)}]{feng_10_1}
\bibinfo{author}{\bibfnamefont{S.}~\bibnamefont{Feng}}, \bibinfo{journal}{Opt.
  Express} \textbf{\bibinfo{volume}{18}}, \bibinfo{pages}{17009}
  (\bibinfo{year}{2010}).

\bibitem[{fn2()}]{fn2}
\bibinfo{note}{These quantities have been introduced in
  Ref.~\onlinecite{feng_10_1} but we use somewhat different terminology. First,
  we refer to $\theta$ as to the ''optical depth'' rather than ''equivalent
  phase shift''. The physical meaning of this parameter should be clear in both
  cases. Second, we do not distinguish between the generalized impedance
  ${\mathscr L}$ and generalized admittance ${\mathscr Y}$. In our terminology,
  both are called ''generalized impedance'' but different formulas are used in
  different polarizations.}

\bibitem[{\citenamefont{Pendry}(2000)}]{pendry_00_1}
\bibinfo{author}{\bibfnamefont{J.~B.} \bibnamefont{Pendry}},
  \bibinfo{journal}{Phys. Rev. Lett.} \textbf{\bibinfo{volume}{85}},
  \bibinfo{pages}{3966} (\bibinfo{year}{2000}).

\bibitem[{fn5()}]{fn5}
\bibinfo{note}{Note that the varition of $X_+$ near the point $X_+ = -1$ is
  quadratic in the variation of ${\mathscr Z}$: $\delta X_+ = - (\delta
  {\mathscr Z})^2$. Therefore, the interplay in Eq.~(\ref{T_Perf_Lens}) is
  between the exponential function $\exp(i\theta)$ and the power function
  $(\delta {\mathscr Z})^2$.}

\bibitem[{\citenamefont{Koschny et~al.}(2003)\citenamefont{Koschny, Markos,
  Smith, and Soukoulis}}]{koschny_03_1}
\bibinfo{author}{\bibfnamefont{T.}~\bibnamefont{Koschny}},
  \bibinfo{author}{\bibfnamefont{P.}~\bibnamefont{Markos}},
  \bibinfo{author}{\bibfnamefont{D.~R.} \bibnamefont{Smith}}, \bibnamefont{and}
  \bibinfo{author}{\bibfnamefont{C.~M.} \bibnamefont{Soukoulis}},
  \bibinfo{journal}{Phys. Rev. E} \textbf{\bibinfo{volume}{68}},
  \bibinfo{pages}{065602(R)} (\bibinfo{year}{2003}).

\bibitem[{\citenamefont{Chen et~al.}(2004)\citenamefont{Chen, Grzegorczyk, Wu,
  Pacheco, and Kong}}]{chen_04_1}
\bibinfo{author}{\bibfnamefont{X.}~\bibnamefont{Chen}},
  \bibinfo{author}{\bibfnamefont{T.~M.} \bibnamefont{Grzegorczyk}},
  \bibinfo{author}{\bibfnamefont{B.~I.} \bibnamefont{Wu}},
  \bibinfo{author}{\bibfnamefont{J.}~\bibnamefont{Pacheco}}, \bibnamefont{and}
  \bibinfo{author}{\bibfnamefont{J.~A.} \bibnamefont{Kong}},
  \bibinfo{journal}{Phys. Rev. E} \textbf{\bibinfo{volume}{70}},
  \bibinfo{pages}{016608} (\bibinfo{year}{2004}).

\bibitem[{\citenamefont{Menzel et~al.}(2008)\citenamefont{Menzel, Rockstuhl,
  Paul, Lederer, and Pertsch}}]{menzel_08_1}
\bibinfo{author}{\bibfnamefont{C.}~\bibnamefont{Menzel}},
  \bibinfo{author}{\bibfnamefont{C.}~\bibnamefont{Rockstuhl}},
  \bibinfo{author}{\bibfnamefont{T.}~\bibnamefont{Paul}},
  \bibinfo{author}{\bibfnamefont{F.}~\bibnamefont{Lederer}}, \bibnamefont{and}
  \bibinfo{author}{\bibfnamefont{T.}~\bibnamefont{Pertsch}},
  \bibinfo{journal}{Phys. Rev. B} \textbf{\bibinfo{volume}{77}},
  \bibinfo{pages}{195328} (\bibinfo{year}{2008}).

\bibitem[{\citenamefont{Joannopoulos et~al.}(1995)\citenamefont{Joannopoulos,
  Meade, and Winn}}]{joannopoulos_book_95}
\bibinfo{author}{\bibfnamefont{J.~D.} \bibnamefont{Joannopoulos}},
  \bibinfo{author}{\bibfnamefont{R.~D.} \bibnamefont{Meade}}, \bibnamefont{and}
  \bibinfo{author}{\bibfnamefont{J.~N.} \bibnamefont{Winn}},
  \emph{\bibinfo{title}{Photonic Crystals: Molding the Flow of Light}}
  (\bibinfo{publisher}{Princeton University Press},
  \bibinfo{address}{Princeton, N.J.}, \bibinfo{year}{1995}).

\bibitem[{\citenamefont{Sakoda}(2005)}]{sakoda_book_05}
\bibinfo{author}{\bibfnamefont{K.}~\bibnamefont{Sakoda}},
  \emph{\bibinfo{title}{Optical Properties of Photonic Crystals}}
  (\bibinfo{publisher}{Springer}, \bibinfo{address}{Berlin},
  \bibinfo{year}{2005}).

\bibitem[{\citenamefont{Menzel et~al.}(2010)\citenamefont{Menzel, Rockstuhl,
  Iliew, Lederer, Andryieuski, Malureanu, and Lavrinenko}}]{menzel_10_2}
\bibinfo{author}{\bibfnamefont{C.}~\bibnamefont{Menzel}},
  \bibinfo{author}{\bibfnamefont{C.}~\bibnamefont{Rockstuhl}},
  \bibinfo{author}{\bibfnamefont{R.}~\bibnamefont{Iliew}},
  \bibinfo{author}{\bibfnamefont{F.}~\bibnamefont{Lederer}},
  \bibinfo{author}{\bibfnamefont{A.}~\bibnamefont{Andryieuski}},
  \bibinfo{author}{\bibfnamefont{R.}~\bibnamefont{Malureanu}},
  \bibnamefont{and} \bibinfo{author}{\bibfnamefont{A.~V.}
  \bibnamefont{Lavrinenko}}, \bibinfo{journal}{Phys. Rev. B}
  \textbf{\bibinfo{volume}{81}}, \bibinfo{pages}{195123}
  (\bibinfo{year}{2010}).

\bibitem[{\citenamefont{Alaee et~al.}(2013)\citenamefont{Alaee, Menzel, Banas,
  Banas, Xu, Chen, Moser, Lederer, and Rockstuhl}}]{alaee_13_1}
\bibinfo{author}{\bibfnamefont{R.}~\bibnamefont{Alaee}},
  \bibinfo{author}{\bibfnamefont{C.}~\bibnamefont{Menzel}},
  \bibinfo{author}{\bibfnamefont{A.}~\bibnamefont{Banas}},
  \bibinfo{author}{\bibfnamefont{K.}~\bibnamefont{Banas}},
  \bibinfo{author}{\bibfnamefont{S.}~\bibnamefont{Xu}},
  \bibinfo{author}{\bibfnamefont{H.}~\bibnamefont{Chen}},
  \bibinfo{author}{\bibfnamefont{H.~O.} \bibnamefont{Moser}},
  \bibinfo{author}{\bibfnamefont{F.}~\bibnamefont{Lederer}}, \bibnamefont{and}
  \bibinfo{author}{\bibfnamefont{C.}~\bibnamefont{Rockstuhl}},
  \bibinfo{journal}{Phys. Rev. B} \textbf{\bibinfo{volume}{87}},
  \bibinfo{pages}{075110} (\bibinfo{year}{2013}).

\bibitem[{\citenamefont{Felbacq}(2000)}]{felbacq_00_1}
\bibinfo{author}{\bibfnamefont{D.}~\bibnamefont{Felbacq}}, \bibinfo{journal}{J.
  Phys. A} \textbf{\bibinfo{volume}{33}}, \bibinfo{pages}{815}
  (\bibinfo{year}{2000}).

\end{thebibliography}

\appendix

\section{Determination of the coefficients ${\mathscr A}_a$,
  ${\mathscr B}_a$, ${\mathscr A}_b$ and ${\mathscr B}_b$ from the
  boundary conditions}
\label{app:A}

Upon substitution of the expressions (\ref{F_eh_def}) into
(\ref{BC}), we obtain the following set of linear equations:

\begin{eqnarray*}
%\label{BC_1}
e^{-i\theta_a} \Big{(}e^{i\phi_a} A_a  + e^{-i\phi_a} B_a \Big{)} +
\Big{(}A_b + B_b \Big{)} = 1 \ , \\
%\label{BC_2}
\frac{\kappa_a}{q_z} e^{-i\theta_a} \Big{(}e^{i\phi_a} A_a - e^{-i\phi_a} B_a \Big{)} +
\frac{\kappa_b}{q_z}  \Big{(}A_b - B_b \Big{)} = 1 \ , \\
%\label{BC_3}
\Big{(}A_a + B_a \Big{)} + e^{-i \theta_b} \Big{(} e^{i\phi_b} A_b +
  e^{-i\phi_b} B_b \Big{)} = 1 \ , \\
%\label{BC_4}
\frac{\kappa_a}{q_z} \Big{(}A_a - B_a \Big{)} +
\frac{\kappa_b}{q_z} e^{-i\theta_b} \Big{(}e^{i\phi_b} A_b - e^{-i\phi_b} B_b
\Big{)} = 1 \ .
\end{eqnarray*}
\noindent
This set is somewhat more complicated than what is encountered in the
ordinary theory of one-dimensional photonic crystals. In the latter
case, the matrix is the same but the right-hand side is zero.
Correspondingly, the task is to find the value of $k_z$ (for a given
$k_x$) such that the equations have a nontrivial solution. This occurs
when $k_z=q_z$, where $q_z$ is defined in Eq.~(\ref{q_z_B}). In this
manner, the natural Bloch wave vector ${\bf q}$ of the medium is
determined. For the case at hand, both $k_x$ and $k_z$ are free
parameters but the right-hand side is nonzero. Therefore, the
current-driven homogenization theory, essentially, replaces the
problem of funding the natural Bloch mode of the medium by a
mathematically unrelated problem of inverting the matrix in the above
set of equations.

The solution to the set stated above is given by (\ref{ABCD_det})
where ${\mathscr D}$ is defined in (\ref{D_def}) and
\begin{eqnarray*}
& {\mathscr A}_a = e^{\frac{i}{2}(\theta_a-\phi_a)}
\left\{ \left( 1 + \frac{q_z}{\kappa_a} \right) \left[
\cos \left( \theta_b + \frac{\theta_a + \phi_a}{2} \right) \hspace*{0mm}\right. \right. \\
& \left. \left. - \cos \phi_b \cos \frac{\theta_a + \phi_a}{2} \right]
+ \left( \frac{\kappa_b}{\kappa_a} + \frac{q_z}{\kappa_b} \right)
\sin\phi_b \sin \frac{\theta_a + \phi_a}{2} \right\} \ , \\
&{\mathscr B}_a = e^{\frac{i}{2}(\theta_a + \phi_a)}
\left\{ \left( 1 - \frac{q_z}{\kappa_a} \right) \left[
\cos \left( \theta_b + \frac{\theta_a - \phi_a}{2} \right)
\hspace*{0mm}\right. \right. \\
& \left. \left. - \cos \phi_b \cos
\frac{\theta_a - \phi_a}{2} \right]
- \left( \frac{\kappa_b}{\kappa_a} - \frac{q_z}{\kappa_b} \right)
\sin\phi_b \sin \frac{\theta_a - \phi_a}{2} \right\} \ , \\
& {\mathscr A}_b = e^{\frac{i}{2}(\theta_b-\phi_b)}
\left\{ \left( 1 + \frac{q_z}{\kappa_b} \right) \left[
\cos \left(\theta_a + \frac{\theta_b + \phi_b}{2} \right)
\hspace*{0mm}\right. \right. \\
& \left. \left. - \cos \phi_a \cos \frac{\theta_b + \phi_b}{2} \right]
+ \left( \frac{\kappa_a}{\kappa_b} + \frac{q_z}{\kappa_a} \right)
\sin\phi_a \sin \frac{\theta_b + \phi_b}{2} \right\} \ , \\
& {\mathscr B}_b = e^{\frac{i}{2}(\theta_b+\phi_b)} \left\{
\left( 1 - \frac{q_z}{\kappa_b} \right)  \left[
\cos\left(\theta_a + \frac{\theta_b - \phi_b}{2} \right)
\hspace*{0mm}\right. \right. \\
& \left. \left. - \cos \phi_a \cos \frac{\theta_b - \phi_b}{2} \right]
- \left( \frac{\kappa_a}{\kappa_b} - \frac{q_z}{\kappa_a} \right)
\sin\phi_a \sin \frac{\theta_b - \phi_b}{2} \right\} \ .
\end{eqnarray*}
It can be seen that ${\mathscr B}_a$ is obtained from ${\mathscr A}_a$
(or {\em vice versa}) by changing the signs of the propagation
constants $\kappa_a$ and $\kappa_b$ and of the related phases $\phi_a$
and $\phi_b$ (but not of $\theta_a$ and $\theta_b$), and similarly for
the pair of coefficients ${\mathscr A}_b$, ${\mathscr B}_b$. Also, the
coefficients are invariant under the the permutation of indices
$a \leftrightarrow b$.

\section{Closed-form expressions for effective medium parameters obtained
  by current-driven homogenization}
\label{app:B}

In this Appendix we give the closed-form solution for the general case
$a\neq b$. If $a=b$, these expressions are significantly
simplified. However, in the numerical codes used to produce the
figures for this paper, we have used the general expressions given
below. 

In addition to (\ref{wv_defs}) and (\ref{eps_mu_hmg}), we need to
introduce the following notations:

\begin{itemize}

\item[(i)] The phase shifts $\phi_a$ and $\phi_b$ computed at $k_x=0$ are
denoted by $\psi_a$ and $\psi_b$:
\begin{equation*}
\psi_a   = \left. \phi_a \right\vert_{k_x=0} = k_a a \ , \ \ \psi_b
= \left. \phi_b \right \vert_{k_x=0} = k_b b \ .
\end{equation*}

\item[(ii)] The refractive indices of each layer are denoted by $n_a =
  \sqrt{\epsilon_a}$ and $n_b = \sqrt{\epsilon_b}$. The branches of
  all square roots are defined by the condition $0 \leq {\rm
    arg}(\sqrt{z}) < \pi$.

\item[(iii)] The dimensional size parameters for the lattice:
\begin{equation*}
x_a = k_0 a \ , \ \ x_b = k_0 b \ , \ \ x = x_a + x_b = k_0 h \ .
\end{equation*}
Note that $\psi_a = n_a x_a$, $\psi_b = n_b x_b$.

\item[(iv)] The following symmetric combinations of the trigonometric
  functions:
\begin{align*}
& \alpha_1 = 2 \left( n_a \sin \psi_a + n_b \sin \psi_b \right) \ , \\
& \alpha_2 = 2 \left( n_a \epsilon_b \sin \psi_a + n_b \epsilon_a \sin \psi_b \right) \ , \\
& \alpha_3 = x_a \cos \psi_a + x_b \cos \psi_b \ , \\
& \alpha_4 = x_b^2 n_a \sin \psi_a + x_a^2 n_b \sin \psi_b \ , \\
\end{align*}
\noindent
and
\begin{align*}
\xi_1 & = n_a \cos\frac{\psi_a}{2}\sin\frac{\psi_b}{2}
      + n_b \cos\frac{\psi_b}{2}\sin\frac{\psi_a}{2} \ , \\
\xi_2 & = n_a \cos\frac{\psi_b}{2}\sin\frac{\psi_a}{2}
      + n_b \cos\frac{\psi_a}{2}\sin\frac{\psi_b}{2} \ , \\
\xi_3 & = n_a \sin\psi_a \cos\psi_b + n_b \sin\psi_b \cos\psi_a \ , \\
\xi_4 & = n_a n_b \frac{\epsilon_a + \epsilon_b}{2} \left(\psi_a \sin\psi_a
  \cos\psi_b + \psi_b \sin\psi_b \cos\psi_a \right) \\
      & + \epsilon_a \epsilon_b \left( \psi_b \sin\psi_a \cos\psi_b +
  \psi_a \sin\psi_b \cos\psi_a \right) \ , \\
\xi_5 & = (\epsilon_a - \epsilon_b)^2
\sin\frac{\psi_a}{2}\sin\frac{\psi_b}{2} \ , \\
\end{align*}
\noindent
and
\begin{align*}
\eta & = \epsilon_a (2 + \cos\psi_b)\sin^2\frac{\psi_a}{2} +
         \epsilon_b (2 + \cos\psi_a)\sin^2\frac{\psi_b}{2} \\
      & + n_a n_b  \sin\psi_a \sin\psi_b \ , \\
\zeta & = \epsilon_a \left(4 \psi_b \epsilon_b - \psi_b \epsilon_a + 3 \psi_a
  n_a n_b \right) \sin^2 \frac{\psi_a}{2}\sin\psi_b \\
      & + \epsilon_b \left(4 \psi_a \epsilon_a - \psi_a \epsilon_b + 3 \psi_b
  n_a n_b \right) \sin^2 \frac{\psi_b}{2}\sin\psi_a \\
      & + 8(\epsilon_a - \epsilon_b)^2
\sin^2 \frac{\psi_a}{2}
\sin^2 \frac{\psi_b}{2} \\
      & - \left(
\psi^2_a\epsilon^2_b \sin^2 \frac{\psi_b}{2} +
\psi^2_b\epsilon^2_a \sin^2 \frac{\psi_a}{2}
\right) \ ,
\end{align*}
\noindent
and the following combinations of the functions introduced above:
\begin{align*}
\sigma  & = \epsilon_a \epsilon_b (\alpha_4 - 4\alpha_3) - \alpha_2 \
, \\
\tau    & = 4\left( \xi_3 \xi_5 + \xi_2 \xi_4 \right) \ .
\end{align*}
Note that by symmetry we mean here invariance with respect to
permutation of indexes $a \leftrightarrow b$.

\end{itemize}

Then the cosed-form expressions for the current-driven effective
parameters of the layered medium considered in this paper can be
stated as follows [we adduce the expressions for $\beta_{xx}$ and
$\beta_{zz}$; $\mu_{xx}$ and $\mu_{zz}$ are given by
(\ref{mu_eff_beta})]:
\begin{align*}
& \epsilon_{yy} = \frac{x}{Z} (\epsilon_a \epsilon_b)^2 {\mathscr D}_0 \ , \\
& \beta_{xx} = \frac{2 \xi_1^2}{Z^2}   (\epsilon_b - \epsilon_a)^2
 \left[ \zeta - x^2 p_a p_b \epsilon_a \epsilon_b \eta \right] \ , \\
& \beta_{zz} = - \frac{\xi_1}{Z^2}(\epsilon_b - \epsilon_a)^2  \nonumber \\
& \times
\left\{ n_a n_b \left[2\rho + x \xi_2 (\sigma - \alpha_1
    \epsilon_\parallel )  \right] - 2\alpha_1 \xi_5 + \tau \right\} \ ,
\end{align*}
\noindent
where
\begin{equation*}
Z = x \frac{(\epsilon_a \epsilon_b)^2}{\epsilon_\perp} {\mathscr D}_0  -
  4 \xi_1 \xi_5
\end{equation*}
\noindent
and ${\mathscr D}_0$ is the determinant (\ref{D_def}) evaluated at
${\bf k}=0$:
\begin{align*}
{\mathscr D}_0 & = 1 - \left. \cos (q_z h) \right \vert_{k_x=0}  \nonumber \\
  & = 1 - \cos\psi_a\cos\psi_b + \frac{1}{2}\left( \frac{n_a}{n_b} +
  \frac{n_b}{n_a} \right)\sin \psi_a \sin \psi_b \ ,
\end{align*}
\noindent
and, finally, 
\begin{align*}
\rho    & = \epsilon_a \epsilon_b \xi_2 x^2 (2 + p_a p_b {\mathscr D}_0) \\
        & - n_a n_b x \sin \frac{\psi_a}{2}\sin\frac{\psi_b}{2}
        \left(\epsilon_a \epsilon_b x^2 + 2 \epsilon_\parallel
          {\mathscr D}_0
          \right) \ .
\end{align*}

\section{S-parameter retrieval techniques used in this paper}
\label{app:C}

S-parameter retrieval is a well-established technique. However, the
vast majority of papers that consider S-parameter retrieval are
focused on normal incidence only. This limitation does not allow one
to access all tensor elements of the effective parameters. In this
Appendix, we will remove this limitation and include off-normal
incidences into consideration.

Consider an anisotropic homogeneous slab characterized by the tensors
$\epsilon = {\rm diag}(\epsilon_\perp, \epsilon_\perp,
\epsilon_\parallel)$ and $\mu = {\rm diag} (\mu_\perp, \mu_\perp,
\mu_\parallel)$. The transmission and reflection coefficients are
given in Sec.~\ref{sec:TR}. It is convenient to rewrite the
expressions given in that section in the following form:
\begin{subequations}
\label{TR_Cp}
\begin{align}
& T = \frac{4 p C}{(C+1)^2 - p^2 (C-1)^2} \ , \\
& R = \frac{(1-p)^2(C^2 -1)}{(C+1)^2 - p^2 (C-1)^2} \ .
\end{align}
\end{subequations}
Here 
\begin{equation}
\label{C_p_def}
C = \frac{q_z \eta_\parallel}{\kappa_0} \ , \ \ p=\exp \left(i q_z L \right) \ .
\end{equation}
\noindent
We refer the reader to Sec.~\ref{sec:TR} for relevant notations. Note
the symmetries $R(p,C) = - R(p,1/C)$ and $T(p,C) = T(p,1/C)$.

If $T$ and $R$ are known at some incidence angle (parameterized by
$k_x$), so are $C$ and $p$. The expressions for $C$ and $p$ in terms
of $T$ and $R$ [inversion of (\ref{TR_Cp})] is unique up to the branch
of a square root and widely known:
\begin{equation*}
C = \frac{\pm D}{T^2 - (1-R)^2} \ , \ \ p= \frac{1 + T^2 - R^2 \pm
  D}{2 T} \ ,
\end{equation*}
\noindent
where
\begin{equation*}
D = \sqrt{(1 + T^2 - R^2)^2 - (2T)^2} \ .
\end{equation*}
\noindent
In the case of low transmission (small $\vert T \vert$), one can use
the approximate formulas
\begin{align*}
 & C = \ \ \frac{1 + R}{1 - R} \ , \ \  p = \frac{T}{1 - R^2} \ \ {\rm
   (first \ branch)} \ , \\
 & C = -\frac{1 + R}{1 - R} \ , \ \  p = \frac{1 - R^2}{T} \ \ {\rm
   (second \ branch)} \ .
\end{align*}
\noindent
Here we have encountered the first instance of a branch ambiguity, but
it can be easily resolved by using the condition of medium passivity,
$\vert p \vert < 1$.  Other than the branch ambiguity mentioned, the
above inversion formulas (that yield $C$ and $p$ in terms of $T$ and
$R$) are well-posed and stable and establish one-to-one correspondence
between the complex pairs $(T,R)$ and $(C,p)$.

Given the above result, we can assume that $C(k_x)$ and $p(k_x)$ are
known as functions of $k_x$ and seek effective tensors $\epsilon$ and
$\mu$ that are consistent with these functions. In general, purely
local tensors $\epsilon$ and $\mu$ that ``reproduce'' two given
functions $C(k_x)$ and $p(k_x)$ may not exist. Therefore, we shall
focus on a more narrow task of finding $\epsilon$ and $\mu$ that
reproduce $C(k_x)$ and $p(k_x)$ at normal ($k_x=0$) and
close-to-normal incidences. 

It is well known that, if we consider normal incidence only (or any
other fixed value of $k_x$), then the solution is not unique due to
the branch ambiguity of the logarithm function. More specifically, the
quantities $C$ and $p$ given by (\ref{C_p_def}) are invariant with
respect to the simultaneous transformation $q_z \rightarrow q_z + 2\pi
m / L$, $\eta_\parallel \rightarrow q_z / (q_z + 2\pi m / L)$, where
$m$ is an integer. We therefore will include both normal and
off-normal incidences into consideration. Unfortunately, the problem
is still ill-posed in this case. Generally, the retrieval problem can
not be formulated as a well-posed system of equations. We shall now
describe three different approaches to obtaining a solution to the
retrieval problem that is optimal in the sense that it yields all
relevant elements of the tensors $\epsilon$ and $\mu$ and reproduces
the angular dependence $C(k_x)$ and $p(k_x)$ [or $T(k_x)$, $R(k_x)$]
in as wide range of $k_x$ as possible.

In what follows, we wasume that the effective medium is a uniaxial
crystal with $\epsilon = {\rm diag}(\epsilon_\parallel,
\epsilon_\parallel, \epsilon_\perp)$ and $\mu = {\rm
  diag}(\mu_\parallel, \mu_\parallel, \mu_\perp)$. Also, the notations
$\eta$ refers to $\mu$ for s-polarization and to $\epsilon$ for
p-polarization.

\subsection*{Method 1}

Let $t=k_x/k_0$ and $x=k_0L$. Given the dispersion equation
(\ref{q_hmg}), we can write
\begin{equation}
\label{c1}
C(t)=\frac{\sqrt{1-t^2}\eta_\parallel}{Q(t)} \ , \ \ p(t) = e^{ix Q(t)}
\ ,
\end{equation}
\noindent
where
\begin{equation*}
Q(t) = \sqrt{n^2 -\frac{\eta_\parallel}{\eta_\perp} t^2} \ .
\end{equation*}
\noindent
Here $n^2 = \epsilon_\parallel \mu_\parallel$ is the squared index of
refraction and recall that $\eta$ refers to $\mu$ for s-polarization
and to $\epsilon$ for p-polarization.

Assuming $C(t)$ and $p(t)$ are known functions, we can express $Q(t)$
by inverting each equation in (\ref{c1}). This yields two solutions
for $Q(t)$:
\begin{equation*}
Q_1(t) = \frac{\eta_\parallel \sqrt{1-t^2}}{C(t)} \ , \ \ Q_2(t) =
\frac{1}{ix} \ln \left[p(t)\right] + \Delta m \ , 
\end{equation*}
\noindent
where $\Delta = 2\pi/x$ and $m$ an arbitrary integer.

If we could find such tensors $\epsilon$ and $\mu$ that $Q_1(t) =
Q_2(t)$ for all values of $t$, then these parameters would describe
transmission and reflection by the slab for all angles of incidence
with perfect precision. This is possible only in the $h\rightarrow 0$
limit. Here we will require that the functions coincide at normal
incidence and have the same second derivative with respect to $t$ (the
first derivative is identically zero). To this end, it is convenient
to introduce the functions
\begin{equation*}
F(t) = \frac{1-t^2}{C(t)} \ , \ \ G(t) = \frac{1}{ix}\ln \left[ p (t)
\right] \ .
\end{equation*}
\noindent
Since these functions are expressed in terms of $C(t)$, $p(t)$ and
known analytical functions (they do not contain any unknowns), we can
view them as directly measurable (or computable in terms of $T$ and
$R$). Then the main equation we wish to fit takes the form
\begin{equation*}
\eta_\parallel F(t) = G(t) + \Delta m \ . 
\end{equation*}
\noindent
Here $\eta_\parallel$ and $m$ (the branch index) are the unknowns. Of
course, this equation has no solutions in general. But we can require
that it holds to second order in $t$ in the vicinity of $t=0$. We can
expand $F(t)$ as
\begin{equation*}
F(t) = F_0 + F_2 t^2 \ , \ \ F_0 = F(0) \ , \ \ F_2 =
\lim_{\tau\rightarrow 0} \frac{F(\tau) - F_0}{\tau^2} \ .
\end{equation*}
\noindent
and similarly for $G(t)$. This results in a pair of algebraic
equations
\begin{align*}
& \eta_\parallel F_0 = G_0 + \Delta m \ , \\
& \eta_\parallel F_2 = G_2 \ .
\end{align*}
\noindent
Even though this is a system of two linear equations with respect to
two unknowns, it still cannot be solved because $m$ is, by definition,
integer. We can, however, solve the first equation exactly (this will
guarantee the correct $T$ and $R$ at normal incidence) and then
minimize the norm of the second equation. This results in the
following inverse solution:
\begin{subequations}
\label{meth_1}
\begin{align}
\label{meth_1_m}
& m = {\rm Nint}\left[ \frac{F_0 G_2 - G_0 F_2}{\Delta F_2} \right] \ , \\
\label{meth_1_eta}
& \eta_\parallel = \frac{G_0 + \Delta m}{F_0} \ ,
\end{align}
\end{subequations}
\noindent
where ${\rm Nint}[z]$ denotes nearest integer to the real part of
$z$. Note that the second expression in the above (for
$\eta_\parallel$) must use the value of $m$ computed using the first
equation.

Once the quantities $\eta_\parallel$ and $m$ are known, we can also
compute the squared refractive index according to
\begin{equation*}
n^2 = (G_0 + \Delta m)^2
\end{equation*}
\noindent
Then, if $\eta_\parallel = \mu_\parallel$ (s-polarization), we can
compute $\epsilon_\parallel = n^2/\eta_\parallel$. Otherwise, if
$\eta_\parallel = \epsilon_\parallel$, we can compute $\mu_\parallel =
n^2/\eta_\parallel$. 

This Method 1 does not give access to $\epsilon_\perp$ and
$\mu_\perp$. To obtain these tensor elements, Method 3 must be used.

\subsection*{Method 2}

This method combines Method 1 with the main idea of
Ref.~\onlinecite{chen_04_1}. Namely, starting from some sufficiently
low frequency $\omega$ or some sufficiently small $h$, we gradually
increase the relevant parameter ($\omega$ or $h$) and apply Method 1
at each iteration of this loop. However, after a few initial
iterations, we change the rule according to which the branch index $m$
is computed. Namely, instead of using (\ref{meth_1_m}), we chose the
index $m$ in such a way as to minimize the jump in the refractive
index $n$. By ``jump'' we mean here $\vert n_i - n_{i-1} \vert$, where
$i$ is the iteration index.

\subsection*{Method 3}

This method is free from the branch ambiguity but does not guarantee
exact fitting of $T$ and $R$ at normal incidence. We will fit the the
index of refraction using second and forth normalized derivatives of
$p(t)$ and then use the function $C(t)$ to find the impedance.

We start by noting the following relations:
\begin{subequations}
\label{meth_3_F2_F4}
\begin{align}
\label{meth_3_F2}
  & F_2 \equiv \frac{p^{\prime \prime}(0)}{p(0)} = -ix
  \frac{\eta_\parallel}{n
    \eta_\perp} \ , \\
\label{meth_3_F4}
  & F_4 \equiv \frac{p^{\prime \prime \prime \prime}(0)}{p(0)} = -3x
  \frac{i + x n}{n} \left(\frac{\eta_\parallel}{n \eta_\perp}\right)^2
  \ .
\end{align}
\end{subequations}
\noindent
We can exclude the ratio $\eta_\parallel/ \eta_\perp$ from the above
set of linear equations and solve for the index of refraction, viz,
\begin{equation*}
n = \frac{i}{x(F_4/3F_2^2 - 1)} \ .
\end{equation*}
\noindent
This gives the index of refraction in terms of the ``measurables''
$F_2$ and $F_4$. We can relate $F_2$ and $F_4$ to ``measurements'' of
$p(t)$ at some small but nonzero values of $t$, say, $\tau_1$ and
$\tau_2$, as follows:
\begin{equation*}
F_2 = \frac{\tau_2^2 b_1 - \tau_1^2 b_2}{\tau_2^2 - \tau_1^2} \ , \
F_4 = 12 \frac{b_2 - b_1}{\tau_2^2 - \tau_1^2} \ ,
\end{equation*}
\noindent
where
\begin{equation*}
b_k = \frac{2}{\tau_k^2} \left[ \frac{p(\tau_k)}{p(0)} - 1 \right] \ ,
\ \ k=1,2  \ .
\end{equation*}
At this point we have found the index of refraction, $n$. We then
compute $\eta_\parallel$ and $\eta_\perp$ from
\begin{equation*}
\eta_\parallel = C(0) n \ , \ \ \eta_\perp =-i \frac{xC(0)}{F_2} \ .
\end{equation*} 
\noindent
In the second equation above, we have used $\eta_\parallel/\eta_\perp
= i nF_2 / x$, as follows from (\ref{meth_3_F2}).

By considering both s- and p-polarizations, we can find all elements
of the effective tensor. 

\end{document}